# Deep Learning Denoising of Real SWOT Sea Surface Height Observations

Gaétan Meis,[a] Anaëlle Tréboutte,[a] Maxime Ballarotta,[a] Marie-Isabelle Pujol,[a] Gérald Dibarboure[b]

[a] *Collecte Localisation Satellites (CLS), Ramonville-Saint-Agne, France*

[b] *Centre National d'Études Spatiales (CNES), Toulouse, France*

*Corresponding author*: Gaétan Meis, gmeis@groupcls.com

ABSTRACT

The SWOT (Surface Water Ocean Topography) mission is currently providing unpreceded high-resolution measurements of Sea Surface Height (SSH), revealing ocean features at finer scales. Nevertheless, the two-dimensional observations of KaRIn altimeter of SWOT suffer from instrumental errors. This noise degradation is altering the high frequencies of SWOT signal, and the small to sub-mesoscale dynamics of interest for some oceanographers. For this reason, Tréboutte et al. (2023) have developed a convolutional neural network (CNN) based on U-Net architecture to separate the noise from the physical signals contained in the SSH. Their approach has demonstrated great potential on simulated SWOT measurements, and Dibarboure et al. (2024) report a positive influence on actual flight data from SWOT. However, degraded denoising performance, and occasional negative side-effects have been observed in atypical conditions (e.g. very high surface waves, internal tides solitons). In this study, we illustrate some of these limitations and we present an improved approach of the CNN-based denoising: we modified the training procedure to obtain a more robust version of the algorithm, to avoid biases and artifacts in the denoised SSH. This paper also presents a more complete validation process with a robust and standardized evaluation benchmark: these metrics could be of interest to assess other SWOT filtering and denoising algorithms.

## 1. Introduction

Satellite altimetry has been completely revolutionized by the arrival of Surface Water and Ocean Topography satellite (SWOT), (Fu et al. 2024). This mission was developed by the Centre National d'Études Spatiales (CNES), the National Aeronautics and Space Administration (NASA) and the United-Kingdom and Canadian Space Agencies. Since December 2022, SWOT has been producing the first 2D high-resolution images of the Sea Surface Height (SSH). These images allow oceanographers to observe global ocean dynamics ranging from large and medium mesoscale and to small and sub-mesoscale, thanks to the new Ka-band Radar Interferometer (KaRIn) instrument that measures 120-km wide 2D swath observations at 250-m and 2-km resolution over the global ocean. KaRIn's unprecedented precision was designed to capture oceanic structures down to 15 km (Fu, 2018) but the first studies shows that even smaller structures could be observed (Wang et al. 2025). This precision also reveals other fine-scale dynamics such as internal waves, submesoscale eddies or coastal flows (Dibarboure et al. 2025; Archer et al. 2025; Fu et al. 2024).

However, KaRIn's noise increases with surface waves (Peral et al. 2024), sometimes altering SSH content at higher frequencies. In this context, even precise KaRIn SSH measurements might benefit from a so-called "denoising" filter that intends to isolate only random noise, and thus to retrieve smaller physical content. This type of problem is well known, and convolutional neural networks (CNN) have shown unpreceded results (Zhang et al., 2022, 2016). A first CNN filter dedicated to SWOT/KaRIn has been developed by Tréboutte et al. (2023). Even if this network was trained with simulated SWOT data, it has demonstrated a good capability to absorb noise on flight data from SWOT in a wide range of ocean conditions (Dibarboure et al. 2025). This U-Net based filter was implemented and operated over the global ocean for 2+ years in the *KaRIn Level-3 v2.0 product* (AVISO/DUACS, 2024a, 2024b). Despite its beneficial denoising properties in nominal conditions, the community reported some negative side-effects in atypical conditions such as unusually high-wave conditions, internal tides solitons, or spurious pixels from rain cells.

In this context, this paper present here a revised version of the U-Net filter from Tréboutte et al. (2023), with the same architecture and an improved training strategy. The revised filter is not only developed to the 2-km KaRIn SSH, but also to the 250-m KaRIn SSH (so-called Unsmoothed L3 Product). Because no ground-truth is available to fully demonstrate that the denoiser behaves as expected, we also developed an extensive evaluation benchmark to assess various properties of SWOT filtering and denoising algorithms (Appendix A).

The paper is organized as follows: Section 2 presents the input SWOT data and the method used to train the neural network. Section 3 gives an overview of our keys results about the denoising properties of SWOT LR data. In Section 4, we discuss the influence of supervised learning based on an ocean simulation model. Then the limitations of denoising filters and various perspectives are discussed in Section 5. After the conclusion in Section 6, a detailed description of the generic evaluation benchmark is given in Appendix A. Lastly, for the sake of traceability, Appendix B provides a side-by-side comparison between our denoising filter and the original one from Tréboutte et al. (2023, 2024) on the evaluation benchmark

## 2. Data & Method

*a. Input data from SWOT/KaRIn*

The objective of the study is to propose an improved denoising strategy for the SWOT LR 2-km and 250-m products. The calibrated and edited Sea Surface Height Anomaly (SSHA) of the SWOT Level-3 Ocean products v2.0 (AVISO/DUACS, 2024a) is used as inputs for the 250-m resolution. The calibrated and edited Sea Surface Height Anomaly (SSHA) of the SWOT Level-3 Ocean products v3.0 (AVISO/DUACS, 2024b) is used as inputs for the 2-km resolution. The editing applied is keeping only the plain ocean SSHA, with eclipses events included because they do not impact the quality of SSHA (AVISO/DUACS, 2025; Dibarboure et al., 2025). Land, ice, coastal pixels and outliers are ignored in *ssha_unfiltered* data by keeping only the flagged data below or equal to 3 in the *quality_flag*.

An arbitrary example of SWOT Level-3 Ocean SSHA without denoising is shown in Figure 1 (2-km resolution) and Figure 2 (250-m resolution). Despite the unprecedented resolution of KaRIn, these examples located in the Gulf Stream clearly illustrate the negative side effect of random noise on velocity and relative vorticity[1] derived from SSHA, especially at the higher 250-m resolution, as explained by Chelton (2024). Note that the example is in a region where the signal to noise ratio is generally much better than other regions where the random noise can be much larger (e.g. Southern Ocean) or the SSH dynamics might be much weaker (e.g. Mediterranean Sea), or both.

Noise-mitigation is an important <u>but optional</u>[2] processing step of the Level-3 sequence from Dibarboure et al. (2024). It comes after the multi-mission calibration and after the editing algorithm. The former mitigates KaRIn systematic errors (e.g. long-wavelength errors from satellite roll) and makes output data more consistent with other sensors. The latter takes out spurious or suspicious pixels caused by spacecraft events, rain cells, sea-ice, boats, land... More details about these two steps are given in Dibarboure et al. (2024) and in the product handbook of the Level 3 product (AVISO/DUACS, 2025).

[1] In this study, relative vorticity is dimensionless, as it is normalized by the Coriolis parameter f.

[2] The L3 product contains one SSHA variant where the denoising filter is applied, and another version where it is not.

The objective of the denoising filter is to isolate and mitigate only random errors: its role is not to mitigate geographically correlated geophysical errors such as tides or wet troposphere residues. Note that the properties of the total measurement error in real SWOT data (i.e. random, systematic, geophysical) might differ from those predicted by the SWOT simulator (Gaultier et al., 2016) also used by Tréboutte et al. (2023). Investigations on the 2-km product have shown that the KaRIn real noise is millimetric and correlated (Dibarboure et al., 2025) whereas the simulated noise used by in Tréboutte et al. (2023) to train their model was centimetric and geographically decorrelated. In the 250-m L3 SSHA products, the noise is almost gaussian and decorrelated (Tréboutte et al., 2024) with an amplitude of a few centimeters. The noise level depends on the Significant Wave Height (SWH) and the cross-track distance (Peral et al., 2014). The edges of both swaths and the regions with higher waves may have a random noise 10+ times larger than the inner swath in low wave conditions.

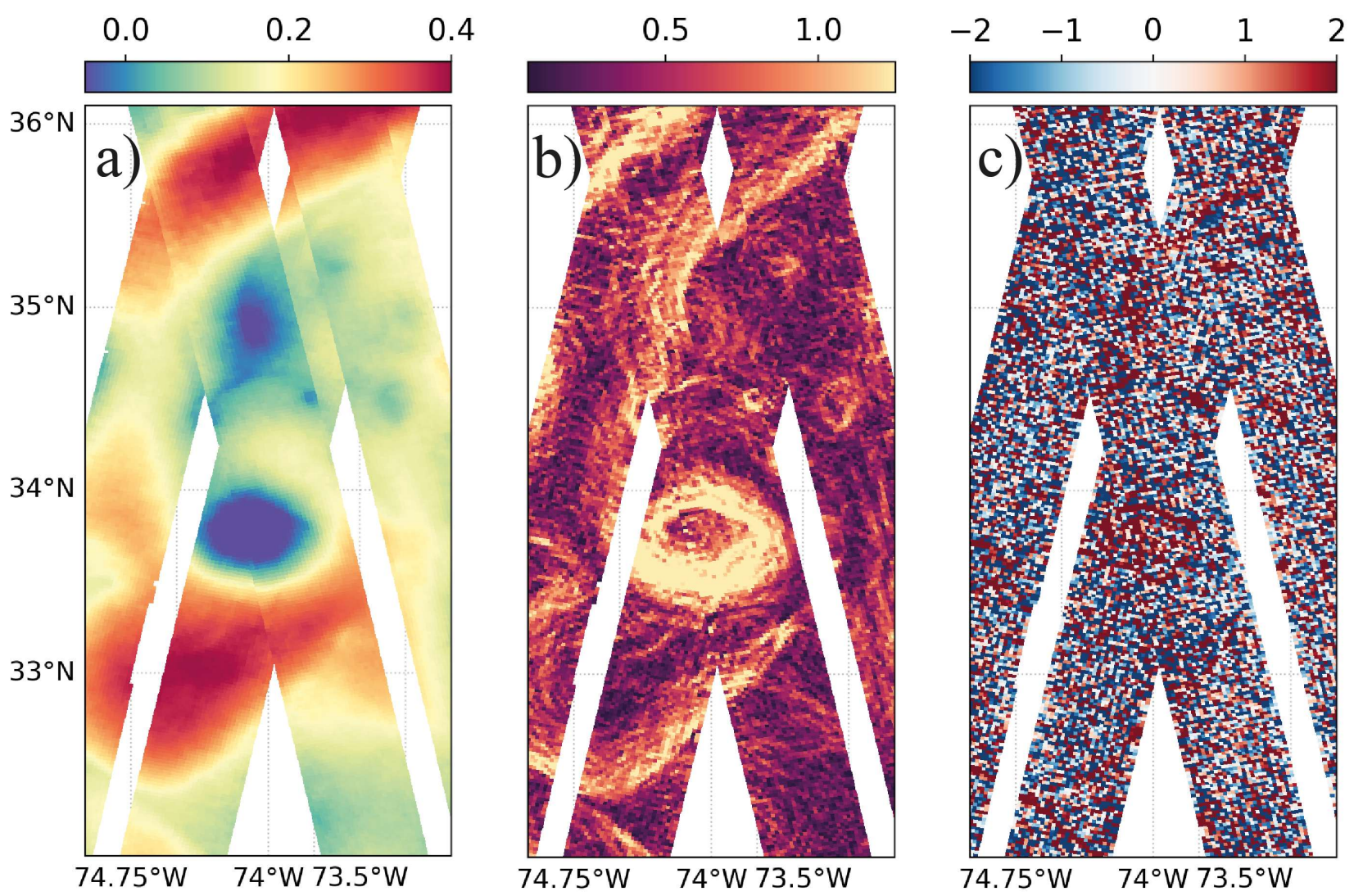

Figure 1. Effect of random noise on 250-m SWOT LR SSHA (unit: m), velocity (unit: m.s$^{-1}$) and relative vorticity. From cycle 489, passes 9 and 22.

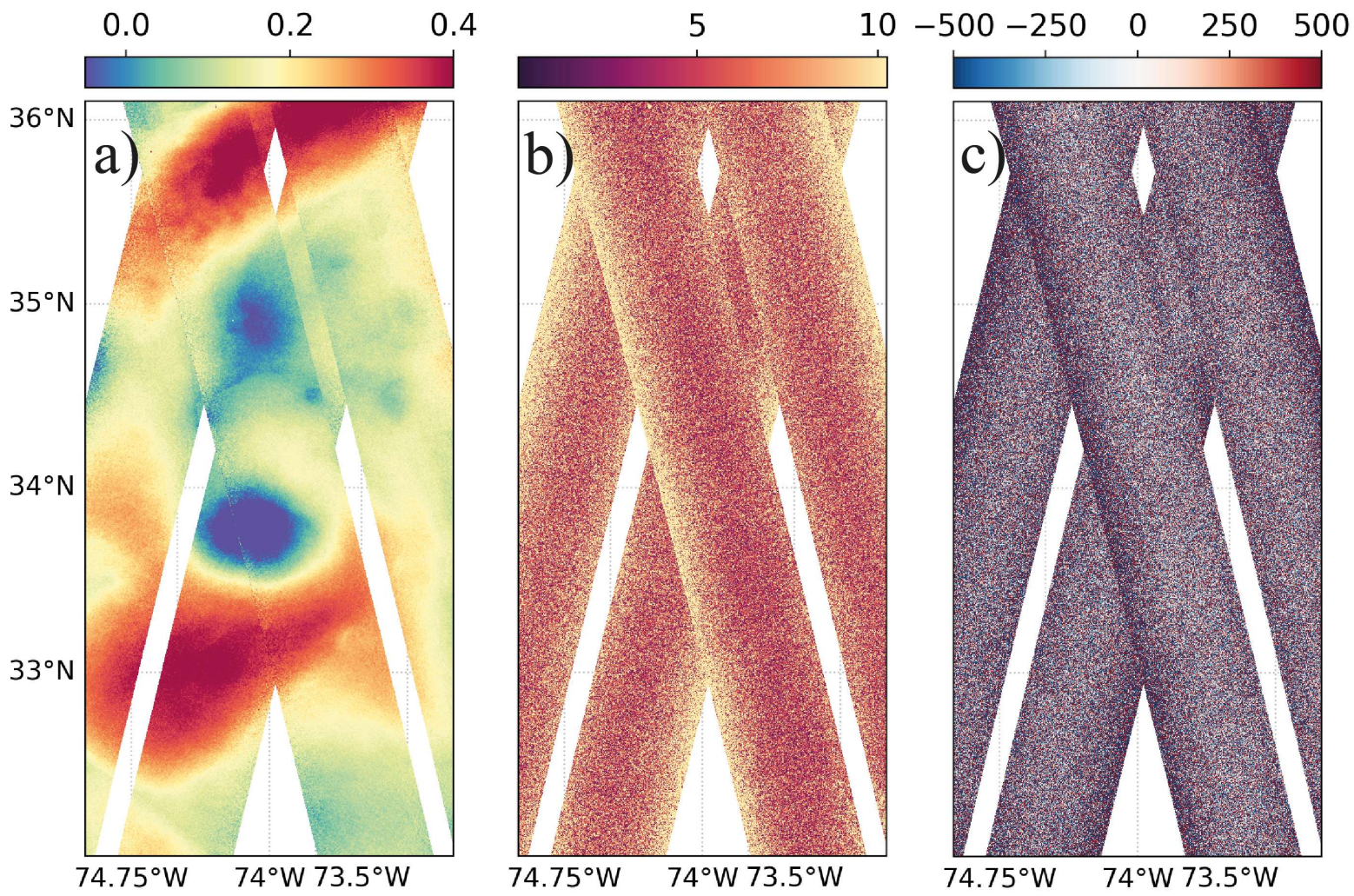

Figure 2. Effect of random noise on 250-m SWOT LR SSHA (unit: m), velocity (unit: $m.s^{-1}$) and relative vorticity. From cycle 489, passes 9 and 22.

*b. Deep learning denoising approach*

This section describes the Convolutional Neural Network (CNN) architecture and the training methodology designed for the denoising of SWOT L3 LR products.

The noise-mitigation algorithm used in this study is a U-Net (Ronneberger et al., 2015). It is composed of an encoder part and a decoder part. Both consist of 3 double-convolution blocks, each comprising two convolutional layers followed by a ReLU activation function (Figure 3). The filter numbers of each block are 16, 32 and 64. The U-Net model for the 2-km denoising consists of 231 000 parameters in total, which weigh only 3Mb. Keeping the historical U-Net in Tréboutte et al. (2023) with 3 levels of depths was chosen because it already proved its robustness and efficiency to treat SWOT data. The interpolation of 2-km dataset to 250-m resolution is multiplying by 8 the size of the total training data. The U-Net model needed to be scaled proportionally too. With two more levels of depths and a bigger convolutional kernel size, the network for 250-m denoising consists of approximately 2.5M parameters (~30 Mb).

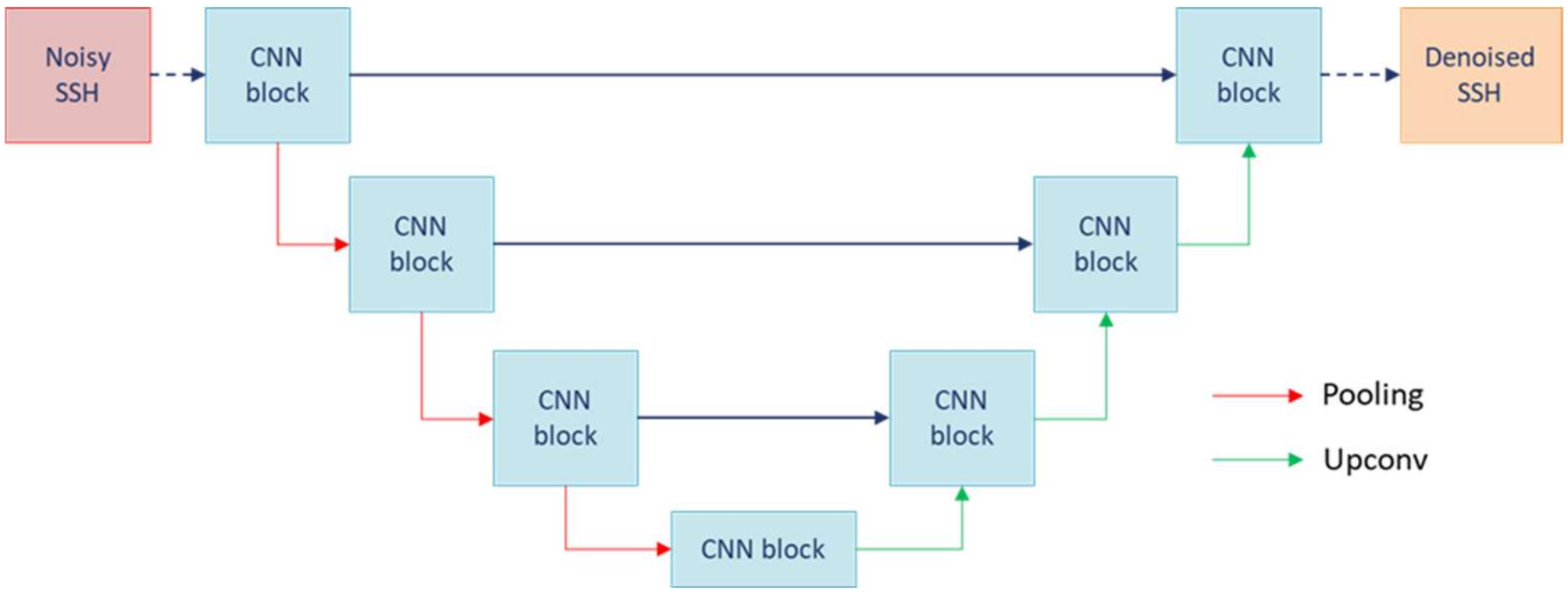

Figure 3. Architecture of the U-Net used for removing the KaRIn noise from a SSH field (from Tréboutte et al., 2023).

To train the network, the learning dataset has been improved to better mimic real SWOT SSH observations at 2 km and 250 m (see Section 1.2.4). The training loss is a L1 loss between the denoised SSH and the ground truth only on valid values above the ocean. First and second derivatives of ground truth (GT) are also added in the L1 loss:

$$loss = L1\left(GT, \widehat{\mathrm{SSHA}}\right) + 2.L1\left(\nabla GT, \nabla\widehat{\mathrm{SSHA}}\right) + 2.L1\left(\nabla^2 GT, \nabla^2\widehat{\mathrm{SSHA}}\right)$$

Where $\nabla$ is the spatial gradient function from open source Kornia library (Riba et al., 2020). These complementary loss terms allow the network to reconstruct more accurately SSHA gradients. For the training dataset of the 2-km denoising, horizontal and/or vertical flips are applied as data augmentation to improve results (Taylor and Nitschke, 2017). Therefore, the size of training dataset is multiplied by 4. For the 250-m denoising, data augmentation is not applied because it was found slowing down the convergence without clear improvements. The ADAM optimizer (Kingma and Ba, 2017) is used to accelerate the training process.

*c. Simulated noiseless SSHA data*

Training a neural network with a supervised strategy requires a ground-truth target, and for this filter we use a simulated SSHA derived from a high-resolution ocean model. In this study, we used the North Atlantic Ocean simulation at 1/60° grid resolution (or eNATL60, from Brodeau et al., 2020). This simulation is an eddy-resolving configuration of the NEMO (Nucleus for European Model of the Ocean; Gurvan Madec et al., 2024) ocean model including explicit oceanic tide motion and forced at the surface with 3-hourly era-interim atmospheric forcings. This ocean model is chosen for its good capability to simulate realistic ocean surface dynamics (Brodeau and Le Sommer, 2019) and its high spatial resolution (grid spacing ranging between 0.8km and 1.6km, which corresponds to a model effective resolution of approximately 10 km). More information about this simulation can be found in Ajayi et al. (2020) and Brodeau et al. (2020).

Simulated SWOT data extracted from this eNATL60 simulation are generated using the SWOT simulator (Gaultier et al., 2016). In order to work on a sea surface height anomaly field, a time-averaged SSH is subtracted to the SSH scene ($SSHA = SSH - \overline{SSH}$).

At this point, it is important to underline that using an ocean model simulation to train a denoising filter of actual measurements from SSH has non-trivial implications for oceanographers (e.g. over-fitting, CNN hallucinations, replication of the ocean model training dynamics). This warrants a particular caution that is discussed in Section 4.

To train the U-Net on simulated SSHA, it is crucial to simulate noise that mimics real SWOT noise as closely as possible, at both 2-km and 250-m spatial resolutions. Thus, the generation of the KaRIn noise in the present study differs from the instrumental noise available in the SWOT simulator to be more realistic. Indeed, the real KaRIn noise on the 2-km product is spatially correlated and the SWOT simulator is not able to produce this type of noise. As the noise on the 250-m data is *almost* white (Tréboutte et al., 2024), our process is to use the SWOT simulator at this full resolution, and to mimic the averaging algorithms of SWOT's ground segment (Stiles and Dubois, 2022) that downscale the 250-m data into a 2-km simulated dataset.

The 250-m KaRIn noise is modulated with the cross-track distance and significant wave height. The significant wave height (SWH) used to parameterize the SWOT simulator are sampled from the global ocean reanalysis wave system (WAVERYS) provided by Météo-France at a resolution of 1/5° and 3 hours sampling.

The noise obtained with this method is shown in the Figure 4b and Figure 4f for 2-km resolution and 250-m resolution data, respectively. The amplitude ranges from a few millimeters at 2-km to approximately 10+ cm for the 250-m observations in high-wave conditions.

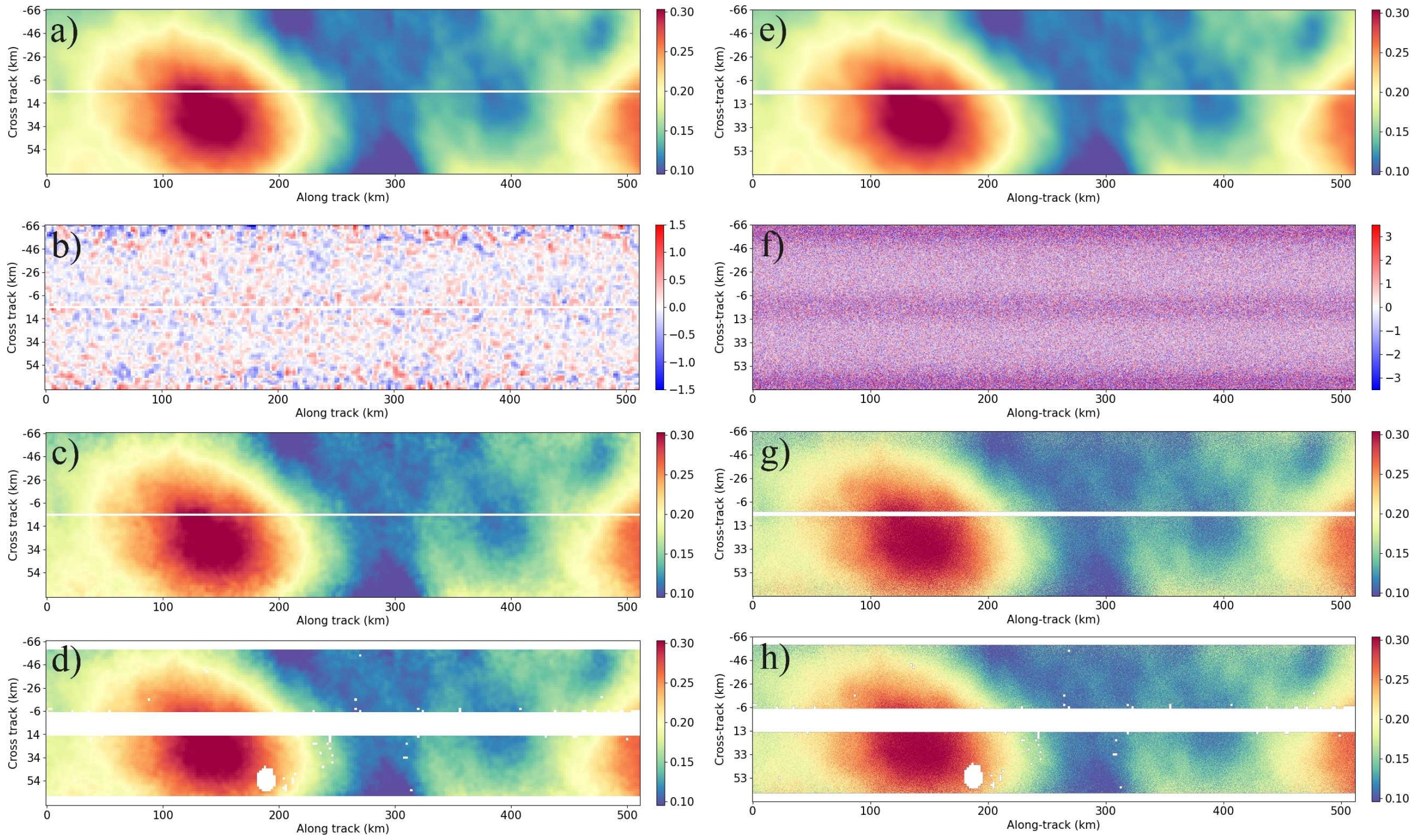

Figure 4. Example of simulated data at both resolutions to generate the training datasets. (a) 2-km ground truth (unit: m). (b) 2-km noise (unit: cm). (c) 2-km noisy SSHA (unit: m). (d) 2-km noisy SSHA with real editing applied (unit: m). (e), (f), (g), (h) are the same quantities respectively for 250-m resolution.

*d. Final simulated realistic SWOT data*

The simulated SWOT noisy SSHA dataset is the sum of the simulated noiseless SSHA and the simulated noise. To be more realistic, a supplemental preprocessing step has also been added. It consists of applying a real editing mask extracted from L3 products to simulate incomplete swaths (Figure 4d, Figure 4h). The masks are taken from the first 24 days of January 2024. On real SWOT swaths, removing data near nadir and on swaths borders is mandatory because the measurements are often too degraded (Dibarboure et al., 2025). Conversely, simulated swaths are clean and defined on the full grid. Editing swaths borders and near nadir is bringing closer the simulation to the real world. This was found to be helping the network adjust its convolutions in the outer edges of each swath.

The last preprocessing step involves the addition of artificial spurious pixels only onto the simulated noisy 250-m SSHA. It is indeed uncommon but not rare that the real L3 v2 editing algorithm misses some outliers (e.g. in coastal regions or rain cells) especially at 250-m. Such spurious pixels can mislead the CNN to propagate incorrect denoised values around it. The rationale (confirmed with sensitivity cases) is that adding simulated spurious pixels to mimic real measured SSHA trains the CNN not to be influenced by their presence.

The simulated dataset is available from 1st July 2009 to 31st July 2010 over the North Atlantic Ocean. The new 2-km and 250-m denoisers are trained on a smaller time span. The dataset is split into training and independent evaluation (75% and 25% respectively). Moreover, the data preprocessing is identical to Tréboutte et al. (2023):

- We remove the time-averaged SSH to get a proxy of the SSH anomaly.
- We split the dataset into 512-km long segments.
- We use a standard scaler on each patch of data.

Because the two models are meant to be in the version 3.0 of SWOT LR products, they will be named accordingly U-Net v3.0 and U-Net v3.0 250-m in this study.

*e. Low-pass filter as baseline for benchmarking*

The difference between denoising and smoothing filters is important to assess in the frame of this study. Smoothing filters absorb and remove high frequency components, following a transfer function. Simple kernel filters (e.g. Gaussian) have a fixed transfer function and smoothing properties in the spectral domain. More sophisticated smoothers (e.g. adaptive) may work in a more complex way (e.g. optimize a local cut-off wavelength based on local signal-to-noise ratio analysis) but the idea remains to remove entirely small wavelength energy where random noise is concentrated. Denoising filters, on the other hand, intends to mitigate the energy of random noise, while keeping the (correlated) signal of interest as untouched as possible. A perfect denoiser would isolate and remove noise entirely and it would not impact the real physical signal. In our context, the goal of our denoising CNN is to keep as much small-scale SSHA content even at high-wavenumbers (small to sub-mesoscale), while removing as much random noise as possible, including their low-wavenumber components.

To provide a baseline for comparison between smoothing and denoising filters, we here use a Lanczos low-pass filter, since it is somewhat common in the oceanography community (e.g. Schlax and Chelton, 1992). This smoothing filter will be used to highlight the substantial differences between denoising and smoothing the KaRIn data. For certain validation metrics, a Butterworth low-pass filter is sometimes preferred because it is significantly faster in terms of computation time. The cutoff wavelength of the filters is constant in time and space. It has been arbitrarily set at 20-km. This value was chosen to be equivalent to the globally averaged cut-off frequency of our denoising filter (Section 3, Figure 7).

## 3. Results

*a. Denoising the real SWOT 2-km product*

This section focuses on the analysis of the key results of U-Net v3.0 denoising performance. We aim to evaluate the U-Net from local and global perspectives, with various qualitative and quantitative metrics. Readers with an interest in supplemental metrics and more technical details of calculation may find them in Appendix A.

For U-Net v3.0, adapting the training dataset made it possible to find removed SSHA content close to random noise, with almost no coherent structure. An example of denoised SSHA swath is shown in Figure 5. No obvious signs of degradation are visible in the denoised SSHA or in the velocity. The large mesoscale eddy in the middle seems well preserved. The 2-km denoising allows velocity and relative vorticity with very little noise to be recovered, revealing the underlying dynamics.

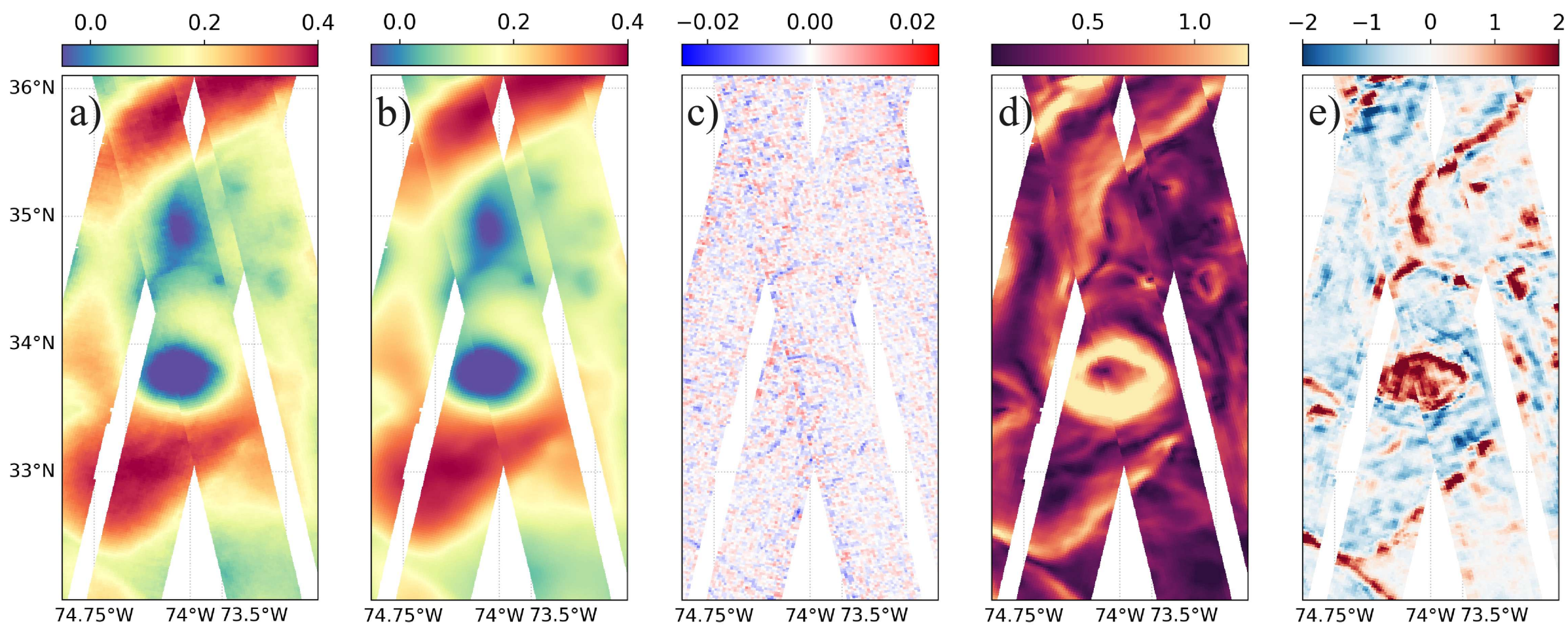

Figure 5. Denoising example of U-Net v3.0 applied on a SSHA crossover, Cycle 489, Passes 9 and 22. (a) SSHA without the denoising filter (unit: m). (b) SSHA with the denoising filter (unit: m). (c) Removed SSHA content (unit: m). (d) Velocity (unit: $m.s^{-1}$). (e) Relative vorticity.

A second example of our denoising filter result is located near South Africa (Figure 6). In their paper, Archer et al., (2025) co-located a very small eddy (approximately 15-km in diameter) observed in SWOT data alongside Sea Surface Temperature (SST) and chlorophyll concentration measurements. In this example, the denoising filter successfully separates the noise from the signal, revealing only noise in the absorbed content but no trace of a physical signal (despite the small size of the structure). This tiny eddy is particularly visible in the velocity field and in the relative vorticity, where its shape and size do not appear altered. This example demonstrates that the denoising filter is capable, at least locally, of restoring oceanic structures up to 15-km in size without absorbing any of their energy.

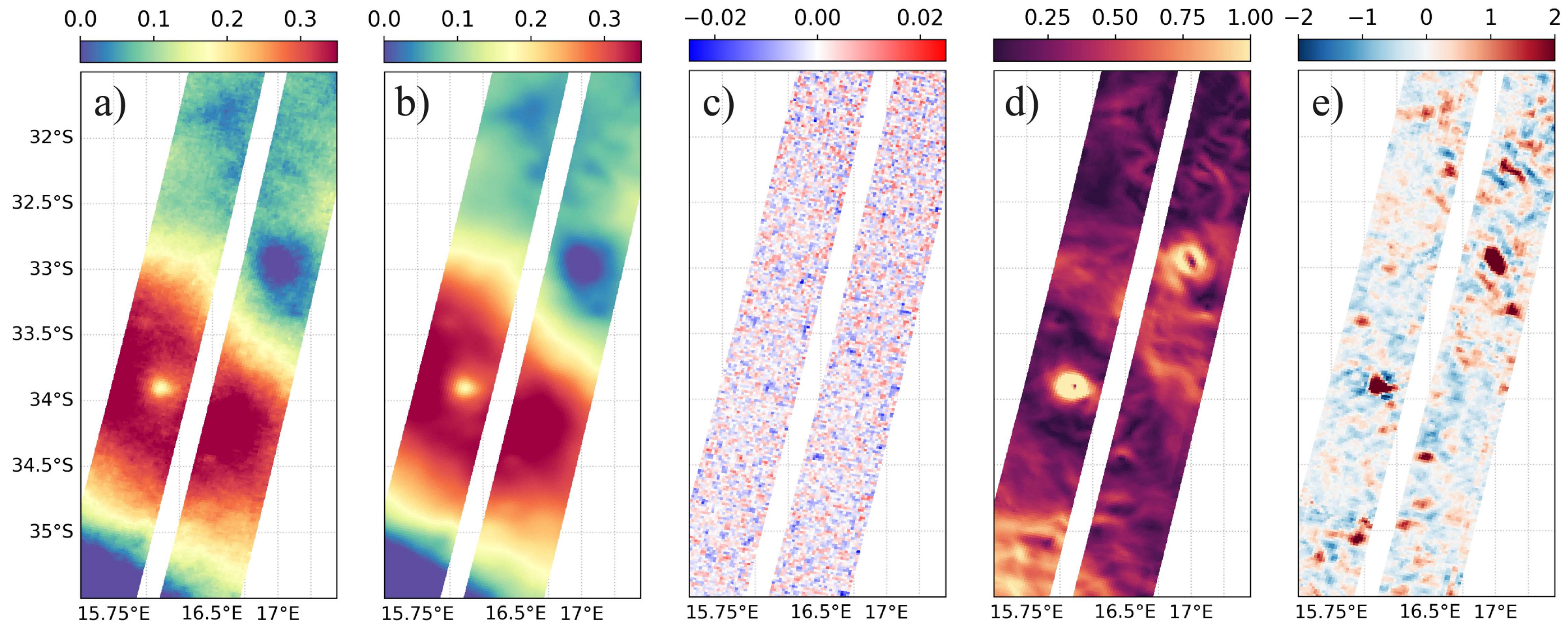

Figure 6. Denoising example of U-Net v3.0 applied on a small eddy in SSHA, Cycle 1, Pass 333. (a) SSHA without the denoising filter (unit: m). (b) SSHA with the denoising filter (unit: m). (c) Removed SSHA content (unit: m). (d) Velocity (unit: $m.s^{-1}$). (e) Relative vorticity.

The main advantage of a well-trained neural network approach is its adaptability to input data. Because L3 denoising adapts to the noise present on the SSHA, denoising effect varies in time and space. Different ocean regions are not denoised to the same extent, as they do not carry the same amount of noise. Moreover, the regional denoising properties might change with SWOT cycle, following the ups-and-downs of random noise and SSHA. To highlight this ability, we calculated locally at what wavelength the denoiser has removed more than half of total SSHA energy. We defined this wavelength as Local Effective Cut-off Wavelength (LECW). The threshold of 50% is somewhat arbitrary: we use this value to be consistent with the definition of a classical smoother cut-off. This approach has already been used by Ballarotta et al., (2019). The LECW is computed over the entire cycle 17 (19 June to 9 July 2024).

The geographical distribution of the LECW (Figure 7) shows that denoising is much stronger in the Antarctic Circumpolar Current (ACC) and in the North Pacific Ocean where waves (and noise) are high in this period of the year. The LECW in these conditions ranges from 25 to 30 km. In contrast, the denoising filter has less impact on the SSHA in the Equatorial Band, where internal tides are most frequent, and waves are low. The LECW ranges from 15 to 20 km in .

The global average of U-Net v3.0 LECW is 22 km (i.e. close to our Lanczos filter reference) [3].

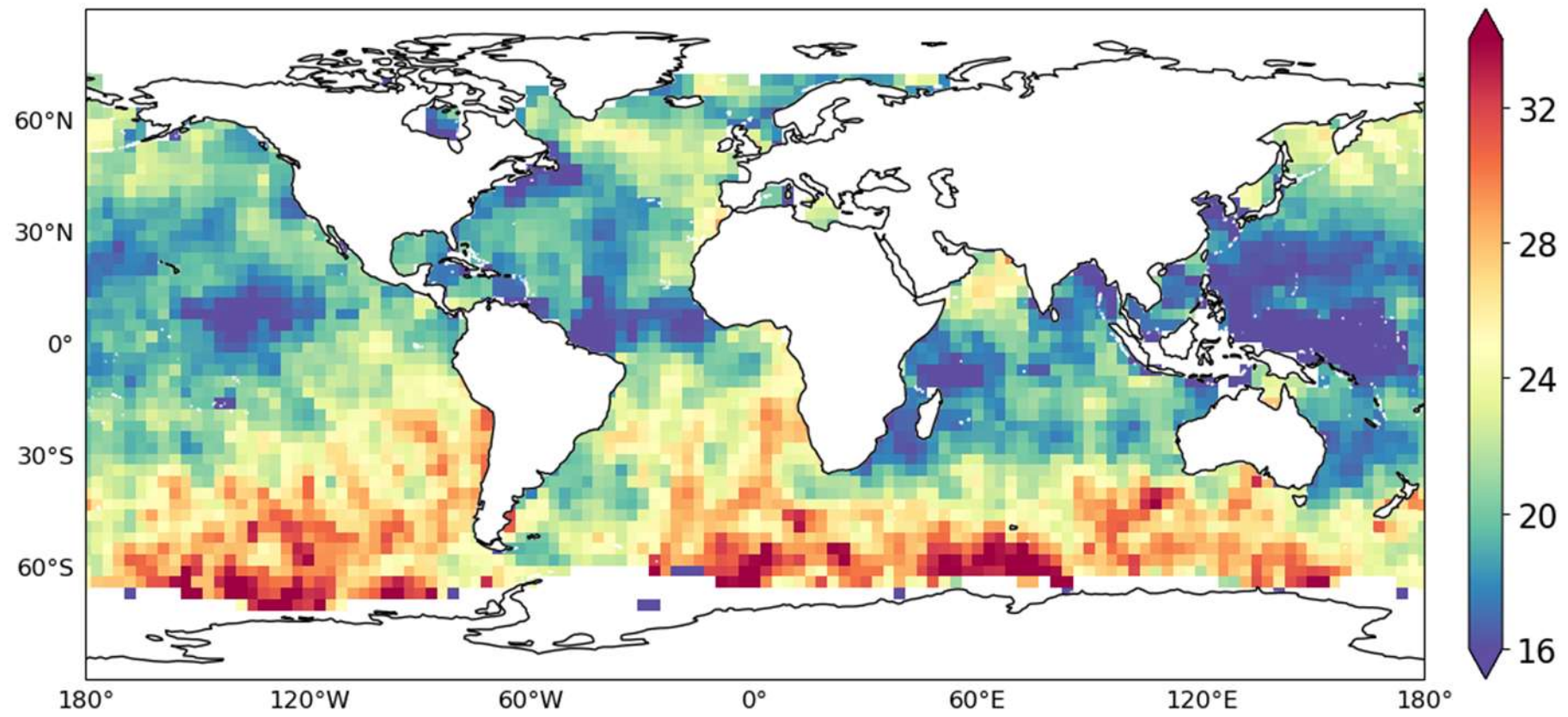

Figure 7. Local Effective Cut-off Wavelength of U-Net v3.0 (unit: km)

To assess the effect of denoising on SSHA in the spectral domain, we also computed its power spectral density (PSD) for the global and regional oceans [4]. The PSDs in Figure 8 are computed globally on cycle 485 of the one-day repeat phase. The KaRIn SSHA power spectrum before denoising (blue curve) exhibits two main properties. At scales above 70 km, the spectral slope follows approxiately a $K^{-3.7}$ power law, close to the Surface Quasigeostrophy theory that describes the ocean surface dynamics (Le Traon et al., 2008). However, below 70 km, the slope approaches $K^{-2}$ potentially due to unbalanced motion energy at these scales (e.g. internal tides (Richman et al., 2012)). Another possible reason for this 70-km slope break could be the presence of red-colored KaRIn measurements errors (Dibarboure et al., 2025).

Our U-Net v3.0 (orange curve) removes a large fraction of the energy from 70 km to 20 km. This is expected when separating the random noise: the energy of random progressively dominates as we go to higher wavenumbers. Note that using a kernel smoother would remove nearly all energy below the cut-off frequency, whereas the denoising filter here leaves a lot of residual content.

[3] The side-by-side comparison with our Lanczos filter with 20 km cutoff wavelength is available in Section d of the Benchmark Appendix.

[4] More details about the PSD calculations are available in the dedicated section in the benchmark Appendix A.

Moreover, the PSD of the content removed by the denoising filter (orange, dashed) is flat at all scales around 15 km, as one would expect from a white noise spectrum, partially correlated due to ground process and the measurement system. It also exhibits a continuous and clean transition from the flat PSD close to 100% of the blue spectrum (i.e. almost all SSHA content removed near 15 km) [5].

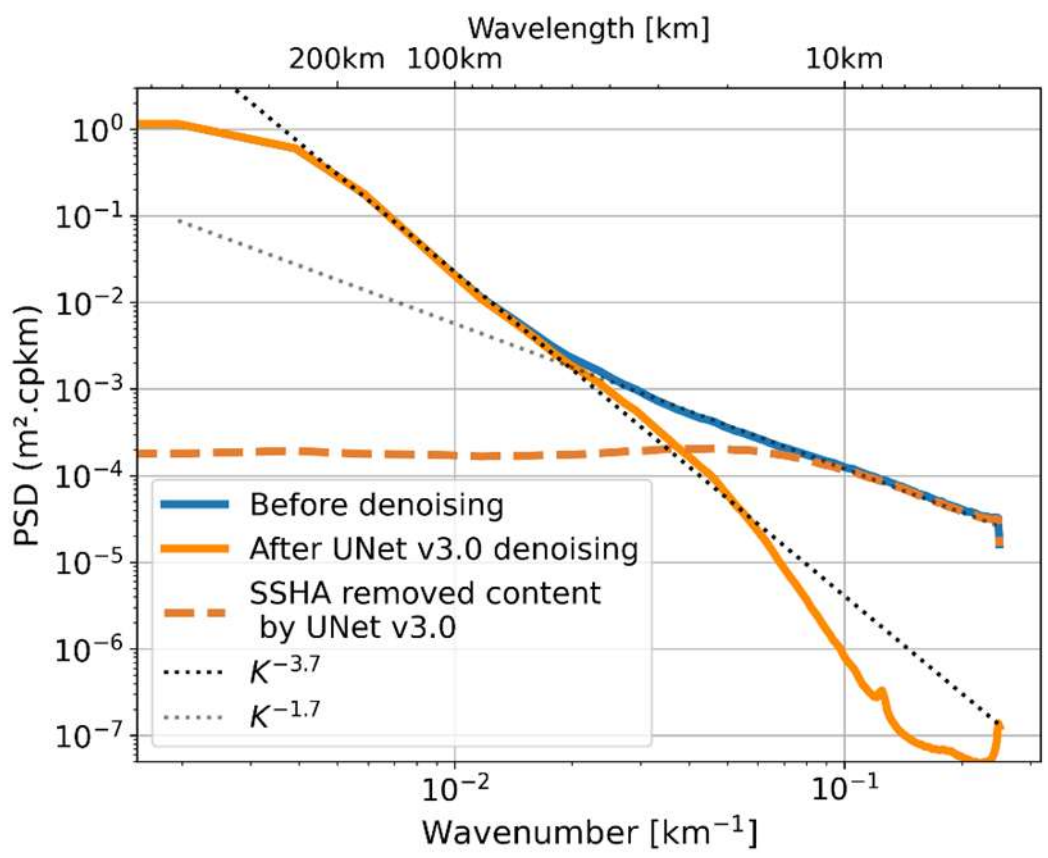

Figure 8. Global PSD of SSHA: PSD of noisy SSHA in blue, PSD of denoised SSHA in v3.0 in plain orange, PSD of U-Net v3.0 absorbed content in dotted orange (unit: $m^2.cpkm^{-1}$).

These global spectra provide an estimate of the frequency distribution of energy removed by denoising, but do not show its spatial distribution. To better understand how denoising is modulated, we computed the geographical variant (i.e. 1°×1° binned standard deviation of removed content by the denoising filter). The maps are computed over the first 15 days of April 2024. To clearly distinguish between the two main denoising regimes, the removed SSHA content is separated into two wavelength intervals. The variance for wavelengths larger than 20 km is shown in Figure 9a, and the variance for wavelengths smaller than 20 km is in Figure 9b.

[5] Note that this PSD does not exhibit any artifact at the 15-km transition, whereas the kernel filter counterpart of Section c (Appendix A) and Figure A8 has stranger features for the removed content.

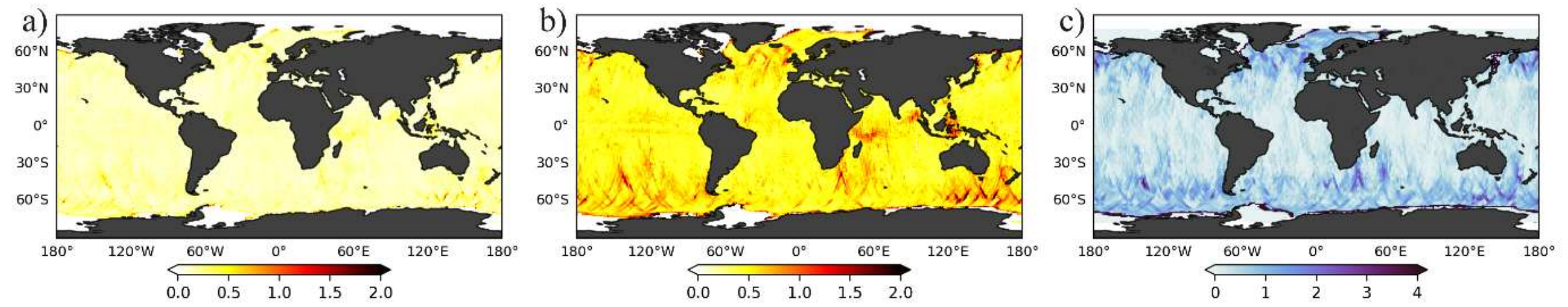

Figure 9. Standard deviation of absorbed content by the denoising filter in cm for wavelengths above 20-km (a) and below 20-km (b). (c) Mean significant wave height in meter, from ECMWF simulation model on the same period.

The former exhibits a background of 1.5 mm, which is consistent with the larger scale of KaRIn's random noise. More importantly, the map is very homogeneous: the denoising filter does not remove more variance in the Gulf Stream or the ACC than in other regions. This is a good indication that the SSH content of interest is not significantly distorted by the CNN since no significant amount of variance was absorbed by the denoising filter.

Furthermore, the latter is strongly correlated with regions and events of higher waves when noise level increases. The comparison with the global map of SWH from ECMWF model on the same period in Figure 9c confirms the strong agreement between variance of removed content and wave level. Specific regions are also apparent: the ITCZ where rainfall can cause residual spurious pixels. During these events, spurious pixels should have been edited out before the filter, and the CNN probably misinterprets remaining pixels as noise.

However, regions known to contain internal tide signatures (north of Madagascar, the Amazon delta, and Indonesia) are also visible in Figure 9b. This is a strong indication that the denoising filter may remove a small fraction of the internal tides (IT) content for scales less than 20 km: the highest nonlinear modes and solitons are partially misinterpreted by the CNN as random noise [6]. By looking at SSHA from variance he north of Madagascar (Figure 10), one

---

[6] Note that the same phenomenon is exactly observed with our reference Lanczos filter (Figure A5). The kernel filter is massively more aggressive with respect to this very small-

can easily notice the blue and red stripes in the SSHA removed content in Figure 10c. This emphasizes that the denoising filter struggles to properly fit the smallest internal tides fronts without absorbing part of their amplitude.

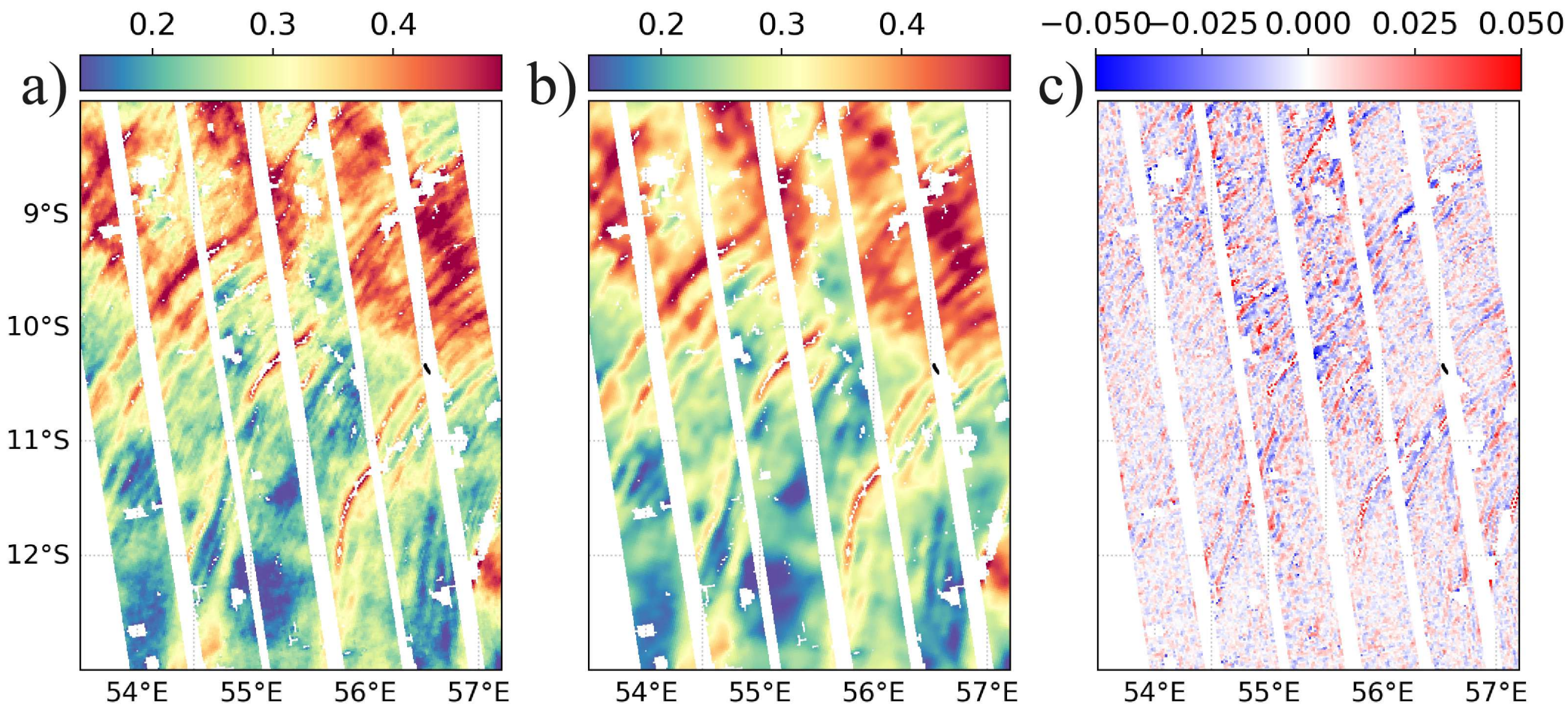

Figure 10. Denoising example of U-Net v3.0 applied on internal tides near Madagascar, Cycle 13, Passes 122, 400 and 428. (a) SSHA without the denoising filter (unit: m). (b) SSHA with the denoising filter (unit: m). (c) SSHA Removed content (unit: m).

Although very intuitive, these spectral and variance maps alone do not demonstrate that the denoising behaves as expected. For this purpose, we have developed a full evaluation benchmark using additional metrics: covariance between the SSHA content removed and kept, day-to-day correlation during the one-day phase, crossovers with other altimetry satellites… The benchmark and the comparison with the Lanczos smoother are presented in Appendix A. Similarly intuitive conclusions are drawn for the v3 denoising filter, with no obvious issues as apparent as the expected limitations of traditional smoothers. Note that when the denoising filter v2 is run through the same benchmark, the metrics capture significant deficiencies that are not seen in v3. This is discussed in Appendix B.

*b. Denoising the real SWOT LR 250-m product*

---

scale ocean content. The differences between the denoising and smoothing filters are presented in the benchmark in Appendix A, Figure A6.

The 250-m denoising appears to successfully separate the random noise from the signal in the SWOT observations, across regions and periods. The gain on velocity and relative vorticity from Figure 2 (before denoising) to Figure 11 (after denoising) is massive. The noise in Figure 11c looks random: no bias or strongly correlated features, no correlation with mesoscale features in the SSHA or its derivatives. After the denoising, the 250-m SSHA derivatives are very similar to those of the 2-km product. The higher resolution seems to capture slightly finer structures of the order of 2 to 10 km, and the larger mesoscale structures are identical. Due to the amount of random noise before the filter, the second SSHA derivative is somewhat noisier than for the 2-km resolution. To display a cleaner result, a Butterworth smoothing with a 2-km cutoff wavelength is applied directly to the vorticity field shown in Figure 11e for visualization purposes only. It does not interfere with physical geostrophic content, whose scales are much larger. The denoising performance seems consistent with the 2-km behavior despite the massive random noise of 250-m measurements.

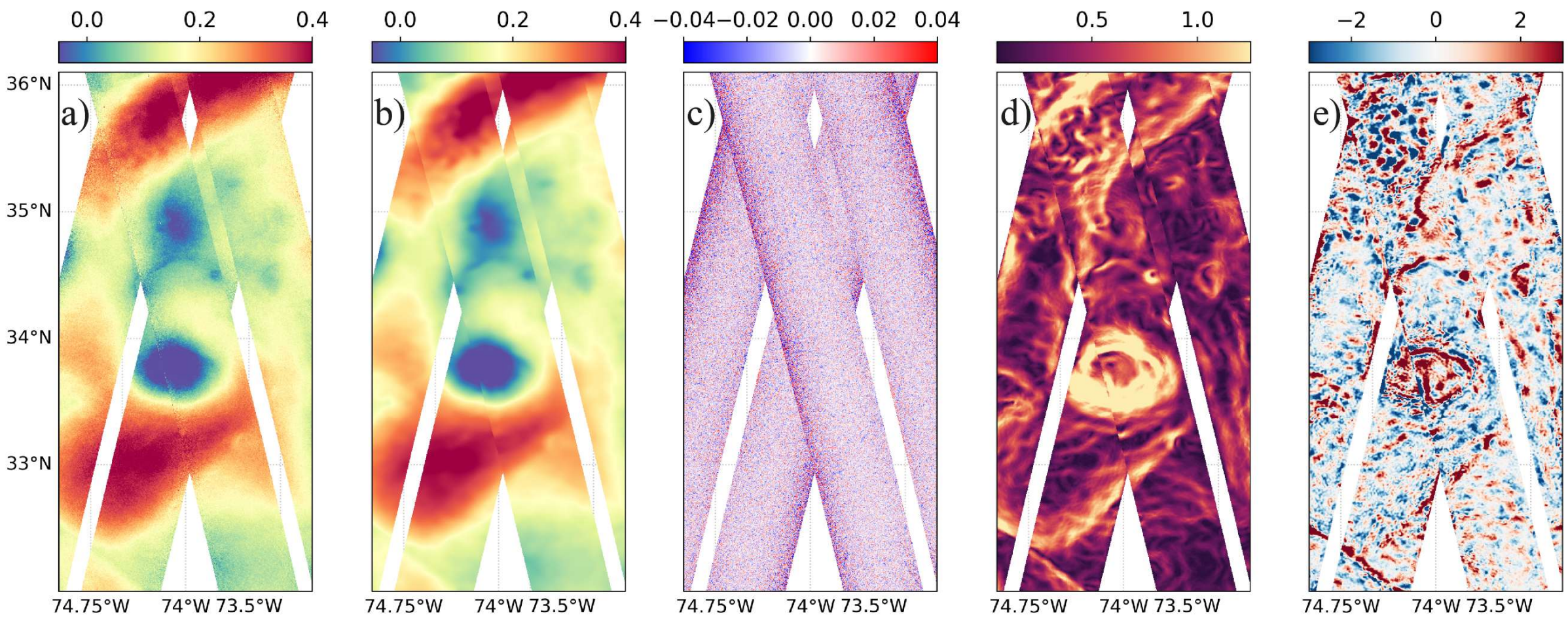

Figure 11. Denoising example of U-Net v3.0 250-m applied on a SSHA crossover, Cycle 489, Passes 9 and 22. (a) SSHA without the denoising filter (unit: m). (b) SSHA with the denoising filter (unit: m). (c) Removed SSHA content (unit: m). (d) Velocity (unit: $m.s^{-1}$). (e) Relative vorticity.

The 250-m SSHA spectra were calculated over the entire cycle 485 of the one-day repeat phase. The results are similar to those for 2-km. The spectral hump from 60 to 10 km is even more visible at the 250-m resolution. The mesoscale energy is mainly carried by geostrophic features (vortexes and fronts) and by internal waves. The PSD of the denoised 250-m SSHA content fits the $K^{-2}$ slope of the 2-km SSHA spectrum from 70 km to 15 km, then it rolls off to very small values (Figure 12a). In other words, because the signal-to-noise ratio is much worse at 250-m, the denoiser captures and removes more high-frequency content, but it also behaves more like a smoothing filter for scales with a 15-km cut-off. The 250-m denoising filter removes very little content for scales above 20 km.

Moreover, the spectrum of removed content is completely flat at the 250-m SSHA noise plateau (Figure 12b) and it aligns very well with the spectral plateau of the unfiltered SSHA PSD. This suggests that the vast majority of what has been absorbed by the denoising filter is pure additive Gaussian white noise. This would be the expected behavior of a filter removing random noise without any spatial correlation. The integration of the removed energy plateau leads to an estimation of noise standard deviation at 2.6 cm. The overall LECW at 250-m resolution on these spectra is around 22 km.

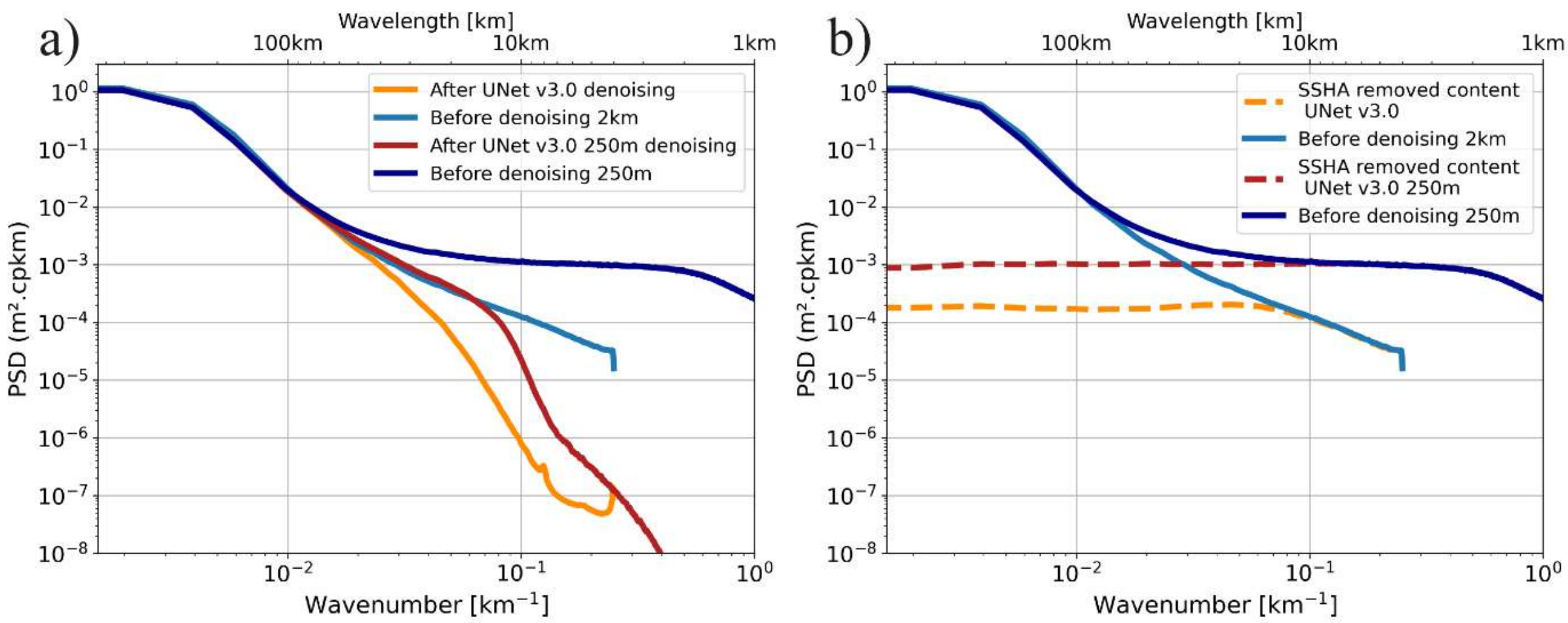

Figure 12. (a) Global PSD of SSHA: PSD of noisy 2-km SSHA in blue and 250-m in dark blue. PSD of denoised SSHA with U-Net v3.0 2-km in orange and 250-m in red (in $m^2$.cpkm-1). (b) Global PSD of removed content: PSD of noisy 2-km SSHA in blue, 250-m SSHA in dark blue. PSD of U-Net v3.0 removed content in orange, PSD of U-Net v3.0 250-m removed content in red (in $m^2.cpkm^{-1}$).

Similarly to the 2-km data, there is still room for improvement in areas dominated by internal tides (Figure 13). It is easy to distinguish in the removed content the presence of internal tides partially absorbed or completely missed by the denoising filter. U-Net v3.0 250-m denoising slightly widens a few wavefronts that are easily noticeable as blue/red stripe patterns in the removed content field (Figure 13c). The same issue is visible using a Lanczos with 15-km cutoff wavelength (Figure 14). The amplitude of wavefronts is clearly reduced by the smoothing effect of the filter.

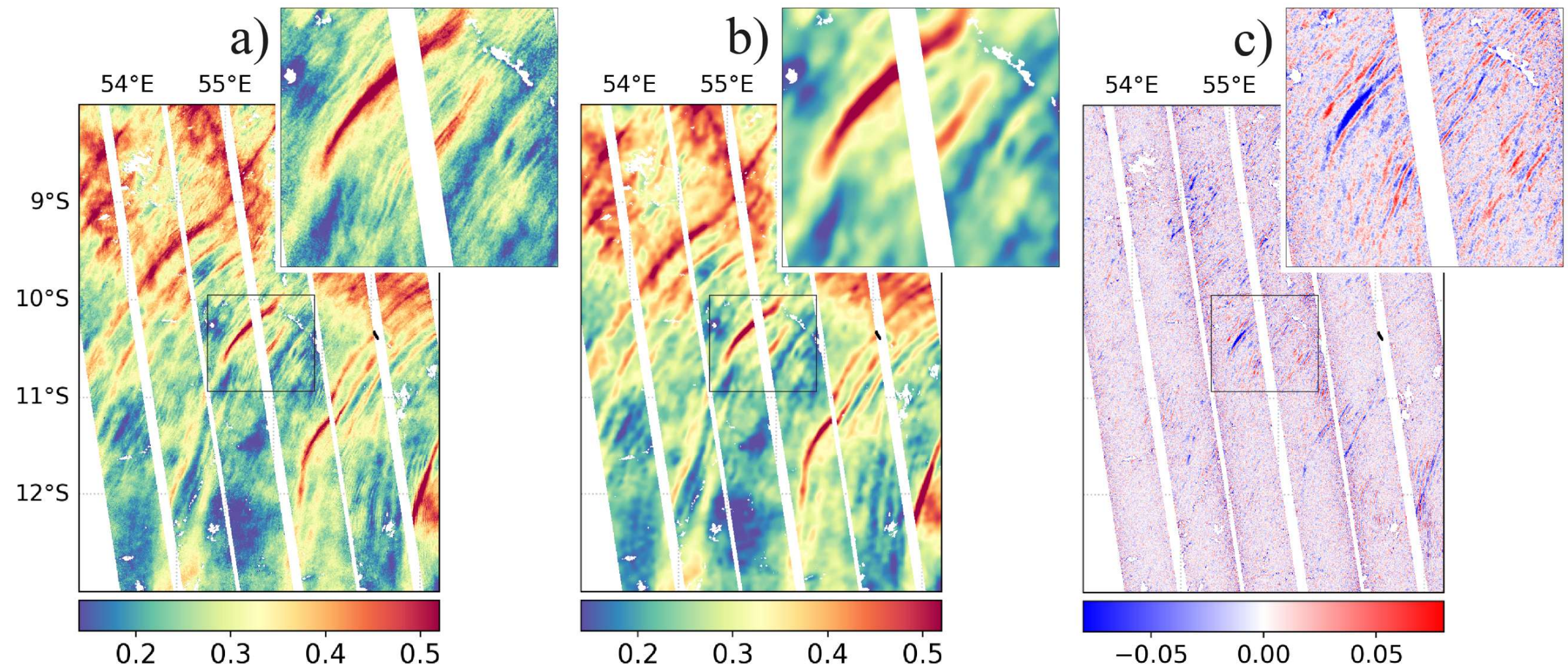

Figure 13. Example of SSHA samples with the U-Net v3.0 250-m denoising filter applied on internal tides near Madagascar, Cycle 13, Passes 122, 400, 428. (a) SSHA without the denoising filter (unit: m). (b) SSHA with denoising filter (unit: m). (c) Removed SSHA content (unit: m).

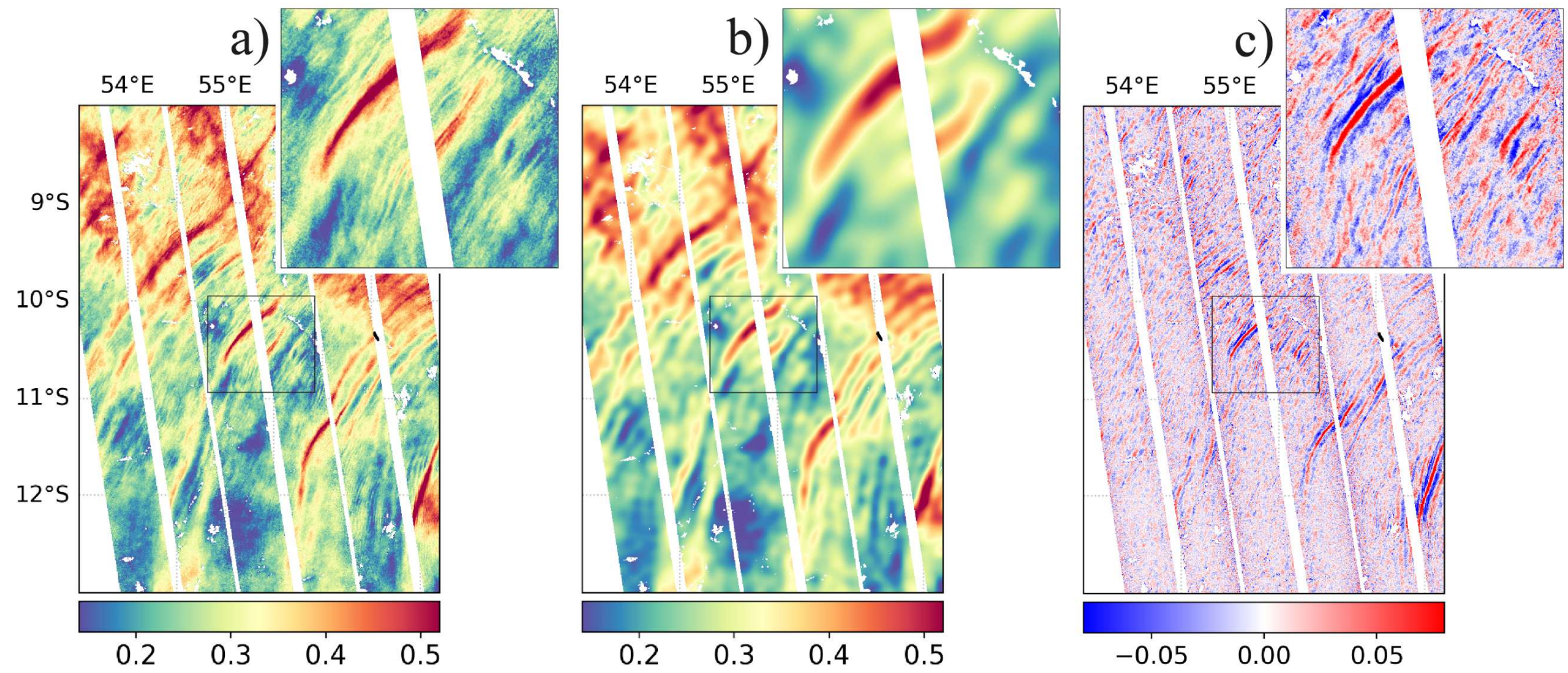

Figure 14. Example of SSHA samples with the Lanczos smoothing filter applied on internal tides near Madagascar, Cycle 13, Passes 122, 400, 428. (a) SSHA without the smoothing filter (unit: m). (b) SSHA with the smoothing filter (unit: m). (c) Removed SSHA content (unit: m).

To summarize, the 250-m denoising properties of the U-Net v3.0 confirm its capability to successfully remove most of the random noise. The signal-to-noise ratio at 250-m is so bad (more than 2 orders of magnitude) that the denoising filter could arguably be considered more as a smoother than a denoiser for scales smaller than 15 km. The results and conclusions of 250-m denoising on the other metrics (Spectral one-day differences, one-day coherence with SSIM, covariance between removed content and denoised signal…) are very similar to 2-km denoising. In particular, both struggle to adapt to the smallest internal tide and soliton signals. Moreover, no major bias or artifact problems have yet been found with this version of denoising, unlike the previous version.

## 4. Does the neural network reproduce ocean dynamics from simulation model to real SWOT data?

Our denoising CNN is trained with a supervised strategy on an ocean model simulation as a proxy of reality. There are two major pitfalls associated with this strategy.

- The representativeness/realism of the training dataset in relation to actual oceanic signals. The model captures ocean dynamics up to a certain resolution. It may not be representative of the global ocean, nor of the smallest scales observed by KaRIn and beyond the reach of such regional ocean models. In this situation, the CNN might behave correctly in "nominal" situations such as the North Atlantic Ocean, but poorly in "non-nominal conditions" (e.g. in the Southern Ocean).
- The finite amount of samples used for the CNN training, and the somewhat large size of the CNN might be prone to overfitting. In this scenario, the denoising filter would actively reproduce a specific trained dynamics from the eNATL60 simulation, rather than deliver the real ocean.

The second pitfall warrants particular caution, and dedicated analyses. The denoising filter output should be consistent in all conditions: the removed content should always look like random noise and not ocean features, and denoised SSHA should always be consistent with the unfiltered observations even for ocean dynamics not found in the Atlantic Ocean. Similarly, the filtered content and the residuals should never exhibit strange features, even on very atypical scenes.

### *a. Impact of the simulation model*

To investigate whether the network tends to reproduce the oceanic variability of its training model, its performance is evaluated on a different ocean model that features physical dynamics distinct from those seen during training. For this purpose, MITgcm LLC2160 model (https://www.nas.nasa.gov/SC21/research/project16.html) is chosen to generate simulated SWOT swaths. As a global model, MITgcm is more representative of the diversity of oceanic signals, particularly in regions that differ significantly from the North Atlantic.

The U-Net denoising network is applied directly at inference (without retraining) on the MITgcm swaths. These swaths were generated by interpolating the model grid onto the SWOT swath. A filtering step using a Butterworth low-pass filter with 10 km cutoff wavelength is used to reduce the interpolation artifacts on the simulated noiseless SSHA. A noise equivalent to that used in the eNATL60 training data is added (Section 2). This noise is also modulated by wave height and cross-track distance.

To test whether the network may struggle to denoise on a different model, in a different region, we place ourselves in South Australia, where wave level is much higher than in Gulf Stream. The PSD of U-Net v3.0 filtered simulated SSHA highlights the network's performance in the region (Figure 15). Indeed, the PSD of SSHA denoised by U-Net v3.0 (orange curve) fits perfectly the MITgcm noiseless spectrum up to 10 km (green curve). The mismatch below 10 km is not an important concern because the energy of the noiseless spectrum at these scales is a mixture of interpolation artifacts and low-passed SSHA.

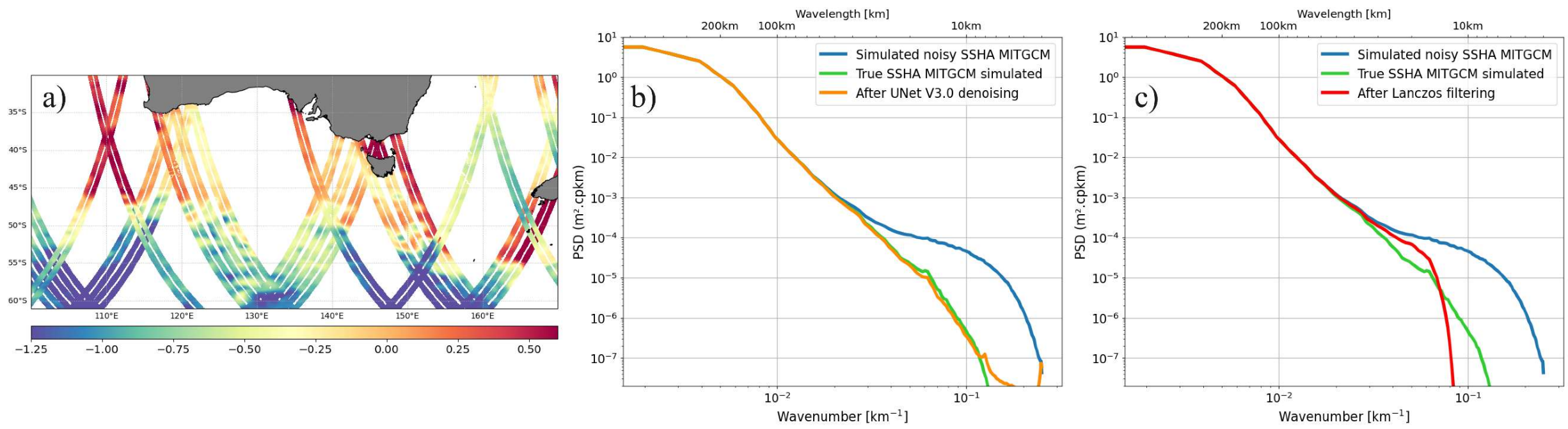

Figure 15. (a) Simulated MITgcm swath samples in high-wave region. (b) PSD of simulated SSHA from MITgcm (noisy SSHA in blue, noiseless SSHA in green, denoised SSHA in orange using U-Net v3.0). (c) PSD of simulated SSHA from MITgcm (smoothed SSHA using Lanczos filter in red).

In comparison, a Lanczos filter with 15 km cutoff wavelength (red curve) mixes noise with the SSHA signal. From 30 km to 15 km, the filter significantly overestimates the energy of noiseless SSHA. Therefore, the oceanic processes at these scales are not accurately retrieved by filtering. Below 15 km, the energy is lost, as expected from the filter. Even if U-Net v3.0 has never seen the oceanic dynamics of this region in its training, it successfully removes the noise from the simulated SSHA. There are no signs in the spectral behavior that indicate a deformation of the data to make it look like to eNATL60 dynamics.

To further ensure that the denoising filter does not behave poorly outside the Gulf Stream region, we also compute the standard deviation of removed SSHA content globally. The simulated truth (i.e. the simulated noise on the swaths) is shown alongside in Figure 16. The content removed by the denoising filter seems very close to the simulated noise reference, which gives us confidence that most of what is absorbed while denoising is only noise. There are no signs of hallucinations from the U-Net v3.0 and the internal tide regions are unseen. A plausible explanation for the absence of internal tide in the removed content in MITgcm swaths is that these signals are not resolved at the fine scales typical of actual SWOT observations. The simulated tides are at sufficiently large scale to be properly denoised, even if the network has never seen such signals.

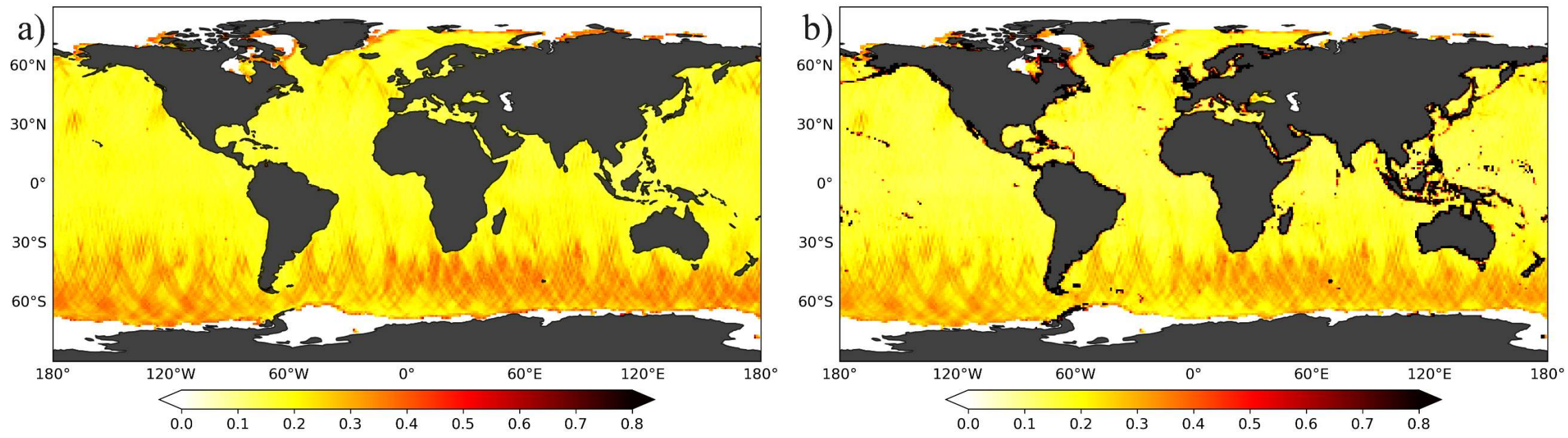

Figure 16. (a) Simulated noise on MITgcm swaths for scales below 20 km. (b) SSHA removed content by U-Net v3.0 filter on simulated noisy SSHA swaths from MITgcm for scales below 20 km.

However, the amplitude of what is removed by the denoising filter is slightly lower than the reference. The denoising filter is probably underestimating the noise level in high-wave conditions. Such very noisy situations were not encountered during training. The neural network fails to match the exact energy of noise. Coastal areas also exhibit higher values, but it is possibly explained by missed pixels in editing.

*b. Impact on the dynamics in presence of internal tides*

To continue our analysis, we turn our attention to northeastern Indonesia, an oceanic region with very different dynamics and very strong internal wave activity. The spectra in Figure 17 reveal that the denoising filter has absorbed a portion of internal tide energy near 20-25 km. Similarly, the small green peaks near 10 and 15 km are also not captured by the U-Net. Conversely, just as in Figure 15, the denoising process retains slightly too much energy below 10 km. All these effects are attributable to differences in simulation models, regions (and thus ocean dynamics). The initial concern was qualitatively justified and warranted discussion. However, these figures clearly show that these effects are quantitatively marginal compared to a Lanczos filter (regardless of its cutoff), which completely eliminates small-scale signals. Even though internal tides are absent from the training dataset, the network generally manages to denoise the associated larger-scale signal. Unfortunately, the finest variations are misinterpreted as noise. In practice, this pitfall is not a major concern but will be addressed in future versions.

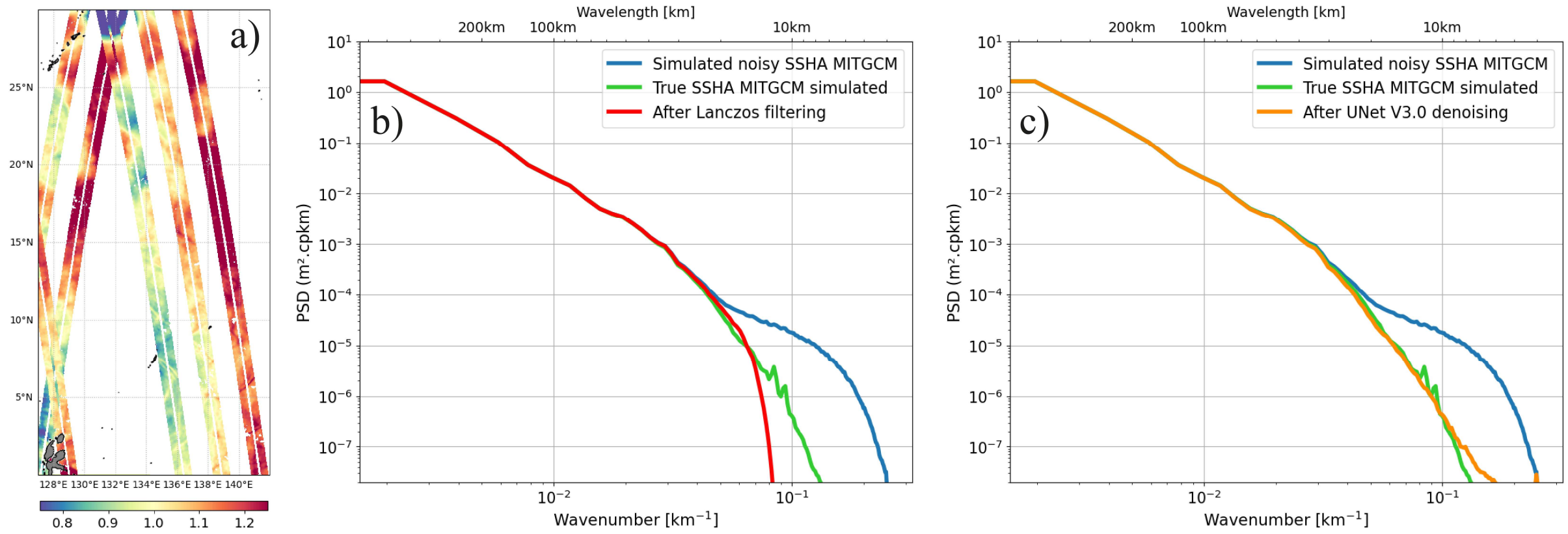

Figure 17. (a) Simulated MITgcm swath samples in internal tides high activity region. (b) PSD of simulated SSHA from MITgcm (smoothed SSHA using Lanczos filter in red). (c) PSD of simulated SSHA from MITgcm (noisy SSHA in blue, noiseless SSHA in green, denoised SSHA in orange using U-Net v3.0).

*c. Addition of non-oceanic signals to real swaths before denoising*

Finally, we performed a sensitivity test to detect potential overfitting to the simulation model. We added an arbitrary, non-physical, unrealistic signal, never seen during the training process, into the SWOT data. If this feature is distorted into an "ocean-like" denoised structure, it is a very strong hint that the CNN is likely distorting actual SWOT measurements in a similar way: in this scenario, the overfitting replicates eNATL60 dynamics instead of retrieving the actual measured SSHA content. Conversely, if the artificial feature is properly denoised and retrieved without any alteration, we can reasonably assume that actual measured ocean features, that were never encountered by the CNN, might not be altered either.

To do this, we inserted the word "SWOT" into real KaRIn swaths, with a scale ranging from a few kilometers (font width) to 50 km (height of each letter). The network was then used to denoise the noisy swaths with the text added, without adapting its weights. The test was run with U-Net v3.0. The result is shown in Figure 18. The word "SWOT" is almost undamaged in the denoised SSHA, and there is no measurable letter-shaped artifact in the removed content. In other words, the CNN filter did not distort the signal to make it look like the real ocean [7].

---

[7] Note that using the Lanczos filter (20 km cutoff wavelength) yields much worse results on filtered letters (Figure S2).

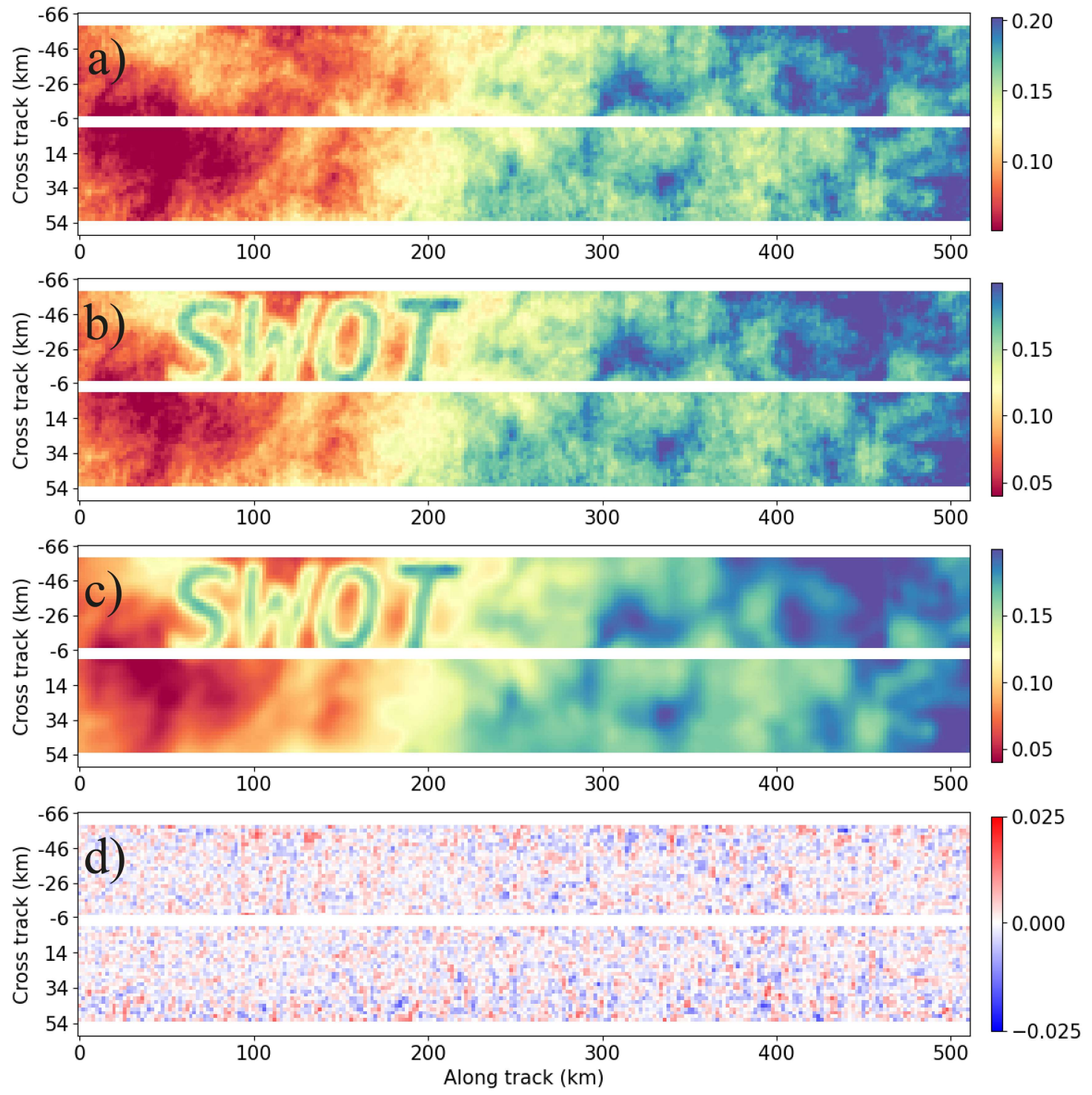

Figure 18. Example of U-Net v3.0 denoising on text data (units: m). (a) Noisy 2-km real SWOT swath. (b) Noisy 2-km real SWOT swath with text added. (c) 2-km real SWOT swath with text added with the denoising filter applied. (d) Removed content by the denoising filter (b - c).

We also performed a similar test where the letters are much more subtly integrated into the swath dynamics (i.e. rounder font, spaced out letters, smaller amplitude, barely noticeable signal). In this scene, the letters look much more like physical content. The objective remains the same: determine whether the network distorts or preserves this signal. This time, the simulated signal is meant to be closer to the distribution learned by the CNN.

In this second test, the results are the same: nothing can be discerned in the removed content (Figure S1). Every letter is retrieved in the denoised SSHA. No physical distortion of artificial signals is observed. The denoising filter does not distort the SSHA to reproduce eNATL60 features[8].

[8] And like for the first test, using the Lanczos smoothing filter yields much worse results than the denoising filter (Figure S2).

## 5. Discussion: limitations and outlook

Lacking a trustworthy ground truth matching SWOT measurements, it is impossible to demonstrate the correct behavior of our denoising filter. In an effort to perform a very robust validation, we developed other metrics to evaluate the denoising performance (benchmark in Appendix A). These metrics all exhibit an intuitive behavior for a good denoising filter. We did not find any anomalous or unexplained results on SWOT data. In contrast, the previous version (v2.0) does exhibit some weaknesses and limitations shown in Appendix B. Even if the v3.0 denoising filter is not perfect, it is clearly a major improvement upon v2.0.

We still observed some limitations in our U-Net v3.0:

- Denoising needs to be more adaptive in high SWH conditions. The filter is probably removing too much real oceanic signal under strong noise conditions. This is likely because our training dataset was not properly balanced when it comes to SWH diversity.
- The supervised training approach requires simulated ocean data, but the current ocean models are not able to realistically reproduce all KaRIn features (e.g. smaller internal waves). It would greatly benefit from training solely on real SWOT data, using self-supervised approaches.
- It is particularly hard to validate the physical realism of small-scale content that is not completely filtered out by denoising. New metrics should be introduced with high-resolution independent data to strengthen the benchmark.

Moreover, the low-pass filter used for reference in Appendix A is admittedly very simple. A fairer comparison would require more sophisticated filters, such as an adaptive filter where the kernel size is adjusted locally according to SWH/noise magnitude, cross-track distance, and the signal-to-noise ratio. The LECW could also serve as a proxy of kernel modulation parameters for the filter. More sophisticated filtering methods would likely yield greater results.

On the denoising side, a more advanced model than the current U-Net architecture might exhibit better results. Similarly, different training strategies (e.g. self-supervised approaches) could take advantage of the larger real-data training set[9].

The smallest scales of 250-m, such as wave groups or very fine solitary waves, are eliminated by the denoising filter. These signals are almost the same scale as the noise and are extremely difficult to preserve. The wavefront dilation (Figure 13) is possibly due to the network being too constrained, predicting an overly regular SSHA field that does not precisely capture ocean anisotropy. These kinds of signals are completely absent from the network's training set because the ocean model used does not have the extremely high resolution needed to resolve such nonlinear internal tide features (the model would need a resolution 5 to 9 times finer than the Unsmoothed pixels from KaRIn). A higher-resolution training dataset with realistic simulations with similar features would very likely solve this weakness. In the absence of a high-resolution ocean model input, an alternative could be to add arbitrary wave-like signals to try and mimic real data from SWOT. This is one clear point of improvement for future versions of L3 LR denoising process. The validation of 250-m denoising filter is still in its early stages. Comparisons with other satellites and measurement techniques are planned and will provide additional information on the qualities and shortcomings of the approach.

In parallel, changes in DUACS products over successive versions also impact denoising performance. With the arrival of version 3.0 products, improvements to SSHA corrections upstream of denoising directly influence the quality of the network's output. Up to this point, the denoising filter is sensitive to various L3 corrections. One of them is the Mean Sea Surface (MSS) model. As discussed in Appendix A, Section f, some high-frequency residual signals are present in the MSS and are systematically absorbed by the denoising filter. Now, the version of MSS used in v3.0 products contains significantly fewer small-scale artifacts than the previous one, which means that the denoising process also absorbs fewer of them. Similarly, the denoising is also sensitive to the L3 editing step. If some spurious pixels (e.g. coastal pixels) are missed by the editing step, they lead the neural network to poor predictions, especially for the 250-m resolution. We can therefore reasonably expect even better denoising results with each new improvement in product version.

---

[9] This was not possible just after launch. Now after 3 years of data collection, self-supervised techniques could be revisited.

## 6. Conclusion

The SWOT Ka-band radar interferometer reveals dynamics that were never observed before in altimetry. Despite the unprecedented precision of the KaRIn instrument, these smaller-scale ocean features can be affected by high-frequency noise, as a combination of random instrumental errors and geophysical correction residues.

To counter the random part of this noise, we used the convolutional neural network from Tréboutte et al. (2023) and trained it on simulated data from the eNATL60 high-resolution ocean circulation model. The simulated noise was carefully designed to reproduce the statistical and spatial properties characteristic observed in real SWOT measurements. The lightweight U-Net architecture employed in this study successfully learned the underlying noise patterns from these synthetic samples and demonstrated robust generalization to actual SWOT observations.

To evaluate the quality of the denoising method, we developed a benchmark composed of several quantitative and qualitative metrics. The U-Net v3.0 filter yields satisfactory results on this benchmark, with no major issues explicitly identified. Qualitatively, we found that overall, the network removes the noise without altering the physical signal underneath, even on structures down to 15 km (Figure 6). We noticed that under high-wave conditions, the distinctive noise signature is not yet perfectly handled. Internal waves are also not perfectly processed. Although it is better than a standard Lanczos filter, the denoising filter absorbs a small portion of their amplitude (Figure 10). The 250-m and 2-km dynamics derived from denoised data are very close to each other. Neither exhibits clear artifacts or biases.

The removed SSHA content is spectrally flat, as expected for pure white noise. The PSD of SSHA after applying the denoising filter maintains energy below 15 km, where a regular Lanczos filter would have eliminated it. We show that the denoising filter is spatially and temporally heterogeneous by computing the Local Effective Cut-off Wavelength. This value ranges from 15 km in internal tide activity areas to 28 km in high-wave regions, with an average 22 km cutoff. We also show that the content removed by the denoising filter is almost fully decorrelated from one day to the next for the same tracks, with only 10% of missing variance in the 24 hours difference (Figure A10). Finally, we calculated that the denoising filter brings

SWOT observations closer to the Sentinel-3A measurements at 20 Hz at the crossover points, with a global mean improvement of approximately 4.5 mm RMS compared to unfiltered SSHA (Figure A17).

However, even if the denoising filter is performing adequately on our metrics, this does not constitute definitive proof of its validity. It will remain impossible to prove the accuracy of the SSHA after the denoising filter without access to ground-truth. We still hope to soon enhance the benchmark with new metrics coming from the scientific community, with other satellite and in situ data. This task has already begun and should bear fruit in 2026, with the aim of continuously improving the quality of SWOT data to provide users with the best possible noise mitigated product.

To conclude, as previously shown, the U-Net v3.0 trained on the eNATL60 model outperforms low-pass filter when denoising MITgcm simulated swaths. The network is not overfitted to eNATL60 dynamics and is able to properly fit the spectral slope of the PSD from noiseless MITgcm swaths, in various oceanic conditions. We attempted to denoise artificial signals (text letters) into real KaRIn swaths to determine whether the denoising tries to modify these unrealistic and unlearned signals to make them resemble simulated ocean dynamics. Unlike the Lanczos filter, the U-Net v3.0 mitigates noise on letters without altering the shapes, removing only the noise component.

Substantial effort has been made to ensure that training on a simulated dataset does not negatively impact real SWOT data denoising. Because the absolute goal of denoising is to remove only the noise on real KaRIn SSHA observations, particular caution is required to avoid any overfitting when training a neural network with a ground truth from a simulation model. Even if such models constantly improve their consistency with real measurements, they will always remain imperfect. We shall continue to study their influence on our approach to denoising.

*Acknowledgments.*

This study was financially and technically supported by CNES through SALP (Service d'Altimétrie et Localisation Précise) research funding to improve the quality of multi-missions altimetric data processing. We thank the High Performance Computing (HPC) center of CNES for enabling us to develop the code, run and train our network on high quality computing resources. We would like to address a special thanks to the SWOT Science Team members who participated in multiple online meetings about denoising. The discussions helped a lot to conceive the benchmark and this collaborative work will be continuously fruitful.

*Data Availability Statement.*

The SWOT L3_LR_SSH product, derived from the L2 SWOT KaRIn low rate ocean data products (NASA/JPL and CNES), is produced and made freely available by AVISO and DUACS teams as part of the DESMOS Science Team project. AVISO/DUACS, 2024. *SWOT Level-3 KaRIn Low Rate SSH Unsmoothed* (v2.0) [Data set]. CNES. https://doi.org/10.24400/527896/A01-2024.003 and *SWOT Level-3 KaRIn Low Rate SSH Expert* (v3.0) [Data set]. CNES. https://doi.org/10.24400/527896/A01-2023.018.

This study has been conducted using E.U. Copernicus Marine Service Information: The Global Ocean Waves Reanalysis product is available freely in the Marine Data Store (MDS) of Copernicus Marine Service Information (CMEMS), DOI: https://doi.org/10.48670/moi-00022. The Global Ocean Gridded L4 Sea Surface Height product is available freely in the Marine Data Store of CMEMS, DOI: https://doi.org/10.48670/moi-00148.

The Sentinel-3A monomission along-track corrected Sea Surface Height (CorSSH/L2P 20 Hz products – NRT, STC) is distributed freely by AVISO+. DOI: 10.24400/527896/A01-2025.004.

# APPENDIX A

## Evaluative Benchmark, results of U-Net v3.0 filter

Evaluating the quality of the denoising process without access to ground truth is very challenging and SWOT swaths are no exception. Various metrics have been developed to approximate where and to what extent the denoiser (or smoother) impacts the oceanic signal. The aim of this appendix is to introduce a comparative benchmark against which other denoising method can be compared.

Version 3.0 of U-Net aims to correct the shortcomings of the previous version while preserving its qualities. This approach is compared with a Lanczos filter through benchmark metrics. Results for U-Net v3.0 are presented here. Readers interested in seeing the results of the previous U-Net (v2.0) on the benchmark should refer to Appendix B.

*a. Visual validation*

The first validation step after denoising SWOT data is to visually inspect the SSHA, its removed content, and its derived quantities on specific examples. This allows for rapid identification of artifacts, network errors, biases, signal absorption, and other issues. To better understand what is absorbed while denoising, the removed SSHA content is studied and defined as the difference between the noisy SSHA before and after applying the denoising filter.

We begin by considering high and large-scale variability regions (such as Gulf Stream, Kuroshio) and especially the ones coupled with strong wave conditions (e.g. the Antarctic Circumpolar Current). In such regions, denoising methods can easily introduce significant biases that can reach several centimeters (e.g. Figure B2). False structures can be created while real ones are hardly smoothed out. U-Net v3.0 does not exhibit such errors. In strong waves conditions, U-Net v3.0 manages to remove noise without visibly damaging the denoised SSHA (Figure A1).

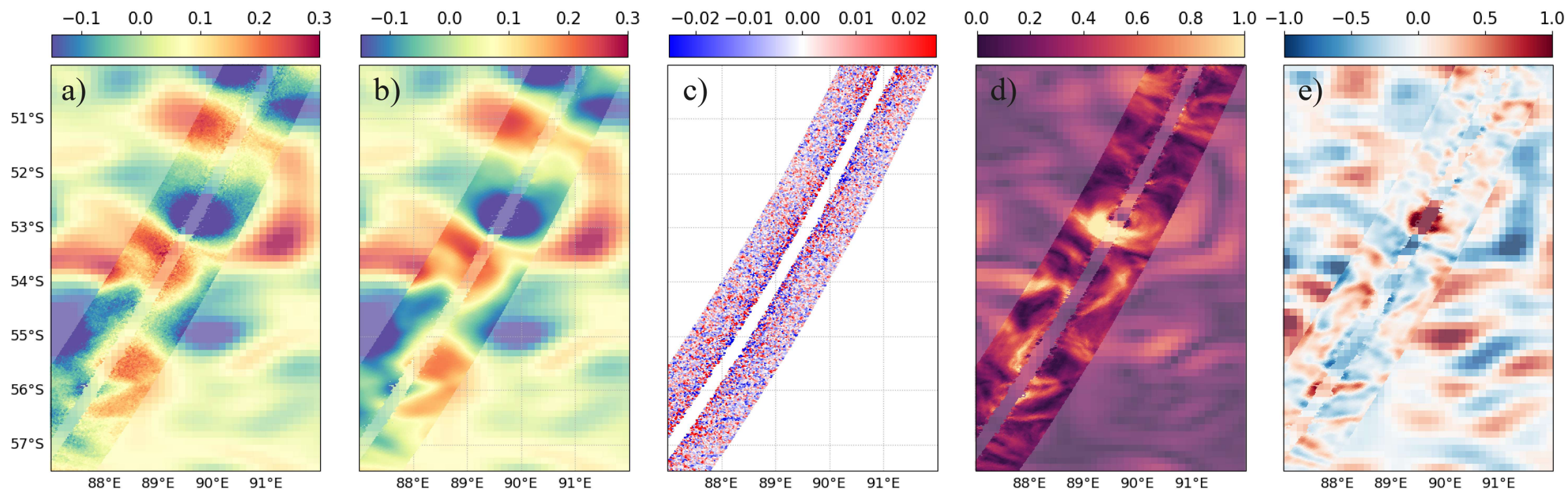

Figure A1. Denoising example of U-Net v3.0 filter in presence of strong noise in the ACC region. Cycle 20, Pass 299. (a) SSHA without the denoising filter (unit: m). (b) SSHA with the denoising filter (unit: m). (c) Removed SSHA content (unit: m). (d) Velocity (unit: $m.s^{-1}$). (e) Relative vorticity.

The larger scales of SSHA with the denoising filter applied are visually very consistent with the L4 gridded Sea Surface Height product in background. The performance of U-Net v3.0 denoising remains imperfect in such conditions. It is still possible to distinguish very small biases (large red and blue patterns of few millimeters) in the removed SSHA content. The noise in regions of high SWH is very different from elsewhere. Additional effort should be put into the noise simulation methodology to better reproduce real noise degradation and thus improve the denoising performance in these conditions.

The second type of region of interest is where small-scale oceanic signals (e.g. solitary waves and internal tides) are most frequent. The amplitude of these signals is often partially truncated by denoising or filtering approaches. Absorbed signals can be seen in the removed SSHA content, where tidal wavefronts can be clearly visible using a Lanczos filter with a 20 km cutoff (Figure A3c). Such overly aggressive filter eliminates massively tide signals. U-Net v3.0 retrieves accurately the small wavefronts, the removed content is closer to pure noise.

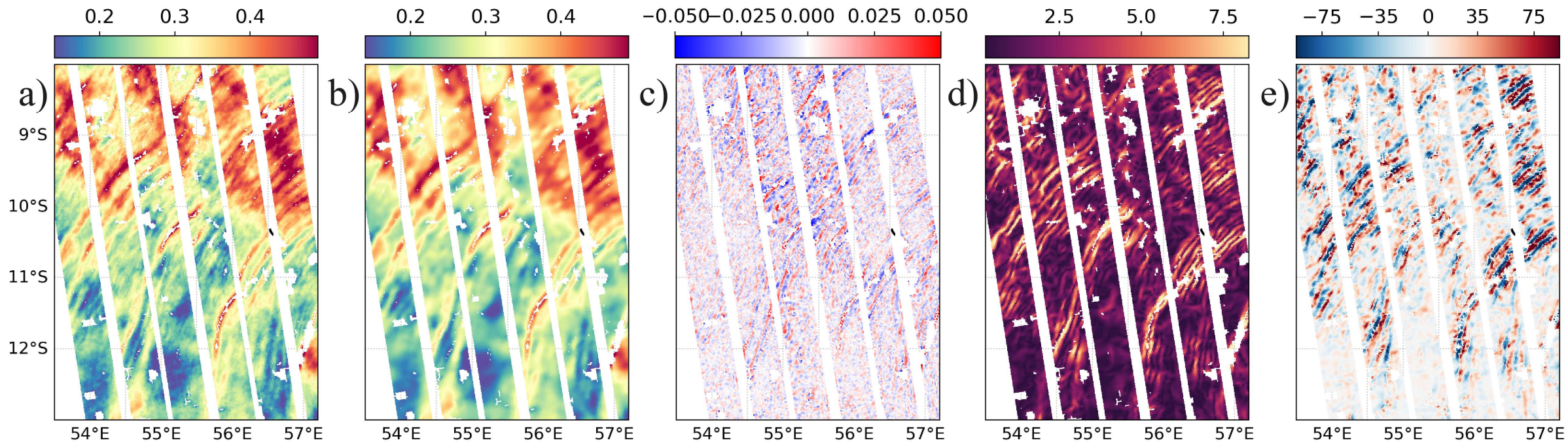

Figure A2. Denoising example of U-Net v3.0 on internal tides near Madagascar, Cycle 13, Passes 122, 400 and 428. (a) SSHA without the denoising filter (unit: m). (b) SSHA with the denoising filter (unit: m). (c) Removed SSHA content (unit: m). (d) Velocity (unit: m.s$^{-1}$). (e) Relative vorticity.

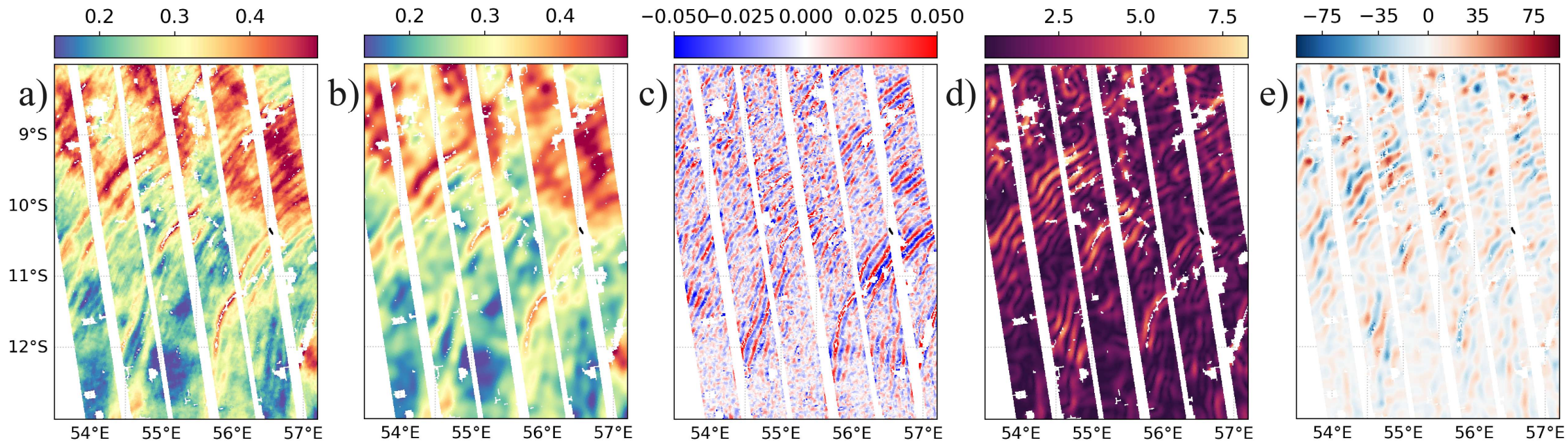

Figure A3. Smoothing example with Lanczos filter on internal tides near Madagascar, Cycle 13, Passes 122, 400 and 428. (a) SSHA without the denoising filter (unit: m). (b) SSHA with the denoising filter (unit: m). (c) Removed SSHA content (unit: m). (d) Velocity (unit: m.s$^{-1}$). (e) Relative vorticity.

This better conservation of internal tide energy can also be seen after differentiation (Figure A2e, Figure A3e). U-Net v3.0 differs in that it preserves higher values and finer details in velocity and relative vorticity. Overall, the smallest oceanic features like internal tide fronts are not always well preserved after denoising. These signals are so subtle that they are particularly difficult to separate from the KaRIn observation noise. Such signals are not simulated in the eNATL60 training dataset. The simulation does not contain any fine-scale signal (the effective resolution of the model is limited to 10 km), which limits the network's ability to denoise them properly. By enhancing the training dataset with various fine-scale motions, this issue could be solved. Even though denoising is imperfect, it remains more conservative than a smoothing approach that deletes such signals.

*b. Binned maps of removed SSHA content*

A simple approach to spatially assess the global behavior of the denoising filter is to plot the standard deviation of the removed SSHA content on a binned map. To identify different regimes of the denoising filter, the content is split into three wavelength ranges: above 60 km, between 60 and 20 km, below 20 km. A low-pass Butterworth filter from the SciPy library is used to separate these scales. The first half of April 2024 is used as input data.

In a perfect scenario, for every wavelength range, one would expect not to see any specific oceanic areas on the maps except high-wave regions, where noise is known to be stronger. A proper denoiser should not absorb any scale of oceanic variability. The previous Lanczos filter with 20-km cutoff wavelength is used as a baseline for comparison.

Above 60 km (Figure A4a), U-Net v3.0 almost achieves the intended behavior. Only very few tracks can be distinguished above 60 km in the ACC region, demonstrating its capability to preserve larger scales of SSHA. As expected, the same effect is seen with the Lanczos filter that has negligible impact on larger scales (Figure A5a).

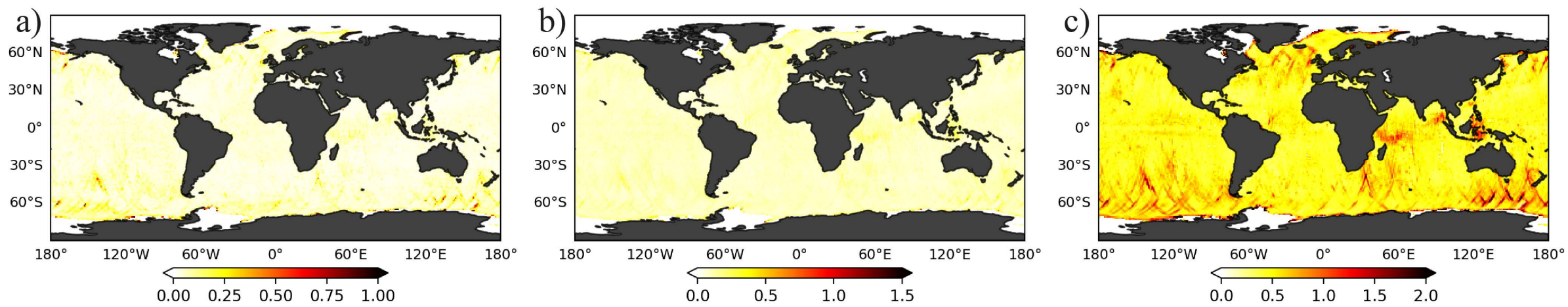

Figure A4. Binning maps of standard deviation of content removed by U-Net v3.0 algorithm on 3 ranges of wavelength (unit: cm). Std of removed content by the denoising filter for wavelengths: above 60 km (a), between 60 and 20 km (b), below 20 km (c).

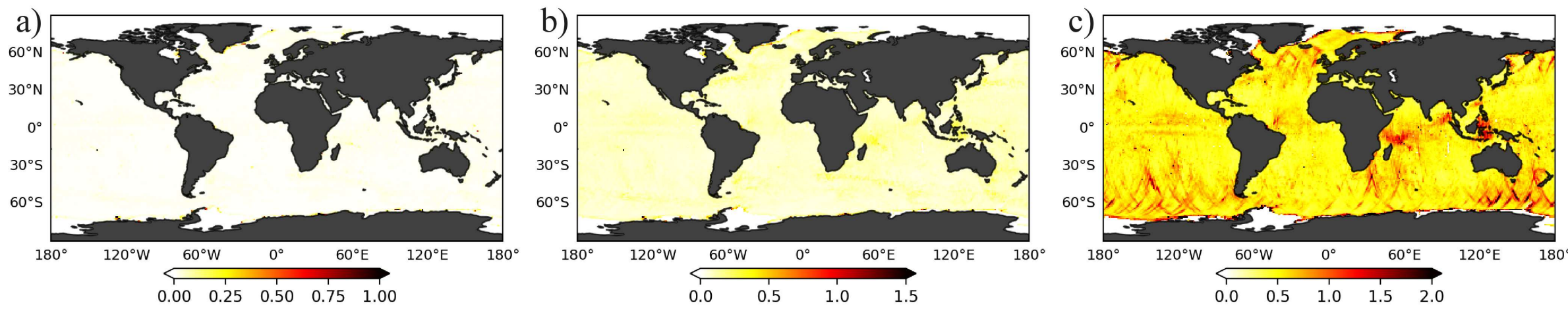

Figure A5. Binning maps of standard deviation of content removed by Lanczos on 3 ranges of wavelength (unit: cm). Std of removed content by Lanczos filter for wavelengths: above 60 km (a), between 60 and 20 km (b), below 20 km (c).

Between 60 km and 20 km (Figure A4b), internal tide regions in the equatorial band are almost indistinguishable in the removed content of the U-Net v3.0 filter. The homogeneity of the map suggests that the network does not remove more variance in specific dynamically active ocean regions. The global background is around 1.5 mm, which seems plausible given the larger-scale component of KaRIn random noise. The Lanczos filter performs very similarly.

Below 20 km, there are still no well-defined differences between denoising (Figure A4c) and a Lanczos smoothing filter (Figure A5c). Internal tide regions are systematically apparent, meaning that both absorb at least part of the variance of these small-scale signals. The network removes more SSHA content in the southern hemisphere and in the North Atlantic Ocean where wave levels are high, as anticipated. Some well-known rainfall regions are also distinguishable (e.g. Intertropical Convergence Zone). This is likely due to the presence of spurious pixels associated with rainfall that have been missed by the L3 editing step and absorbed by the filters.

Overall, the maps are very similar when using a Lanczos filter with a cutoff wavelength of 20 km. To contrast the two, the difference in standard deviation of the removed content between denoising and smoothing filters is calculated for wavelengths greater than and smaller than 20 km in Figure A6. Where pixels are blue (i.e. negative values), the content removed by the denoising filter has lower variance compared to what is absorbed by the Lanczos filter. The opposite holds for red pixels.

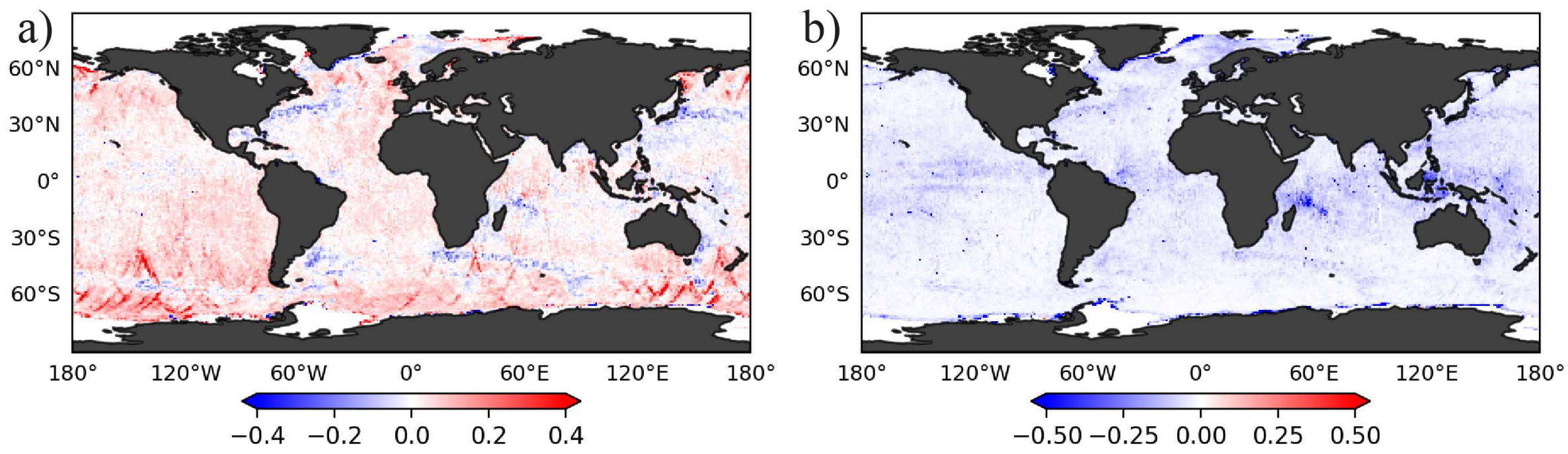

Figure A6. Difference of removed content standard deviation between U-Net v3.0 and Lanczos filters (unit: cm). (a) Scales above 20 km. (b) Scales below 20 km.

Below 20 km, the figure shows only negative values, mainly located in regions with strong baroclinic tide signals and along major western boundary currents. This suggests that the denoising filter absorbs less dynamical signal (-1.1 mm on average) everywhere at scales below 20 km (Figure A6b). However, there is no guarantee that the preserved signal is physical and not just residual noise or errors. Beyond 20 km, the denoising filter retains more variance in major ocean currents (and internal wave regions) and absorbs more variance in areas of strong waves. This result is consistent with the LECW maps in Section A.d. The denoising filter has a stronger effect in high-wave regions, whereas smoothing is spatially uniform.

*c. SSHA Power Spectral Densities (PSDs)*

To assess the effect of the denoising filter, the SSHA power spectral density (PSD) is computed. The interpretation of spectra is critical because such a representation of the signal encompasses all kinds of oceanic processes. Without a reference noiseless PSD, it is not trivial to disentangle unbalanced motion (internal tides, internal waves) from red-colored KaRIn measurement errors (e.g. from sea state or atmospheric sources) and pure instrumental noise. Simulation models might give an idea about what the noiseless SSHA spectral response should look like, but they will remain imperfect, especially when modelling unbalanced motions.

Spectra are computed using Welch's method over cycle 12. The SSHA is split into non-overlapping segments of 512 km. Each segment is detrended and tapered with a Hann window to reduce spectral leakage. Their periodograms are averaged to yield a one-sided PSD.

As previously discussed in Section 3, the noisy SSHA power spectrum displays two main tendencies. At scales below 70 km, the slope approximately follows a $K^{-2}$ trend. Above 70 km, the spectral slope follows a $K^{-3.7}$ power law. Unbalanced motion energy at these scales (e.g. internal tides) combined with the presence of noise and red-colored KaRIn measurement errors remaining in KaRIn observations could explain the break to $K^{-2}$. The SSHA filtered by U-Net v3.0 (Figure A7, orange curve) appears to retain a small share of the $K^{-2}$ slope break from 70 km to 20 km. The Lanczos filter eliminates all SSHA energy below 20 km, as expected (green curve). It constitutes a reasonable option for a user who is not interested in the smallest scales of SSHA or simply does not have confidence in what the denoising filter is preserving.

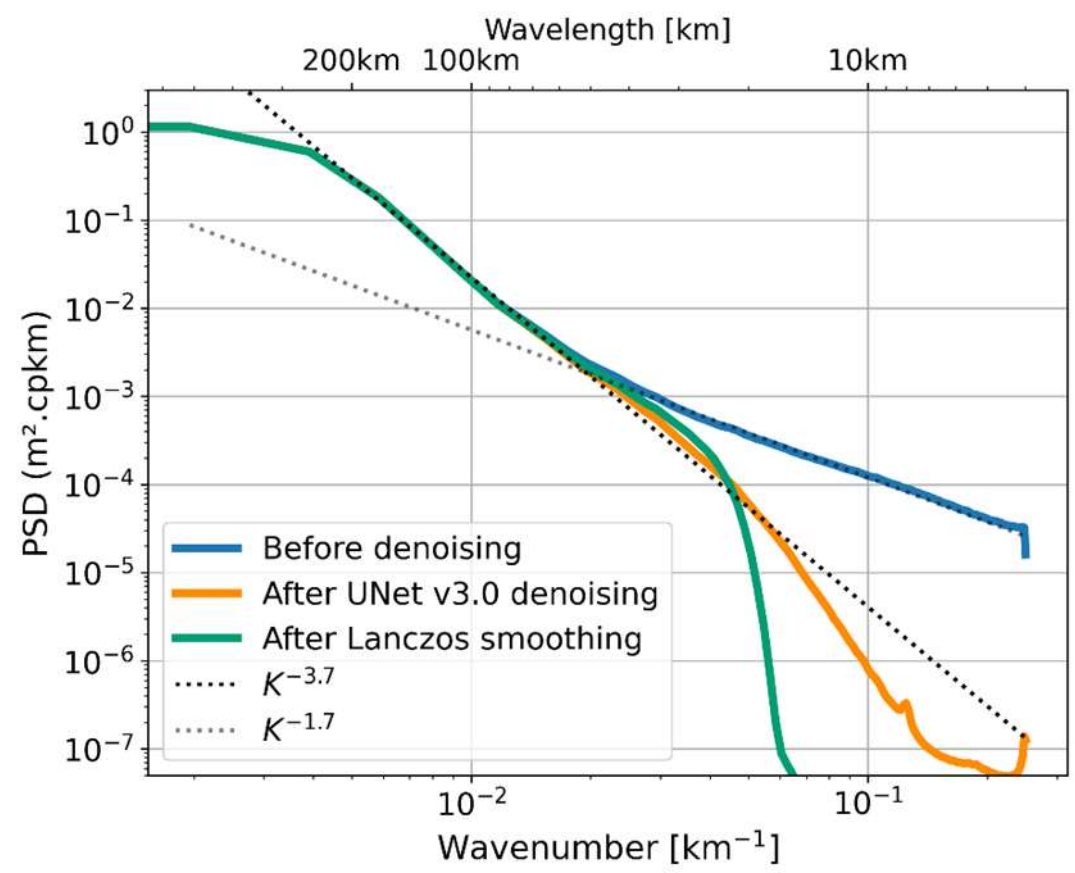

Figure A7. Global Power Spectral Densities of SSHA: PSD of noisy SSHA in blue, PSD of SSHA after UNet v3.0 filter in orange, PSD of SSHA after Lanczos filter in green (in $m^2.cpkm^{-1}$).

In addition to the SSHA PSD, the spectra of the removed content are also computed. Signal processing theory suggests white noise has a characteristic constant PSD. Thus, the PSD of the content removed by the filters is expected to be mainly flat across scales. This behavior is seen with the U-Net v3.0 filter in Figure A8. It is also very similar for the low-pass filter (green curve) although the smoothing appears to eliminate more energy at every scale. One can presume that both methods mostly remove the noise component from SWOT observations. An inappropriate denoising method would exhibit non-flat PSD of the removed content (as it was the case for previous denoising, Figure B5b). The spectrum associated with the Lanczos filter (green curve) reaches the spectrum of KaRIn SSHA observations at 20 km, with a little bump (i.e. all SSHA content below this wavelength is removed by the filter). Conversely, the PSD of the removed content by the U-Net v3.0 filter converges slowly across scales to 100% of the total SSHA energy. The spatial distribution of what is preserved by the denoising filter was already illustrated in Figure A6b (mainly internal tides and some details in western boundary currents).

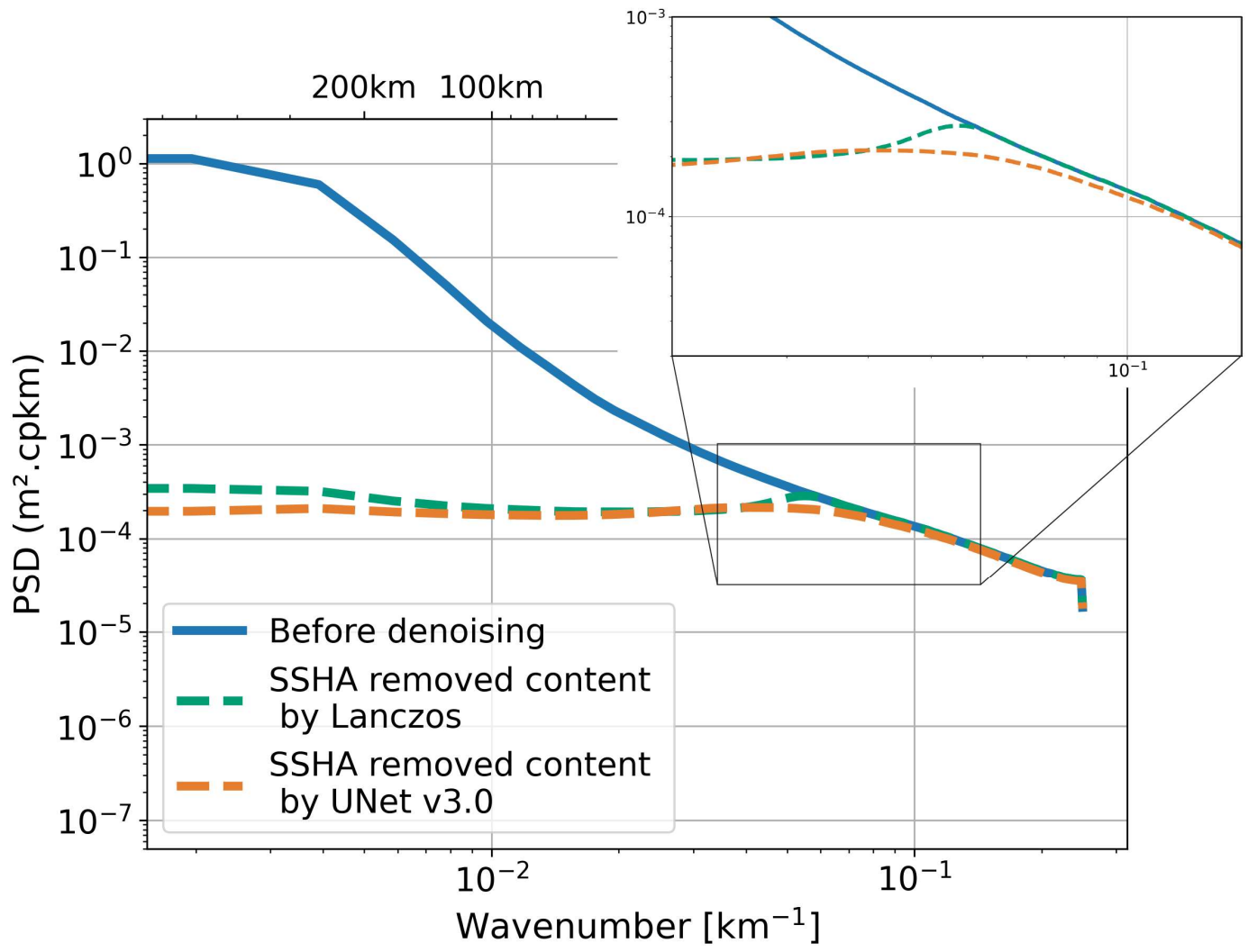

Figure A8. Global Power Spectral Densities of SSHA removed content: PSD of noisy SSHA in blue, PSD of removed content with v3.0 denoising filter in orange, PSD of removed content with Lanczos smoothing in green (in $m^2.cpkm^{-1}$).

To better evaluate spectrally the effect of denoising in the spectral domain, we assume that, ideally, the total signal seen by SWOT can be formulated as the sum of two contributions, a purely oceanic signal and a measurement error. In other words, the total signal of KaRIn observations ($KaRIn_{obs}$) can be separated into a physical part ($SSHA$) and an additive error ($\epsilon_{KaRIn}$), composed of geophysical correction residuals and instrumental errors:

$$var(KaRIn_{obs}) = var(SSHA + \epsilon_{KaRIn}) = var(SSHA) + var(\epsilon_{KaRIn})$$

In the case of denoising, the total KaRIn signal is separated into a SSHA field after applying the denoising filter ($SSHA_{filtered}$) and the removed SSHA content ($RC$):

$$var(KaRIn_{obs}) = var(SSHA_{filtered}) + var(RC) + 2Cov(SSHA_{filtered}, RC)$$

If the denoising is perfect, we should have $Cov(SSHA_{filtered}, RC) = 0$, indicating that pure oceanic signal and errors are properly disentangled. This explains why minimizing this covariance term is one of the objectives of denoising. However, this metric must be interpreted carefully. Indeed, if the denoising filter removes all the signal (i.e. $RC = KaRIn_{obs}$, $thus\ SSHA_{filtered} = 0$) the covariance would be 0. A negative covariance implies that the denoising filter is creating additional energetic signals that do not exist in KaRIn measurements. Conversely, if the covariance is positive, the denoising filter is absorbing true oceanic signal into the removed content. The same equations hold for power spectra:

$$PSD(KaRIn_{obs}) = PSD(SSHA) + PSD(\epsilon_{KaRIn})$$

The computation of the covariance term between the SSHA after filtering and the associated removed content ($RC$) is given by:

$$Cov(SSHA_{filtered}, RC) = (1/2) * (\mathrm{PSD}(KaRIn_{obs}) - \mathrm{PSD}(\mathrm{SSHA}_{\mathrm{filtered}}) - PSD(RC))$$

The total energy of noisy SSHA should be equal to the sum of the energies of SSHA after filtering and the filtered-out content. In other words, the sum of the orange curve of Figure A7 with the orange dotted curve of Figure A8 should be equal to the blue curve. However, the sum of the two PSDs does not equal the total PSD of unfiltered KaRIn observations. This remaining dependence between removed content and filtered SSHA is represented by the covariance term. To better emphasize covariance at small scale, the covariance is normalized by the spectrum of $KaRIn_{obs}$ (Figure A9).

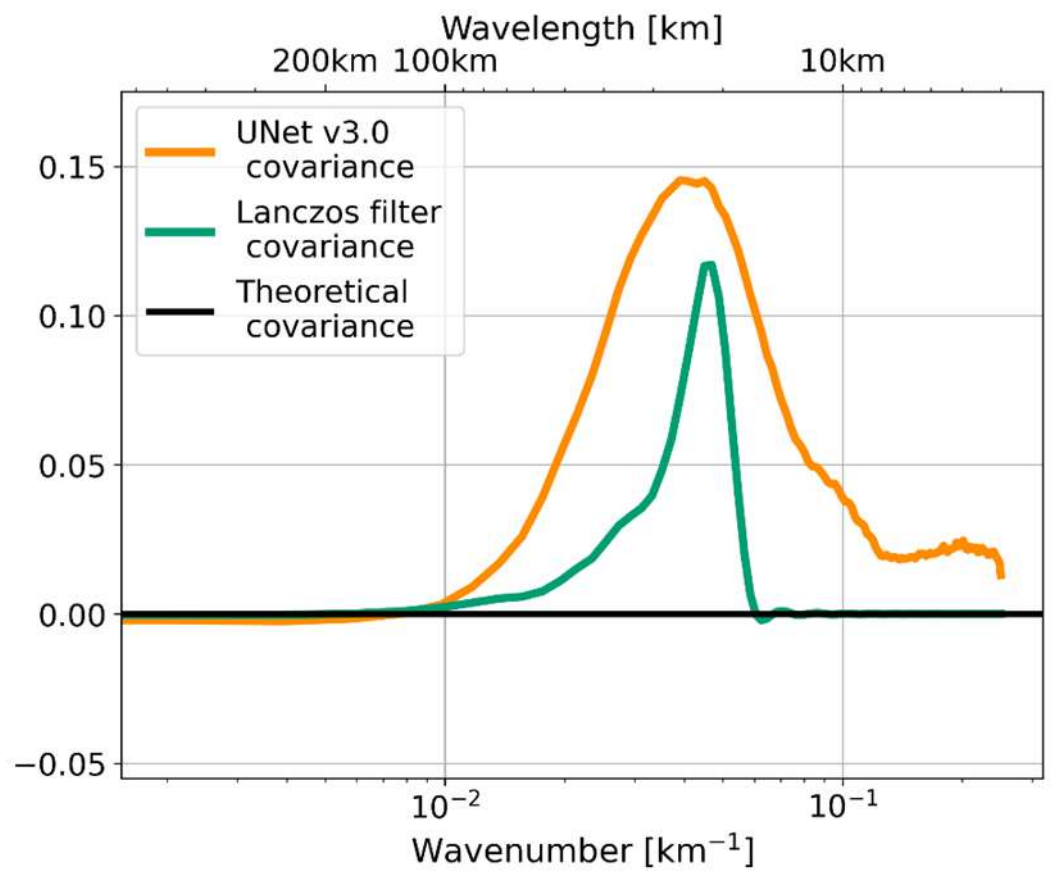

Figure A9. Global covariance between SSHA after filtering and the associated removed content, estimated from power spectral densities: Covariance of U-Net v3.0 filtering in orange, Covariance of Lanczos filtering in green.

If the covariance is negative, this implies:

$$PSD(KaRIn_{obs}) < PSD(SSHA_{filtered}) + PSD(RC)$$

which means that there is more energy in the output of the network than in the KaRIn observations; the denoising filter has invented artificial SSHA content. Conversely, if the covariance is positive, this implies that the energy of the filtered SSHA added to the energy of removed content does not reach the energy of KaRIn measurements; denoising has necessarily absorbed a share of true oceanic variability. In the best-case scenario, covariance should be zero; the denoising filter succeeds in disentangling noise from real oceanic features.

The covariance is positive at all scales for the U-Net v3.0 and Lanczos filters. Below 80 km, the network absorbs a significant share of total KaRIn signal. Near 20 km, almost 15% of the total KaRIn energy is contained in the covariance. This correspo ds to the amount of signal that is not properly retrieved. The sharp transition of the Lanczos filter covariance to 0 near 20 km is consistent with the fact that all signal is filtered out below this cut-off. On this specific metric, the Lanczos exhibits lower covariance values across wavelengths. As previously noted, if no signal remains, the covariance will be zero. The filter, by cutting sharply the signal, does not distribute covariance energy across frequencies. The covariance will be high at the cutoff, but zero elsewhere. The denoising filter, by trying to retrieve the finest possible scales, therefore shows a higher covariance on a wider scale range.

Thanks to the SWOT 1-day revisit phase, it is possible to estimate spectra for 24 h differences. The PSD of the difference between two consecutive SSHA separated by 24 hours is computed. An equivalent calculation is performed for removed content and SSHA with the denoising filter applied (Figure A10). Such analysis was already done by Dibarboure et al. (2025) for the previous U-Net algorithm and most of the interpretations remain unchanged:

- A large fraction of mesoscale signal is canceled out in 24h (blue curve)
- $K^{-2}$ slope is no longer discernible in 24 h, particularly at finer scales, which makes sense if the slope break is indeed mainly due to unbalanced motions
- The PSD of SSHA after denoising follows the same energy loss as unfiltered SSHA at larger scales after 24 h differences. In addition, smaller scales are expected to decorrelate quickly, within 24 h. This is consistent with the fact that both curves slowly reach their respective denoised SSHA PSDs (same colors but in dotted lines) at these scales.
- The mean PSD of the removed content (dashed red line) almost merges with the PSD of the 24 h difference in removed content (red line for v3.0). This indicates that the content removed by U-Net v3.0 is decorrelated within 24 h as expected from measurement errors and other physical effects. The filter seems to mainly remove random noise.

As already highlighted by Dibarboure et al. (2025) there is a missing variance gap between removed content PSD and 24h difference removed content PSD, even with new the U-Net. The two red curves do not yield the same amount of energy. This variance gap is on the order of 10% missing variance (previously 30% in Dibarboure et al., (2025)). However, the energy gap is uniformly distributed across wavelengths and suggests that the U-Net v3.0 might not have absorbed a specific oceanic signal (e.g., higher modes of internal tides or a small fraction of the stationary mesoscale), but something else. In fact, this missing variance is mostly explained by the presence of Mean Sea Surface (MSS) artifacts in the SSHA (Appendix A, Section f). The Lanczos filter is affected similarly. For comparison, the missing variance for a Lanczos filter with a 20-km cutoff wavelength is close to 18%. The smoother absorbs much more temporally coherent signals.

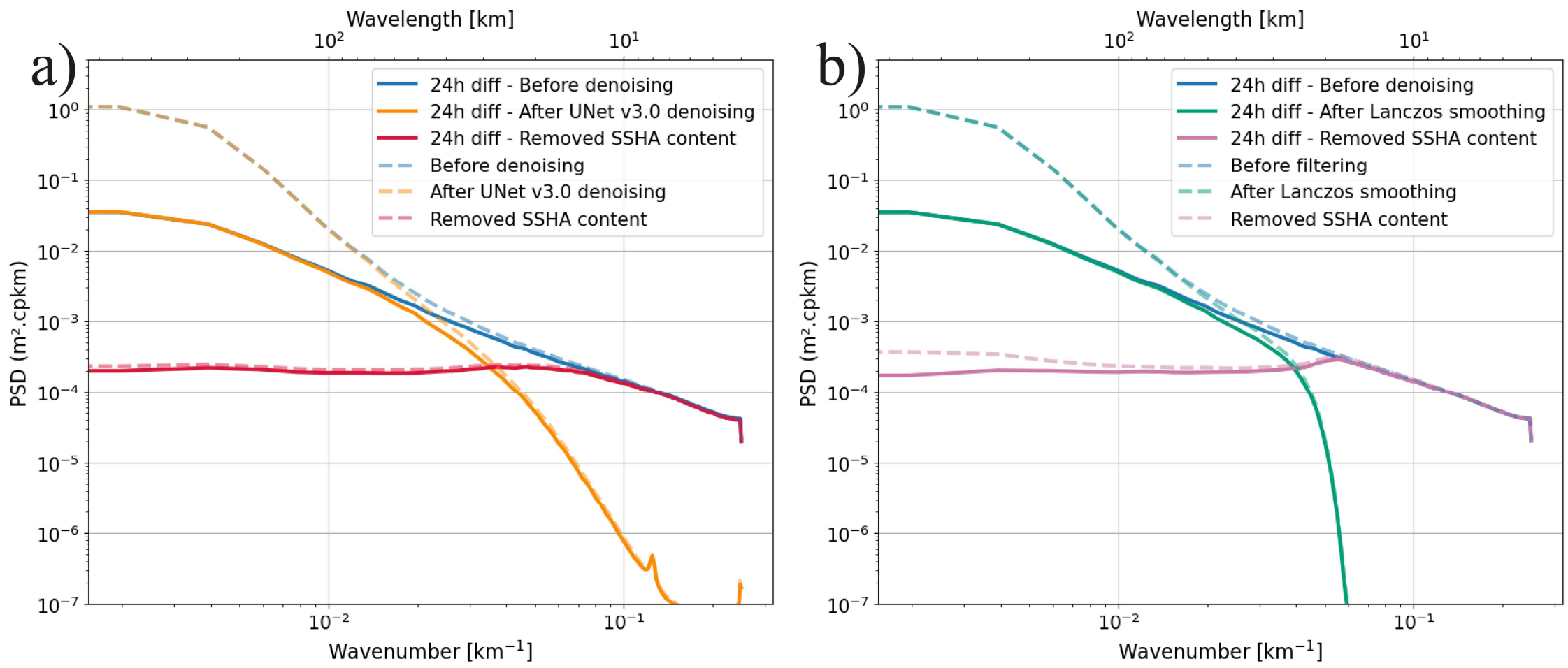

Figure A10. Global PSD of the KaRIn SSHA. Noisy SSHA (blue curve). (a) SSHA with U-Net v3.0 filter applied (orange curve). Removed content by U-Net v3.0 filter (red curve). (b) SSHA with Lanczos filter applied (green curve). Removed content by Lanczos filter (pink curve). Dotted curves are the same variables but for 24h differences from one-day phase (unit: $m^2.cpkm^{-1}$). The input data is taken from cycles 518 and 519.

*d. Local Effective Cut-off Wavelength (LECW)*

The KaRIn instrumental noise level is geographically heterogeneous and is modulated by oceanic conditions (e.g. higher waves correspond to higher noise levels). Therefore, a proper denoising approach should be able to adapt its filtering strength spatially, according to the local noise amplitude. Conversely, a basic smoothing approach affects the same wavelength everywhere and at all times, which makes it unsuitable for KaRIn swath denoising. A more suitable smoothing technique would require at least a convolutional kernel size that adapts geographically.

We defined the Local Effective Cut-off Wavelength (LECW) as the threshold wavelength at which the signal-to-noise ratio is considered to be dominated by noise. To remain consistent with the classical definition of a cut-off wavelength, the limit is set where more than 50% of the total SSHA energy is absorbed by the denoising approach:

$$LECW\colon\ log\left(\frac{PSD(KaRIn_{obs})}{PSD(RC)}\right) \leq 0.5$$

To spatially assess the effect of denoising, this value is calculated in longitude-latitude boxes and displayed on a map (Figure A11). The results of U-Net v3.0 are already analyzed in Section 3. In addition, the same metric is computed for the Lanczos filter with a 20 km cut-off in larger boxes. As expected, the map is overall spatially homogeneous with an LECW close to 20 km everywhere.

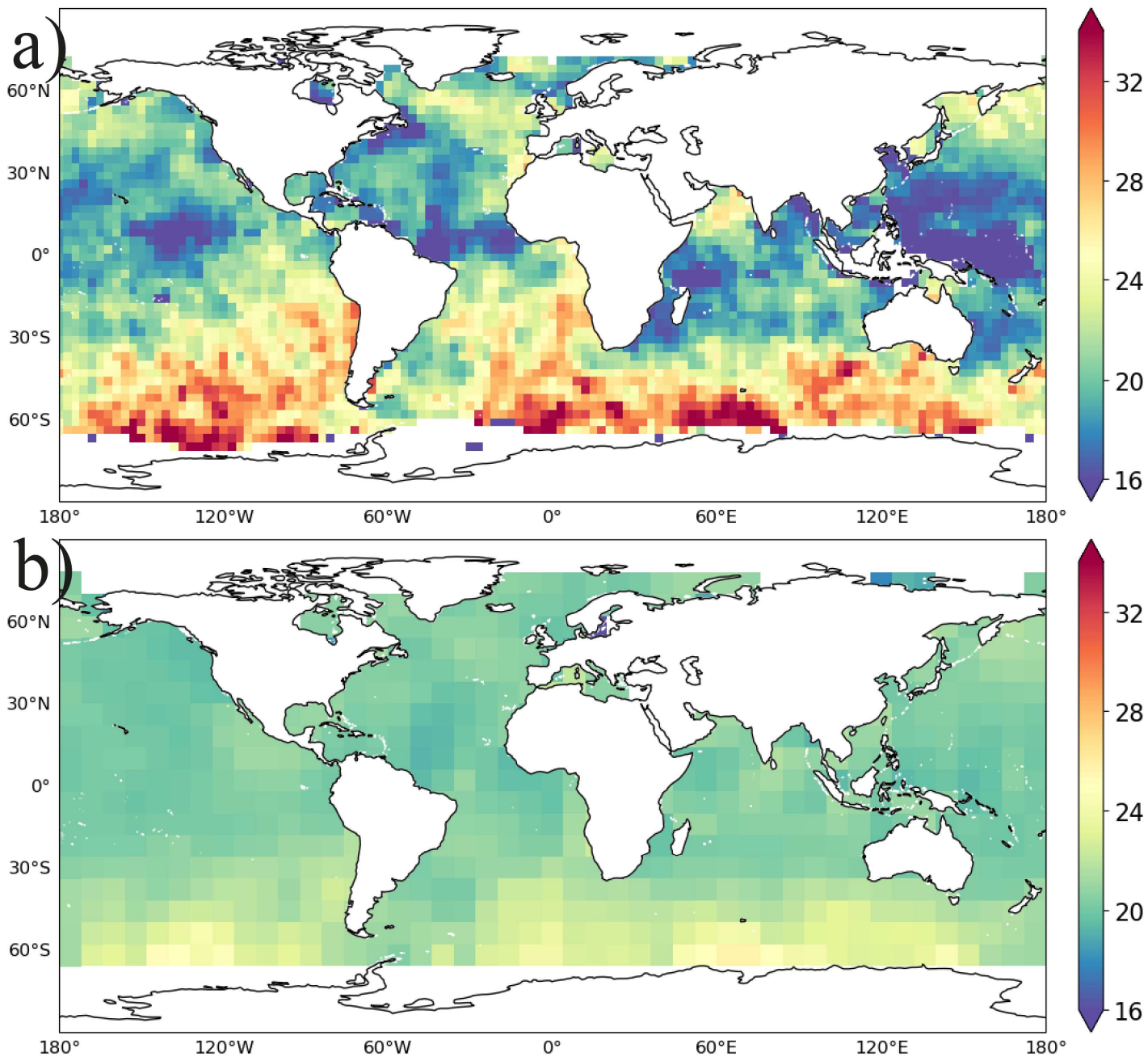

Figure A11. Local Effective Cut-off Wavelength (unit: km). (a) U-Net v3.0. (b) Lanczos filter. The 17th cycle of SWOT is used as input data.

The values of this metric cannot be used as an absolute measure. Changing the threshold would lead to a different LECW. This metric is mainly employed as a visualization tool to assess spatial behavior of a denoising method. Reducing the LECW as much as possible does not guarantee better results. An optimal value of the LECW exists but can only be found with true noiseless SSHA reference.

*e. One-day regional SSIM*

Without access to noiseless real SSHA, full-reference image quality assessment metrics like Peak Signal-to-Noise Ratio (PSNR) are unusable. However, it is possible to repurpose these metrics for our case study. The SWOT one-day phase provides an opportunity to compare the filtered swaths from one day to the next. Mesoscale oceanic structures have changed little in 24 hours. It is therefore possible to compare the swaths from consecutive one-day cycles.

SSIM (Structural Similarity Index Measure; Wang et al., 2004) can be used to evaluate the similarity between two images. Unlike PSNR, which estimates the similarity between two images by calculating differences in intensity between pixels, SSIM is a measure that is designed to match human visual perception. The method is based on the structural information of an image (the idea that nearby pixels have a strong dependence on each other). SSIM is computed as a function that combines contrast, luminance, and local structure of images. It typically varies between 0 and 1, with 1 representing a perfect match between the reference image and the image compared, and 0 representing two completely different images.

To avoid masking local variability by computing a global score, segments of swaths in different oceanic regions are selected. For the following results, the selected segment is in the ACC, where SWH is high. This analysis could be conducted on crossover diamonds to obtain an even smaller revisit window.

Two SSHA samples separated by 24 hours are expected to look very similar because mesoscale events do not move quickly enough to be different one day to the next (as previously seen in Figure A10). However, fast small-scale oceanic features can be spatially displaced and unseen in less than a day. This explains why we should observe a high SSIM (close to 1) between two consecutive filtered swaths but not exactly 1. Otherwise, it would mean that denoising has absorbed all the small-scale variability and that the two swaths are identical. There is no prior knowledge of what the theoretical optimal SSIM between two noiseless swaths should be. One can reasonably expect that two filtered swaths should have a higher SSIM than two unfiltered swaths because the noise is totally decorrelated within 24 hours.

The U-Net v3.0 denoising filter and Lanczos smoothing consistently slightly improve spatial coherence between consecutive swaths as one would expect from a proper noise-mitigation approach (Figure A12). The improvement is identical for both methods.

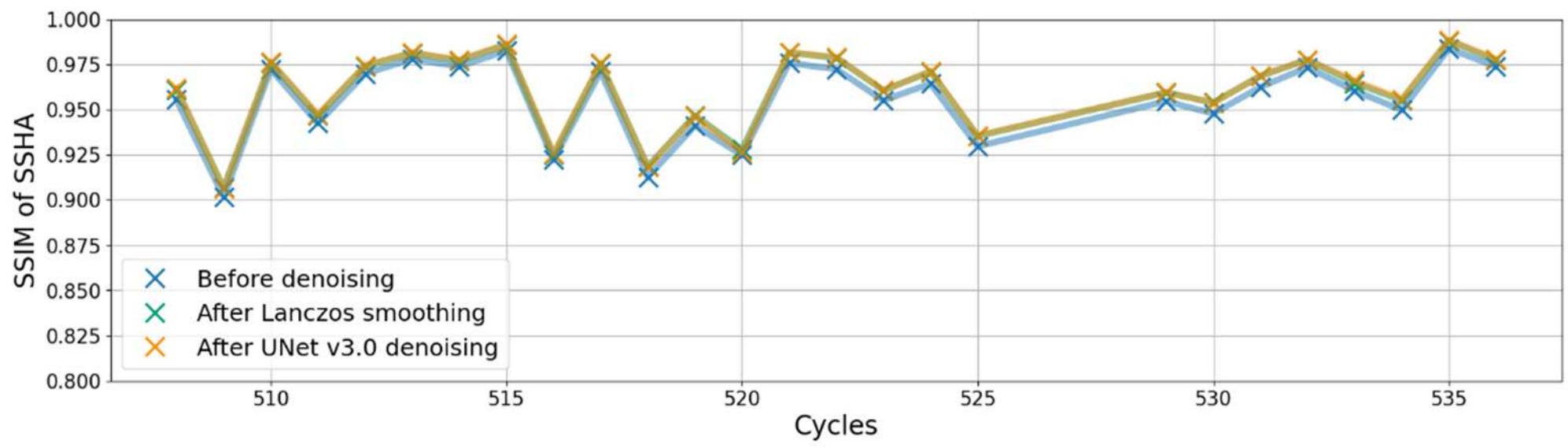

Figure A12. Local SSIM between consecutive SSHA swaths. Noisy SSHA in blue, SSHA after U-Net v3.0 filter in orange, SSHA after Lanczos 20 km cut-off filter in green. The input data is taken from 1 May 2023 to 31 May 2023 (Pass 2).

Because noise decorrelates within one day, SSIM between two consecutive removed-content fields should be zero (if there are no temporally coherent errors in KaRIn measurements). Unfortunately, real SWOT data can contain daily recurring errors and correction residuals that are absorbed by denoising or smoothing (see Appendix A, Section f). This partially explains why the content removed by the U-Net v3.0 filter, although its SSIM is flat and very low, keeps a small amount of coherence (Figure A13, orange curve).

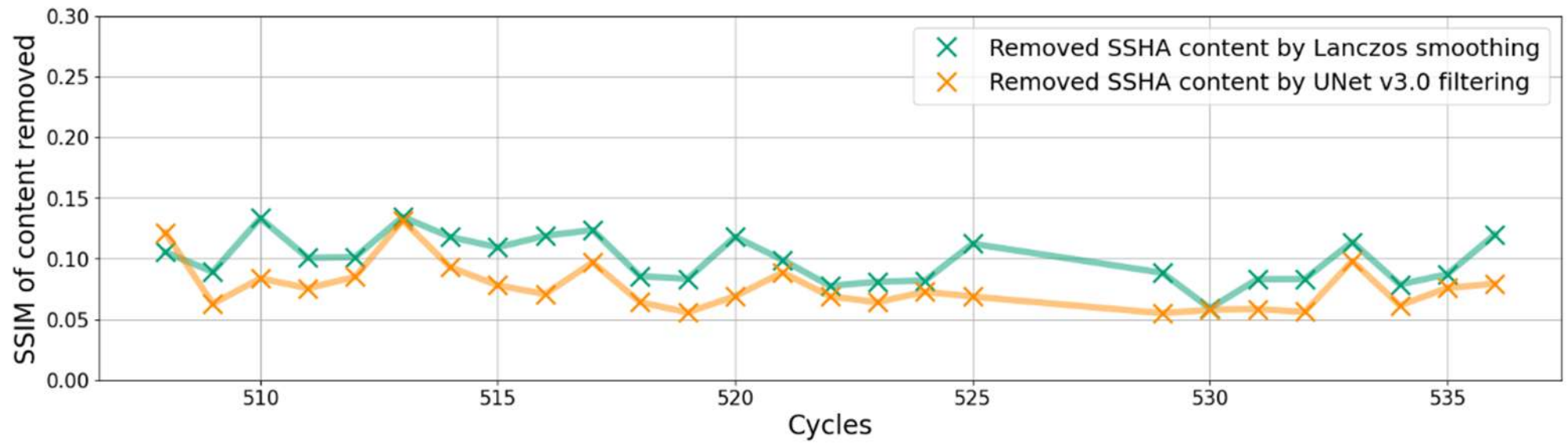

Figure A13. Local SSIM between consecutive removed content swaths. Removed content of the U-Net v3.0 denoising filter in orange, Lanczos 20 km filter in green. The input data is taken from 1 May 2023 to 31 May 2023 (Pass 2).

The network removes the same quantity of random noise from day to day. The same conclusions can be drawn from the Lanczos filter. In a more general case, when this score is high, the noise-mitigation method has oversmoothed the SSHA by absorbing part of the temporally coherent oceanic features.

Following the same rationale as for SSHA samples separated by 24 hours, the SSIM of velocities separated by 24 hours can be computed. The right balance needs to be found: The SSIM of two filtered and derived swaths should increase after denoising but not excessively. The U-Net v3.0 and Lanczos filters exhibit flat and consistent SSIM of velocities (Figure A14). Both significantly improve the coherence of consecutive swaths.

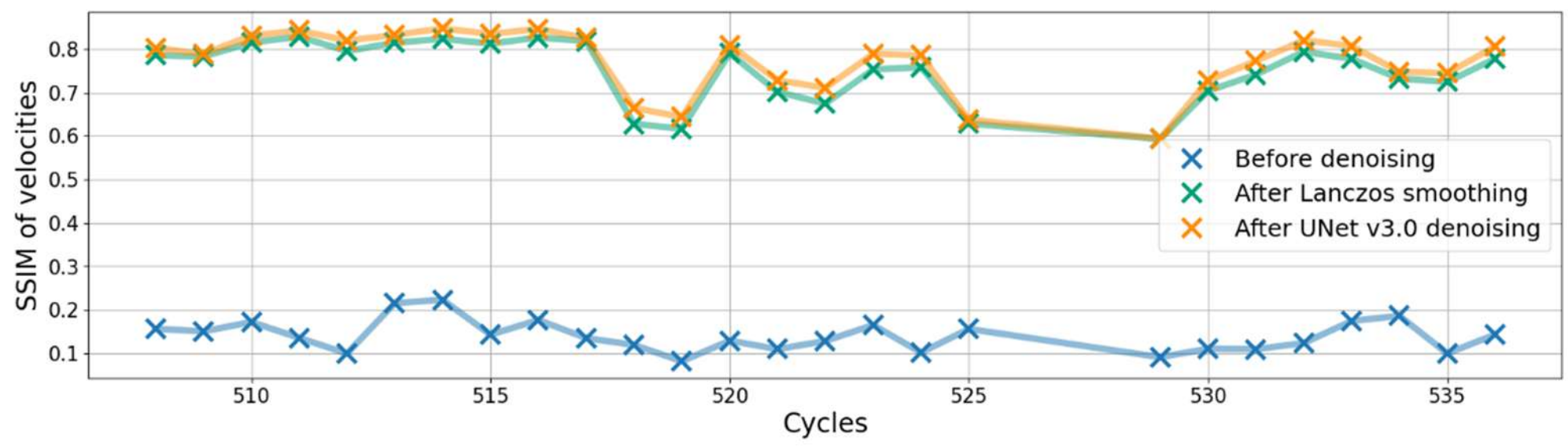

Figure A14. Local SSIM between consecutive geostrophic velocity swaths. Velocity derived from noisy SSHA in blue, from SSHA after U-Net v3.0 filter in orange, from SSHA after Lanczos 20 km cut-off filter in green. The input data is taken from 1 May 2023 to 31 May 2023.

One-day regional SSIM can highlight the structural coherence of SWOT signals but it should be interpreted carefully. This metric should always be supported by a thorough understanding of the input data. For the sake of clarity, the focus has been placed on a unique swath sample. Such a metric could be computed over all oceanic regions of the entire one-day repeat phase and displayed on a map.

*f. Mean Profile of removed SSHA content*

A potential drawback of denoising could be the absorption of the same signal repeatedly over time in the same region. As previously seen in Figure A10, there is a small amount of coherence remaining between consecutive 24 hours removed-content fields. Proxies of mean profile of the filtered swaths and the removed content are computed to understand more precisely what the data contain. The proxies are estimated over the entire 1-day Cal/Val phase of SWOT by averaging the same passes across cycles. No oceanic patterns should be noticeable on the swaths of mean removed content. Otherwise, it would suggest that the denoising approach is removing some repetitive physical signal or temporally correlated errors.

As illustrated in Figure A15, this metric enables us to detect what repetitive signals the denoising filter has absorbed in the ocean. These signals are mainly artifact residuals in the high-frequency components of Mean Sea Surface correction applied in the L3 product. They are static, small-scale and blended with 2-km noise in the SSHA. For visualization purposes, The MSS is filtered using a 20-km high-pass filter and is displayed as background in Figure A15. The presence of small blue and red dots in the MSS and on the swath indicate us that the U-Net v3.0 filter is absorbing part of these residuals. This was not initially expected from the denoising filter but does not appear to be an issue since it helps eliminate an error. However, this confirms our previous intuition from PSD that very small-scale signals are partially absorbed by the network.

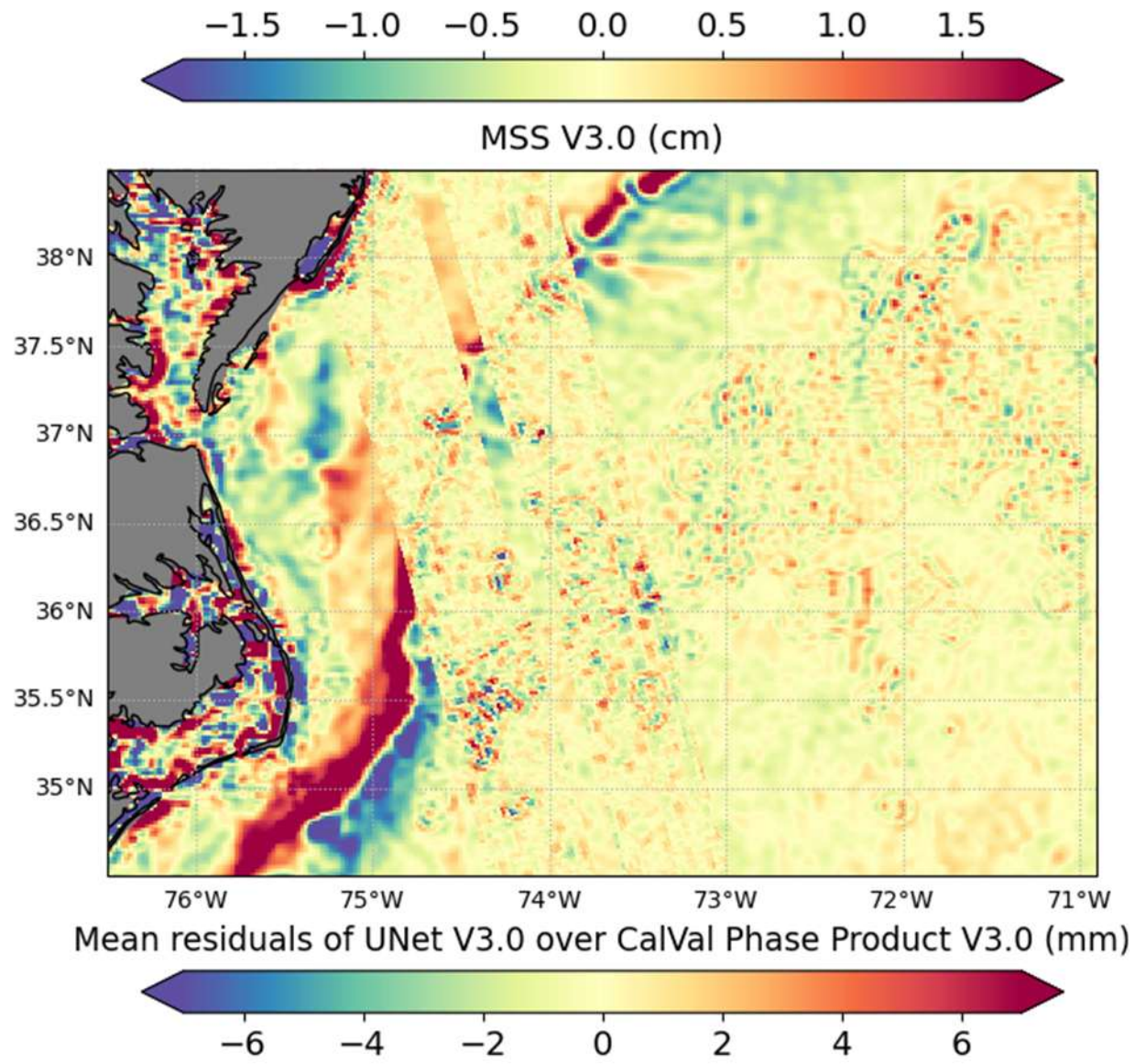

Figure A15. Mean of removed content of U-Net v3.0 filter over the one-day repeat phase above high-frequencies of MSS.

A poorly tuned denoiser (or smoother) that is repeatedly absorbing real oceanic features would exhibit them on the mean swath of removed content (Figure B8). The distribution of the U-Net v3.0 mean removed content is centered and concentrated around zero which is what one would expect from a pure noise field. The absolute mean of U-Net v3.0 removed content profiles is 0.6 mm and with a 2.5 mm standard deviation. These values are driven by the quality of the MSS. Thanks to the new MSS CNES_CLS_2025 (AVISO/DUACS, 2025) in version 3.0 of the L3 LR products, most of these artifacts are mitigated.

Any low-pass filter will yield very similar results because it would also absorb MSS residues. This metric has not been computed using the Lanczos filter due to the very slow computation of the filter and the large amount of data that needs to be filtered. However, any other competitive noise-mitigation method that wishes to compare itself to the method presented here should calculate this metric. With the exact same base LR product, it would then be possible to see which method absorbs less temporally coherent signals.

*g. SWOT KaRIn/S3A crossovers*

To compare denoising performance with an independent source of data, the coherence of KaRIn wide-swath observations with other satellite measurements can be computed at crossover points. In this benchmark, S3A satellite (at 20 Hz sampling) is selected arbitrarily. This comparison requires several strong assumptions. The metric is initially based on Nencioli et al. (2025) methodology. Even if the data differ slightly, the approach is identical. The main arguments and hypotheses needed to use it as a denoising metric are written below but the reader is invited to read their publication for further details.

It is necessary to assume that the total signal seen by a satellite can be formulated as the sum of two contributions, a true oceanic signal observed and an instrumental error. The total signal of KaRIn observations can be disentangled into a physical part ($SSHA$) and an additive error ($\epsilon_{KaRIn}$), composed of geophysical correction residuals and instrumental errors:

$$KaRIn_{obs} = SSHA + \epsilon_{KaRIn}$$

Second, it is necessary to suppose that S3A and SWOT measurements are totally independent, which may be questionable because they might share atmospheric and geophysical corrections (e.g. Dynamic Atmospheric Correction, ocean tide…). This implies that S3A altimeter and SWOT KaRIn noise components are also independent. Thus, as formulated by Nencioli et al. (2025):

$$\Delta Obs_{KaRIn/S3A} = (\mathrm{SSHA} + \epsilon)_{\mathrm{KaRIn}} - (\mathrm{SSHA} + \epsilon)_{\mathrm{S3A}} = \Delta\mathrm{SSHA}_{\Delta\mathrm{t}} + \epsilon_{KaRIn} + \epsilon_{S3A}$$

where $\Delta SSHA_{\Delta\mathrm{t}}$ is the difference of SSHA between the two measurements (i.e. how much the ocean has changed during $\Delta_t$). For a reasonably small time gap ($\Delta_t$) between the measurements of the two satellites, the ocean state should not have changed significantly:

$$if \Delta_t \longrightarrow 0, \Delta SSHA_{\Delta\mathrm{t}} \approx 0$$

With all these assumptions, the KaRIn/S3A variability difference is:

$$\begin{aligned} var(KaRIn_{obs} - S3A_{obs}) &= var(\Delta\mathrm{SSHA}_{\Delta\mathrm{t}} + \epsilon_{KaRIn} + \epsilon_{S3A}) \\ &= var(\Delta\mathrm{SSHA}_{\Delta\mathrm{t}}) + var(\epsilon_{KaRIn}) + \mathrm{var}(\epsilon_{S3A}) \end{aligned}$$

Which can be simplified for a sufficiently small $\Delta_t$ :

$$var(KaRIn_{obs} - S3A_{obs}) \approx var(\epsilon_{KaRIn}) + var(\epsilon_{S3A})$$

The improvement of coherence with S3A by denoising KaRIn data can be formulated by:

$$\begin{aligned} var\left(SSHA_{filtered} - S3A_{obs}\right) &= var\left(SSHA_{filtered} - (SSHA_{S3A} + \epsilon_{S3A})\right) \\ &= var\left(\Delta SSHA_{filtered/S3A}\right) + var(\epsilon_{S3A}) \end{aligned}$$

Where $\Delta SSHA_{filtered/S3A}$ is the difference between SSHA observed by S3A and the filtered SSHA of SWOT. Finally, the coherence gain from denoised swaths against noisy KaRIn swaths can be expressed as:

$$\begin{aligned} gain &= var(KaRIn_{obs} - S3A_{obs}) - var(SSHA_{filtered} - S3A_{obs}) \\ &= var(\epsilon_{KaRIn}) + var(\epsilon_{S3A}) - (var(\Delta SSHA_{filtered/S3A}) + var(\epsilon_{S3A})) \\ &= var(\epsilon_{KaRIn}) - var(\Delta SSHA_{filtered/S3A}) \end{aligned}$$

Maximizing the gain associated with denoising implies minimizing $var(\Delta SSHA_{filtered/S3A})$. The best denoiser (or smoother) is the one that brings the KaRIn SSHA the closest to the S3A measurement. In a perfect denoising scenario, the metric would give an estimation of KaRIn error. Nevertheless, this metric is based on strong assumptions and is very dependent on the time difference between the measurements of the two satellites.

A total of 5,951 crossovers over almost one year are used, from 1 January 2024 to 31 December 2024. The S3A 20 Hz nadir data come from the L2P LR nadir product available on Aviso+. For the following analysis, the case where $\Delta t_{max} = 60$ minutes is considered. The crossovers are located on each KaRIn half-swaths, the inter-nadir space is not interpolated to ensure continuous crossover segments. An S3A track that intersects a KaRIn swath therefore gives two crossovers, one on each half-swath. The crossovers are restricted to latitudes between -70° and 70°. The KaRIn data under the S3A 20 Hz crossovers are upsampled at the nadir sampling frequency. Crossovers that are too short (<7 km) are excluded from further analysis. With this methodology, the average crossover length is 141 km (median is 100 km) and the crossovers are distributed all over the globe (Figure A16).

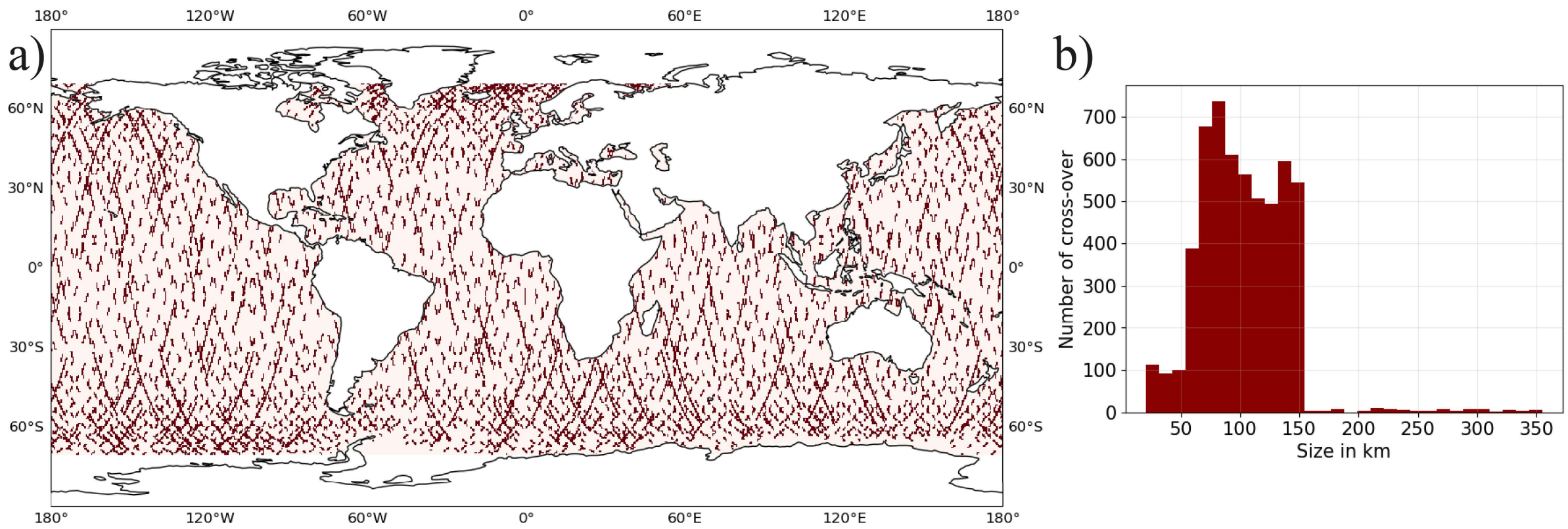

Figure A16. (a) Spatial coverage of S3A/SWOT crossovers separated by less than 1 hour. (b) Distribution of S3A/SWOT crossovers lengths separated by less than 1 hour. The input data is taken from 1st January 2024 to 31st December 2024.

The coherence gain is estimated by the difference between the variance of the denoised KaRIn SSHA relative to S3A and the variance of raw KaRIn SSHA relative to S3A. It can be plotted on a binned map to assess where and to what extent the denoising method improves the consistency between S3A/KaRIn measurements (Figure A17). The U-Net v3.0 filter homogeneously reduces the variance between the filtered SSHA and the S3A measurements compared to noisy KaRIn observations. This leads to an overall median RMS reduction of 4 mm (the mean can be strongly impacted by outliers). Some spurious red pixels remain near Antarctica and coastal areas, which is not surprising because the observations from the two satellites may differ substantially or spurious values may remain in the analyzed data.

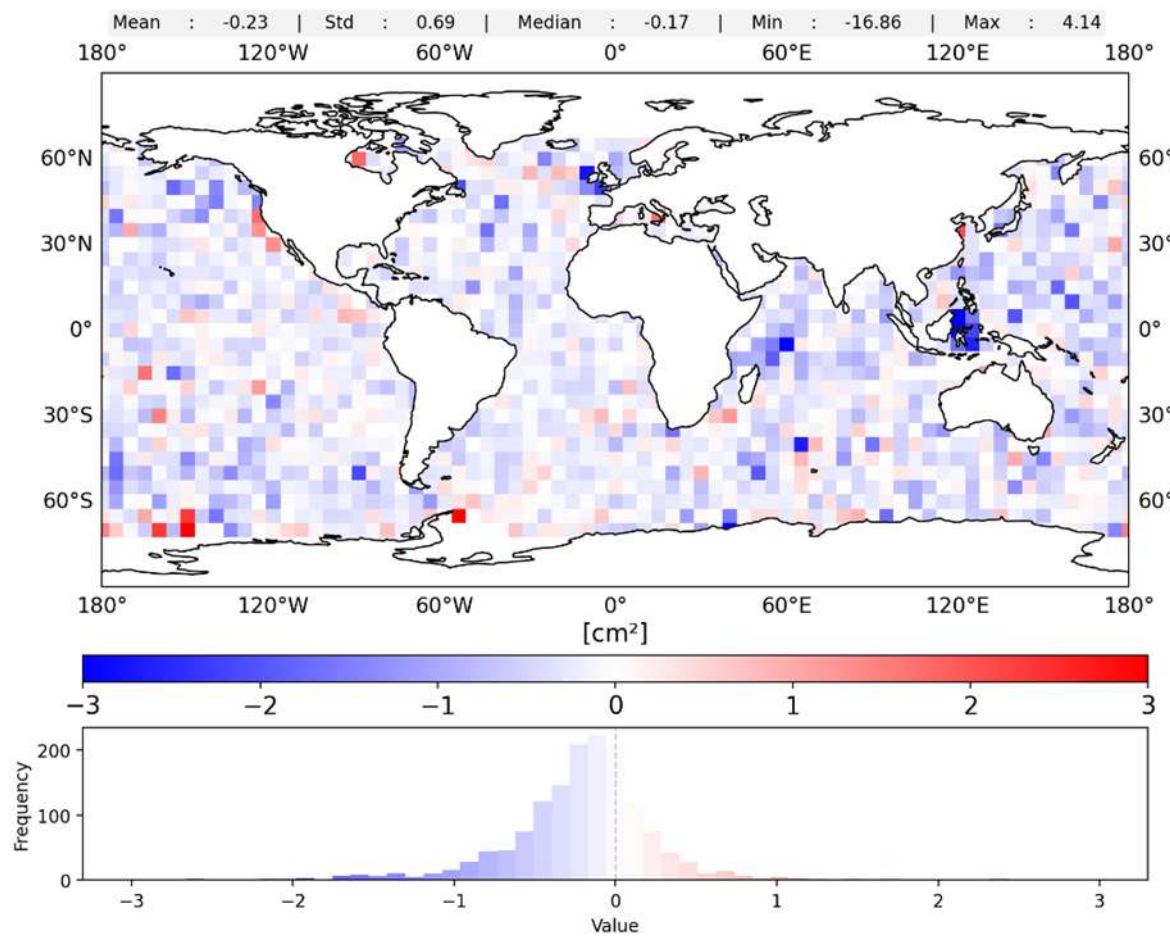

Figure A17. KaRIn/S3A coherence gain from the U-Net v3.0 denoising filter (unit: cm²).

Instead of using noisy KaRIn observations as a basis for comparison, the metric is also adapted to compare two noise-mitigation approaches. The U-Net v3.0 filter and a low-pass filter (here, a Butterworth filter with 20 km wavelength cutoff) reach almost the same gain (Figure A18). The improvement of the low-pass filter is mainly localized in internal tide regions and in the low latitudes. It does not constitute a clear improvement compared to the network. Theoretically, the lower the $\Delta t_{max}$, the more similar the ocean seen by the two satellites is. For $\Delta t_{max} \geq 60min$, smoothing small-scale dynamics will bring the KaRIn SSHA closer to S3A measurements, due to small-scale oceanic displacement between the two measurements. By smoothing the KaRIn SSHA through low-pass filtering, only the slow mesoscale features are kept, which are the ones shared with S3A.

Furthermore, from an instrumental perspective, S3A cannot detect the same oceanic motions as SWOT. Internal tides, coastal tides and submesoscale eddies are not detected by conventional nadir altimetry (Archer et al. 2025) but are seen by SWOT for the first time. Preserving these small-scale signals as much as possible will reduce the gain according to this metric. This explains the presence of red pixels in Indonesia and north of Madagascar where small, fast unbalanced motions that are not detected by S3A are predominant.

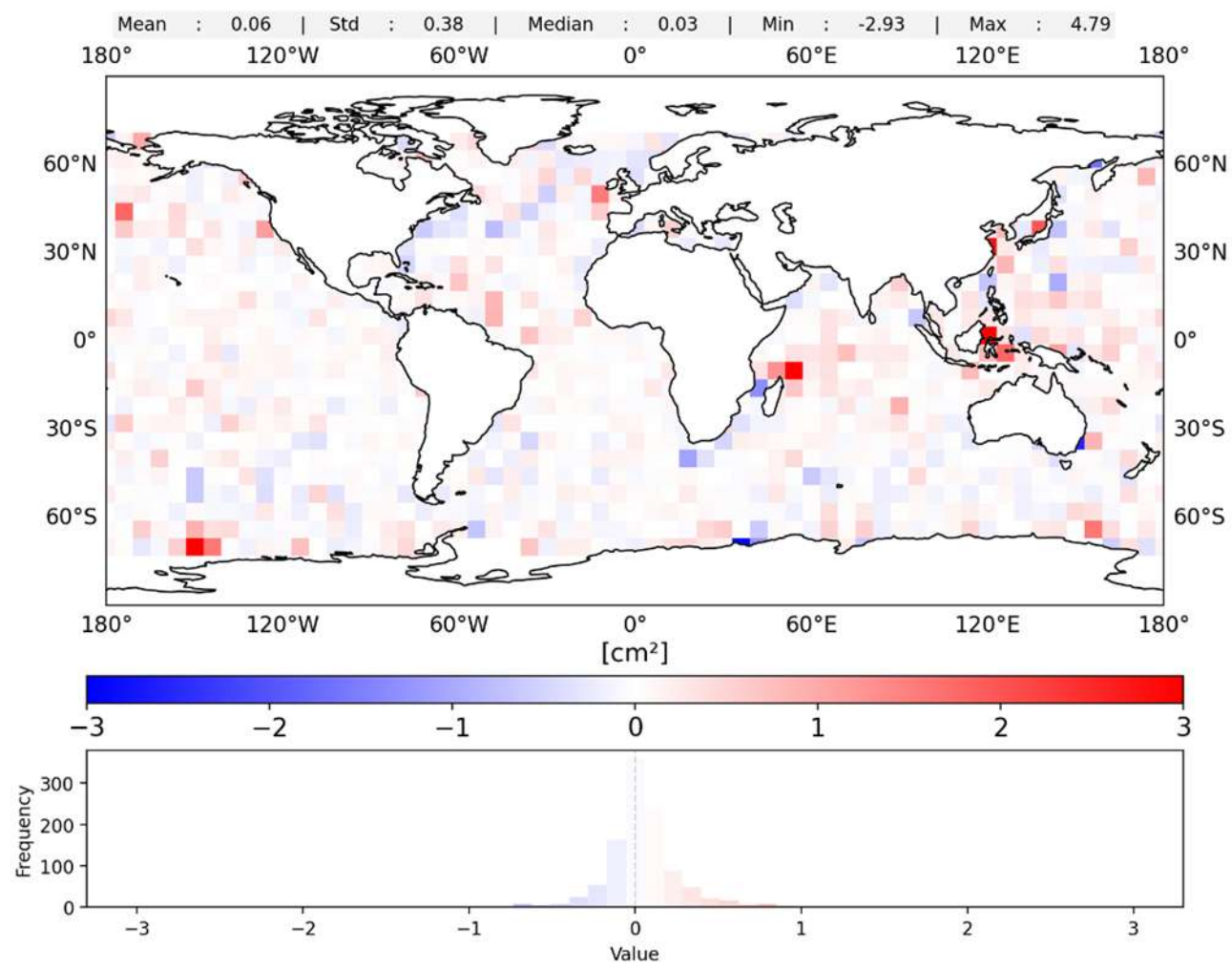

Figure A18. Low-pass filtering gain in coherence with S3A compared to U-Net v3.0 filter (unit: cm²).

## APPENDIX B

### Evaluative Benchmark, results of U-Net v2.0 filter

Readers should note that the previous version of U-Net for the 2-km resolution was run on a previous product version (v2.0). Although the product version has an impact on the overall quality of the results, it does not change the conclusions about previous CNN performance. The calibrated and edited Sea Surface Height Anomaly (SSHA) of the SWOT Level-3 Ocean products v2.0 (AVISO/DUACS, 2024b) is used here as input for all metrics.

For U-Net v2.0, the noise was simulated using a style transfer method. This method was originally developed by Gatys et al. (2015) to create an image painted in the style of an artist. The idea is to extract the correlated noise in the real data (treated as the ‘style’ by this approach) and to apply it to the simulated SWOT data (considered as the ‘content’). During the style transfer, a real editing mask is applied to improve robustness of the network during inference. A training dataset is generated using this method over the North Atlantic Ocean (Figure B1).

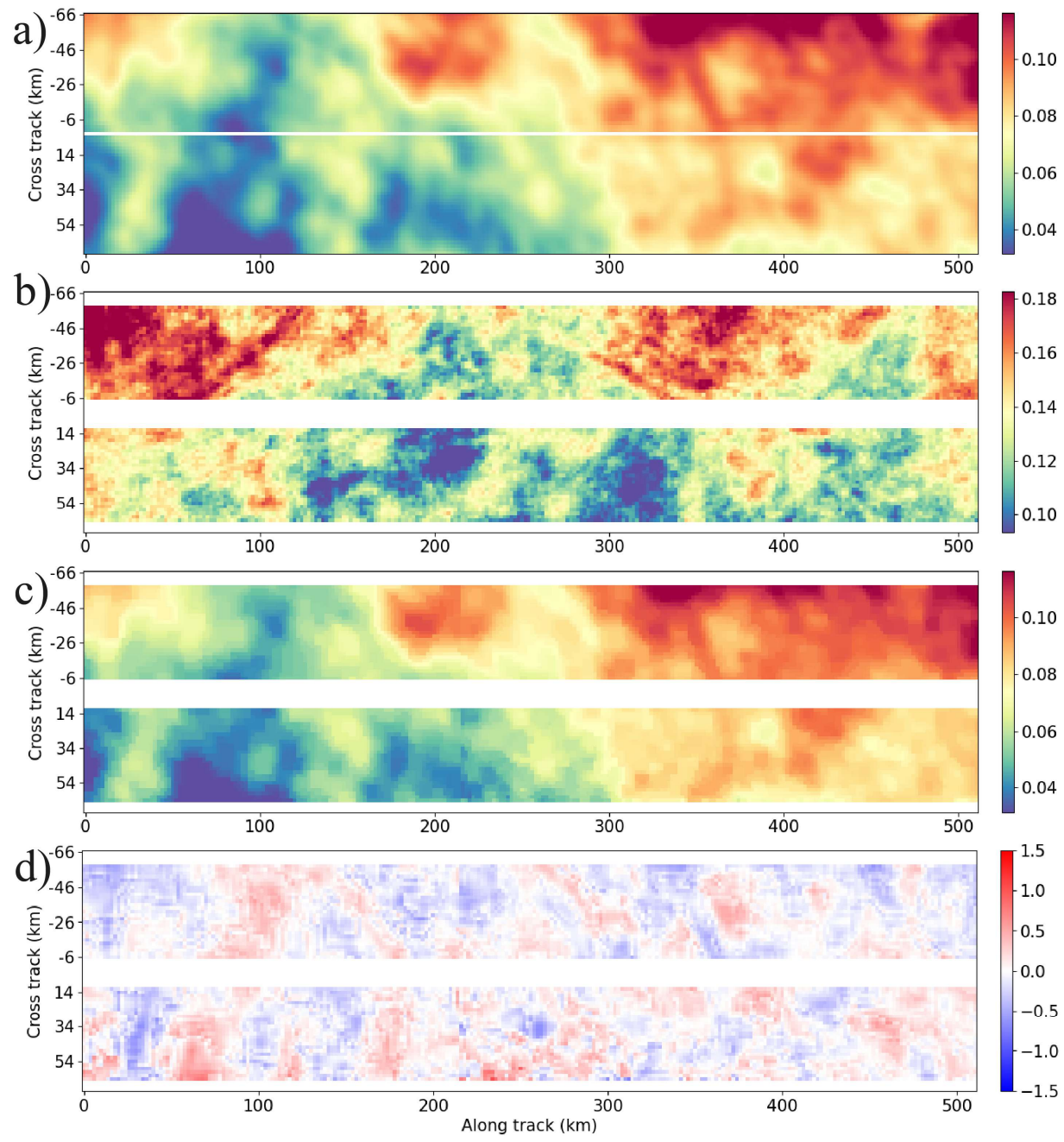

Figure B1. Example of style transfer on simulated SSHA for U-Net v2.0 training procedure. (a) Simulated data ("Context", unit: m). (b) Real data ("Style", unit: m). (c) Result of the style-transfer (a + b, unit: m). (d) Noise added by style-transfer.

*a. Visual validation*

Under high-wave conditions (such as in the Antarctic Circumpolar Current) large biases are visible with U-Net v2.0. These biases can reach several centimeters at swath borders (blue and red patterns in Figure B2c). False structures are generated while real structures are strongly smoothed out. This problem also appears in westward edge currents (such as Gulf Stream, Kuroshio…). This artifact was due to the method used to generate simulated noise (in this case, style transfer). It was designed to approximate real SWOT KaRIn noise by mimicking its characteristics. However, this proxy of real KaRIn noise was selected from a random region and then applied to a simulated eNATL60 swath. This approach degrades the relationship between noise characteristics and oceanic conditions. For example, the process could add a high-wave specific noise to a simulated swath with very small waves or, conversely, add a very small noise to a simulated swath with high SWH (Significant Wave Height). This led the network to hallucinate when denoising real SWOT swaths in strong wave conditions. Moreover, these artifacts can lead to cross-track and along-track discontinuities that are even more visible in the velocity and relative vorticity fields. These poor predictions are greatly improved in v3.0. The hallucinations are avoided by abandoning the style transfer approach which proved too sensitive.

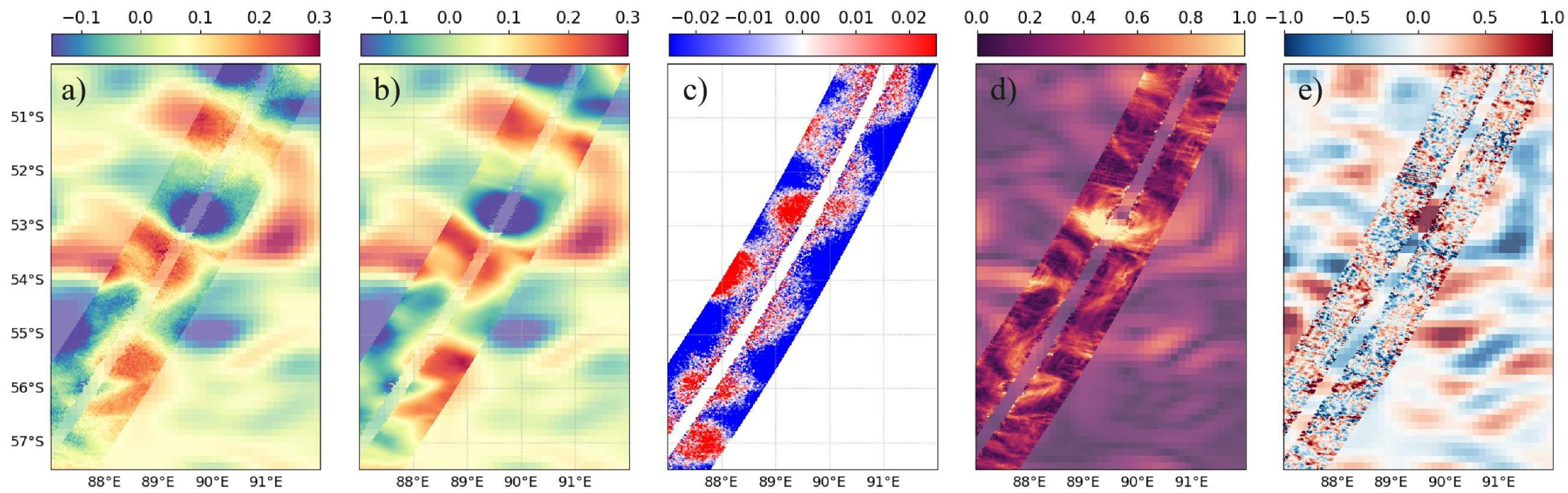

Figure B2. Denoising example of U-Net v2.0 filter in presence of strong noise in the ACC region. Cycle 20, Pass 299. (a) SSHA without the denoising filter (unit: m). (b) SSHA with the denoising filter (unit: m). (c) Removed SSHA content (unit: m). (d) Velocity (unit: m.s$^{-1}$). (e) Relative vorticity.

The U-Net v2.0 also absorbs more internal waves, destroying part of the wavefronts (Figure B3).

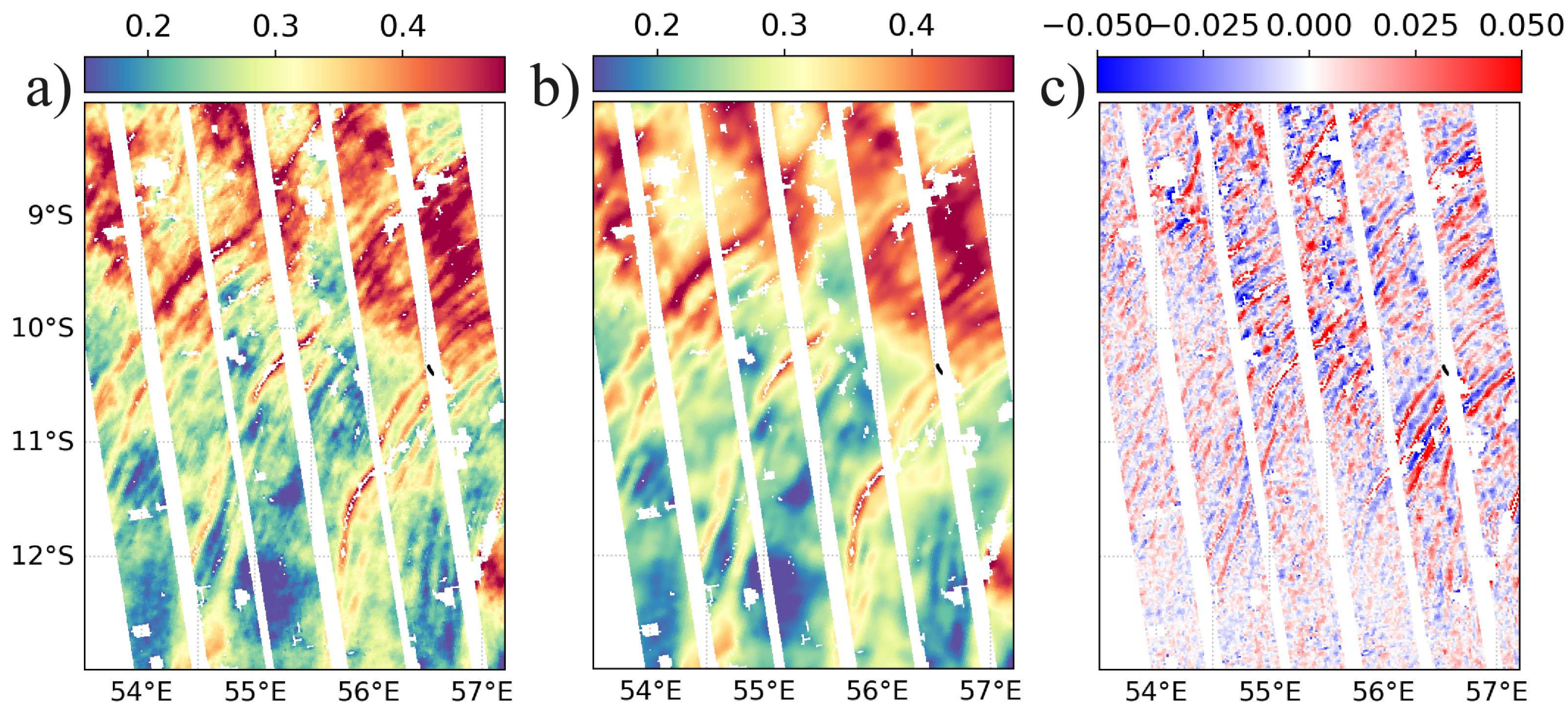

Figure B3. Filtering example using U-Net v2.0 on internal tides near Madagascar, Cycle 13, Passes 122,400 and 428. (a) SSHA without the denoising filter (unit: m). (b) SSHA with the denoising filter (unit: m). (c) Removed SSHA content (unit: m).

*b. Removed SSHA content binned maps*

Above 60 km (Figure B4a), the hallucinations of the U-Net v2.0 detailed in the previous Section are so strong (>1cm) and sufficiently large-scale to be visible above 60 km. The dark oceanic regions in the binned map are a clear illustration of the issue. Between 60 km and 20 km (Figure B4b), the map is homogeneous, similarly to the U-Net v3.0 or the Lanczos smoother. Below 20 km (Figure B4c), there are no clear differences from the other filters.

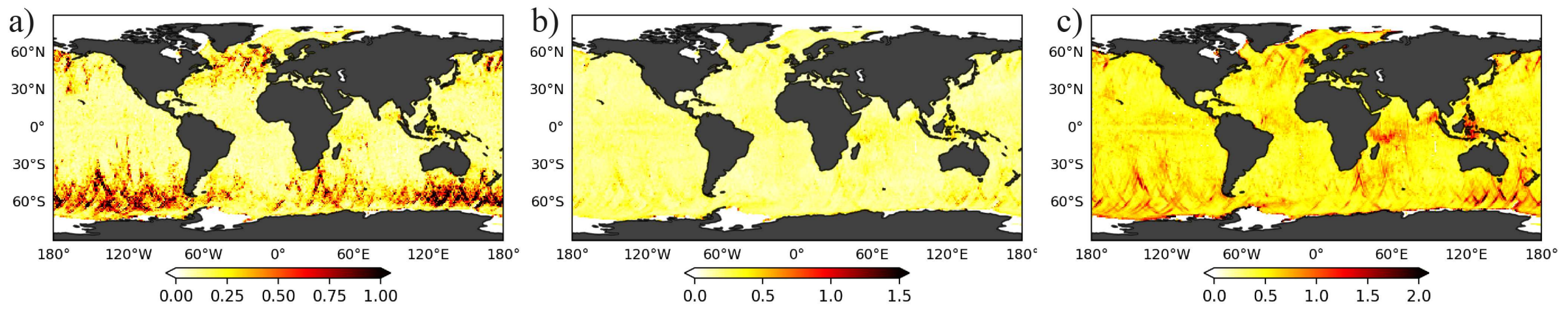

Figure B4. Binning maps of standard deviation of content removed by U-Net v2.0 algorithm on 3 ranges of wavelength (unit: cm). Std of removed content by the denoising filter for wavelengths: above 60-km (a), between 60 and 20-km (b), below 20-km (c).

*c. SSHA Power Spectral Densities (PSD)*

The PSD of SSHA denoised by U-Net v2.0 (purple curve) does not follow the slope break and almost perfectly fits the initial $K^{-3.7}$ slope of the KaRIn observations PSD down to 15 km (Figure B5a). If the hypothesis about the causes of the slope break is correct, this would indicate that U-Net v2.0 either smooths out unbalanced motions or red-colored KaRIn errors.

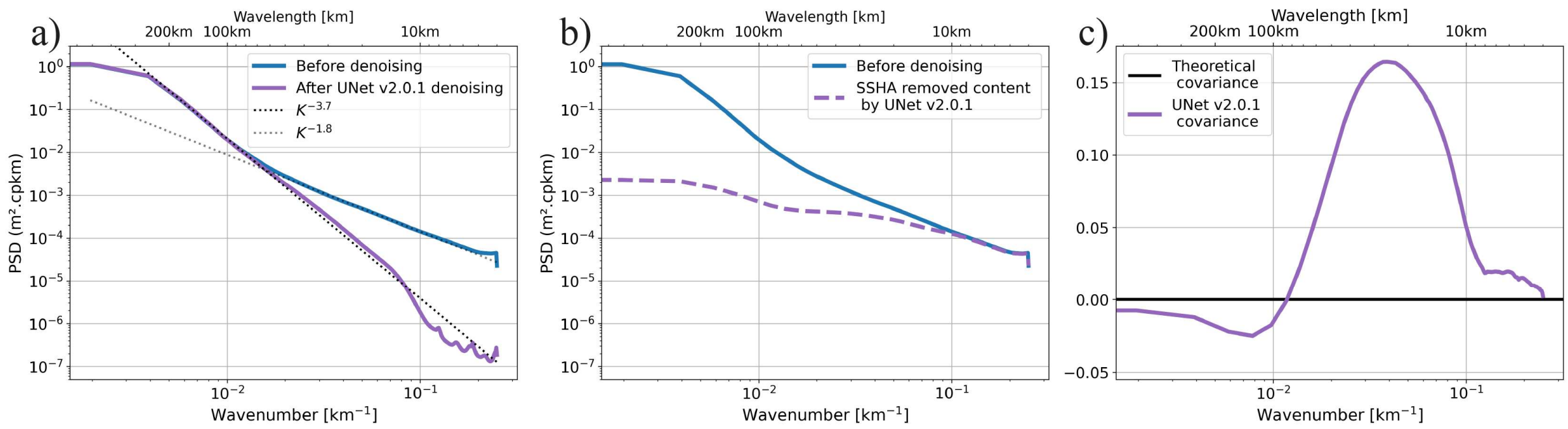

Figure B5. Global Power Spectral Densities of SSHA. (a) PSD of noisy SSHA in blue, PSD of denoised SSHA by UNet v2.0 in purple (in $m^2.cpkm^{-1}$). (b) PSD of noisy SSHA in blue, PSD of removed content by UNet v2.0 in purple (in $m^2.cpkm^{-1}$). (c) Covariance of U-Net v2.0 between denoised SSHA and the removed content.

As highlighted by Dibarboure et al. (2025), the PSD of content absorbed by U-Net v2.0 seems to fit a $K^{-0.5}$ slope (Figure B5b). They proposed two potential causes. The first is the absorption/creation of oceanic features at these scales. The second is the mitigation of red-colored KaRIn errors such as wet troposphere, atmosphere or sea-state residuals. This excess energy above 100 km of U-Net v2.0 compared to the PSD of removed SSHA content of U-Net v3.0 comes from hallucinations in the previous denoising version. When the network is making poor predictions in high-wave conditions, it either absorbs substantial signal energy or amplifies existing mesoscale oceanic structures (e.g. Figure B2). Both kinds of hallucinations inflate the removed content of the network.

The U-Net v2.0 denoising filter shows negative covariance until 80 km (Figure B5c). The denoising algorithm is inventing large-scale biases. This issue no longer exists with U-Net v3.0.

Below 80 km, U-Net v2.0 also absorbs a significant portion of KaRIn energy. This positive covariance at lower scales could hide local hallucinations and artifacts of the network.

The one-day difference PSD is already calculated by Dibarboure et al. (2025). They evaluated that 30% of the variance is missing in the 24 h difference residuals PSD. This is much higher than what was found with U-Net v3.0 (around 10%). The U-Net v2.0 filter absorbs more small-scale MSS error residuals. These residuals were significantly more present in MSS of v2.0 products.

*d. Local Effective Cut-off Wavelength (LECW)*

The U-Net v2.0 denoising filter absorbs less SSHA content in major western boundary currents and in the Mediterranean Sea with an LECW around 15 km. High-wave regions, where noise is known stronger, exhibit much higher LECW values (Figure B6). The average LECW of U-Net v2.0 is 22 km. This metric does not reveal any clear issues.

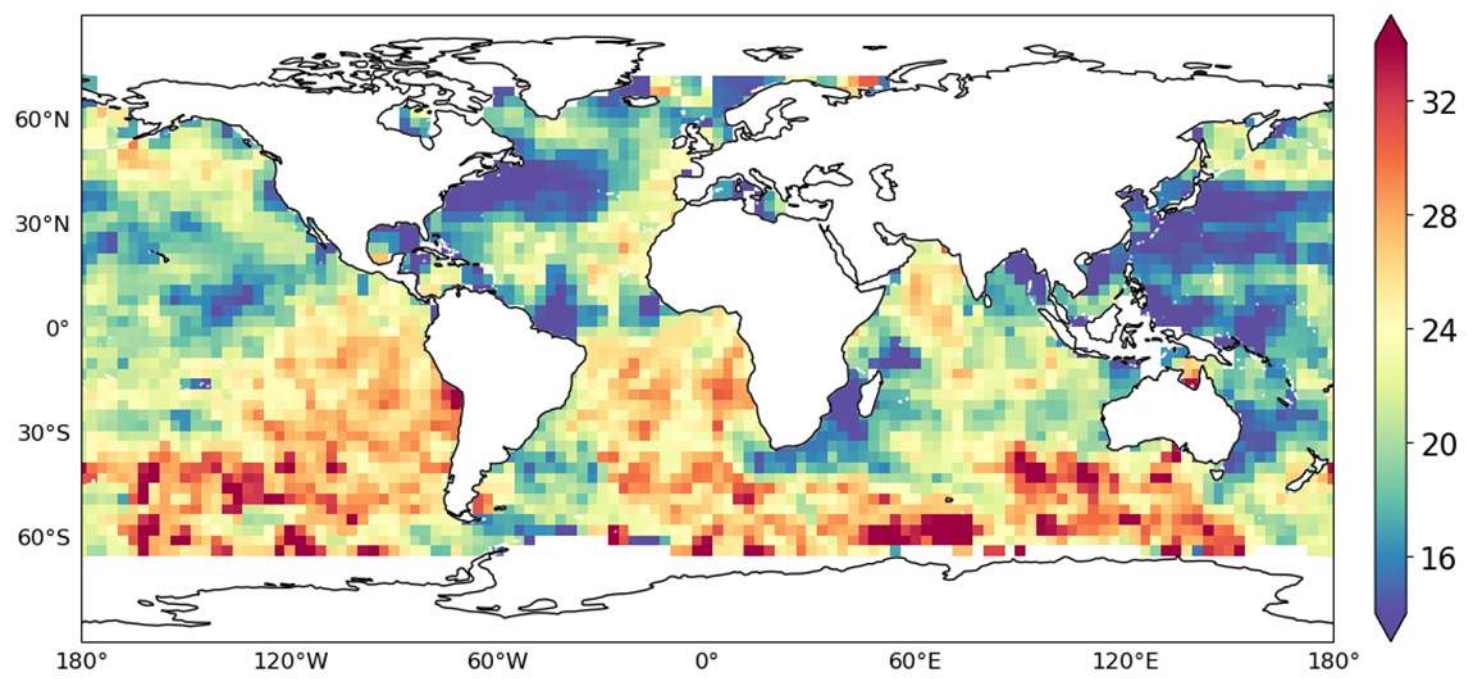

Figure B6. LECW of U-Net v2.0, Cycle 17 (unit: km)

*e. One-day regional SSIM*

For cycle 518 and 530, the U-Net v2.0 SSIM score of the denoised swaths is lower than for the undenoised ones (Figure B7a). This is one of the consequences of the network hallucinations in high-SWH regions. The algorithm degrades the coherence between consecutive SWOT measurements more than noise itself does. As shown in Figure B7b, with U-Net v2.0 (purple curve), sometimes the SSIM reaches 0 and sometimes much higher values. This is again explained by the false generated features on this segment of SSHA. Because the network can hallucinate completely differently between two consecutive swaths, there is no temporal consistency in what is removed and the SSIM can therefore reach very low values. When hallucinations are aligned from one day to the next, SSIM rises (e.g. cycle 516 or 534). The SSIM of velocity are slightly less consistent compared to the Lanczos smoother (Figure B7c). This is very likely due to the network poor predictions.

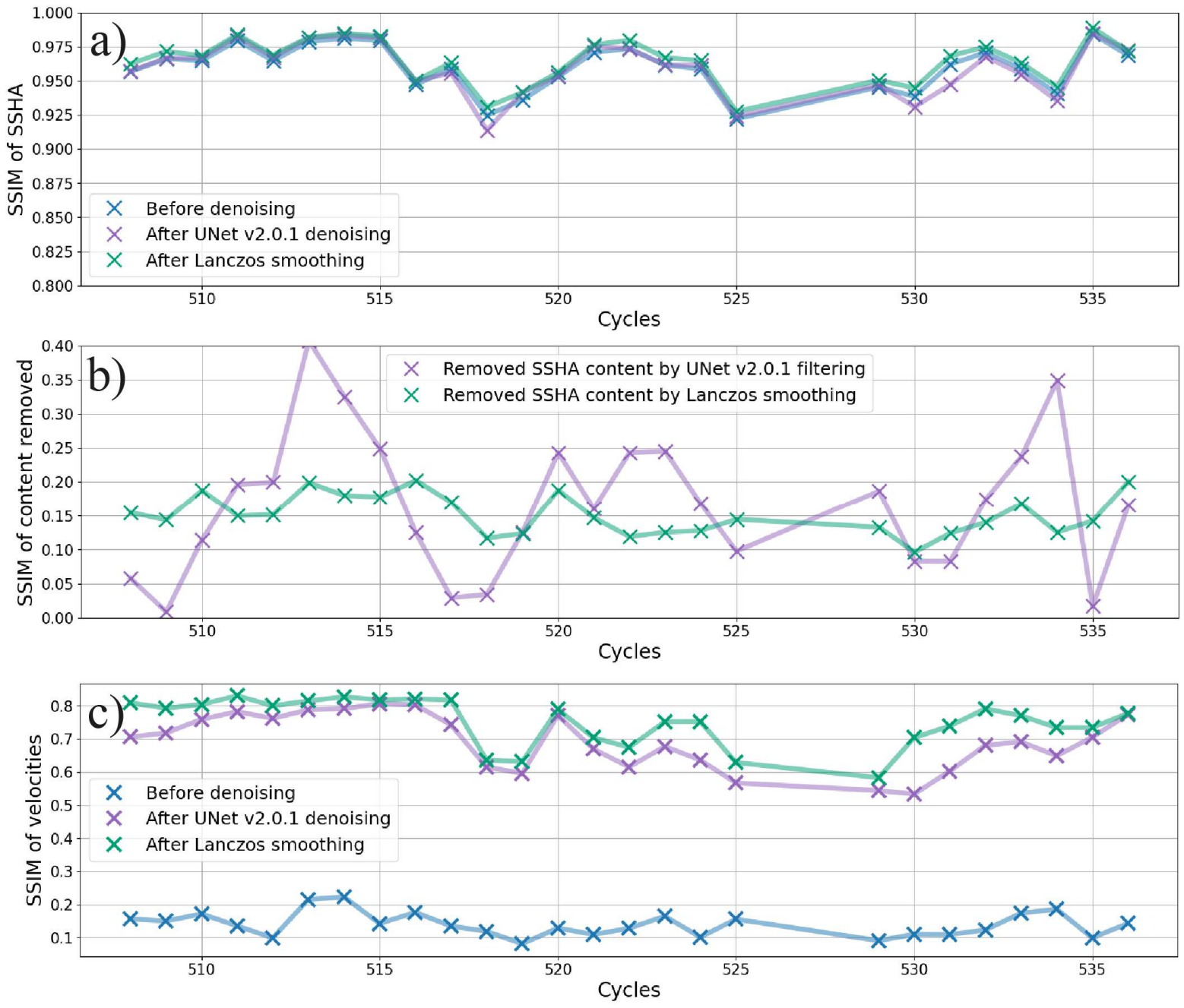

Figure B7. Local SSIM, Lanczos smoothing filter in green, the U-Net v2.0 denoising filter in purple. (a) Local SSIM between consecutive SSHA swaths. Noisy SSHA in blue (b) Local SSIM between consecutive removed content swaths. (c) Local SSIM between consecutive velocity swaths.

### *f. Mean Profile of SSHA removed content*

U-Net v2.0 also absorbs small-scale MSS residuals. The presence of biases in the mean profile of removed SSHA content implies that the U-Net v2.0 filter is removing part of the real oceanic signal (e.g. the blue pattern that crosses the swath near 36°N in Figure B8). It could be a recurrent absorption/creation of signal each cycle or a very strong isolated event that it is still visible in the mean. In each case, this must be avoided. The U-Net v2.0 filter is clearly removing more than just MSS residues.

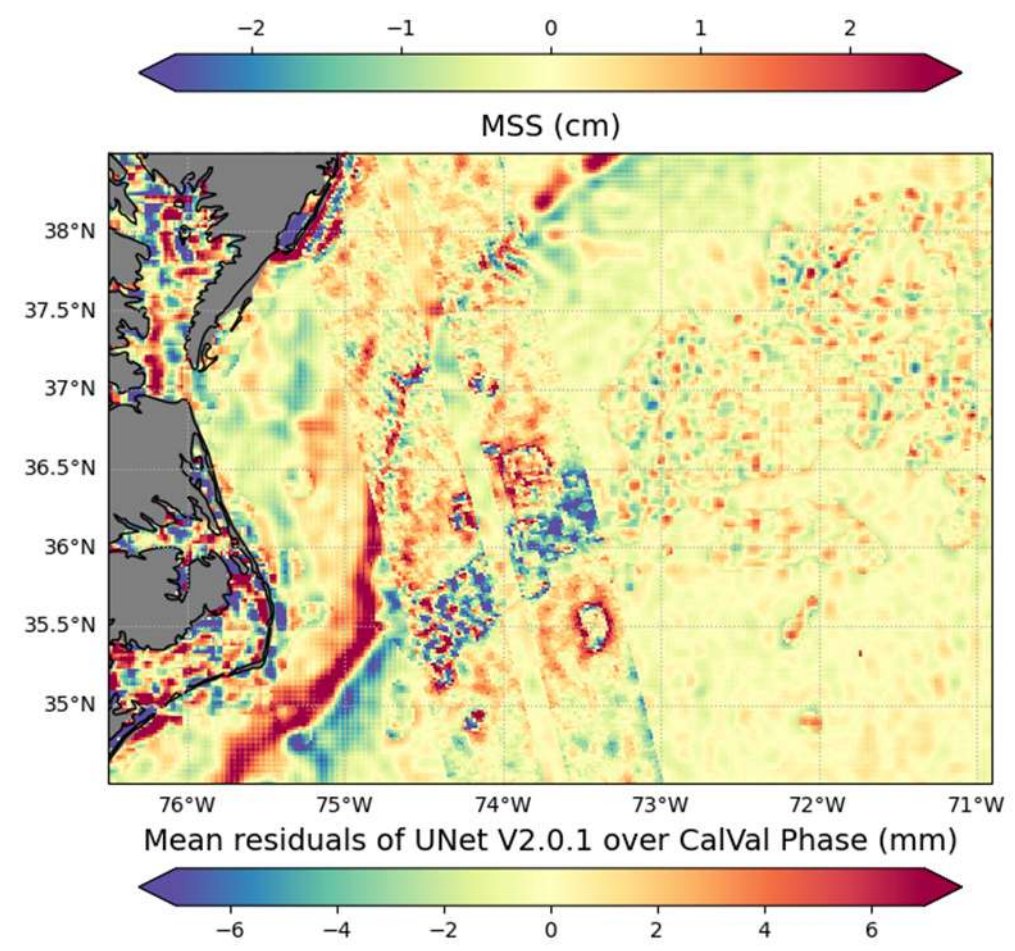

Figure B8. Mean of residuals of U-Net v2.0 over the one-day repeat phase above high-frequencies of MSS.

*g. SWOT KaRIn/S3A crossovers*

For U-Net v2.0, ocean regions with high waves again appear in red pixels, meaning that the large-scale network hallucinations reduced the coherence between SWOT and S3A (Figure B9). U-Net v2.0 demonstrates lower agreement with S3A than the new denoiser or the basic smoother.

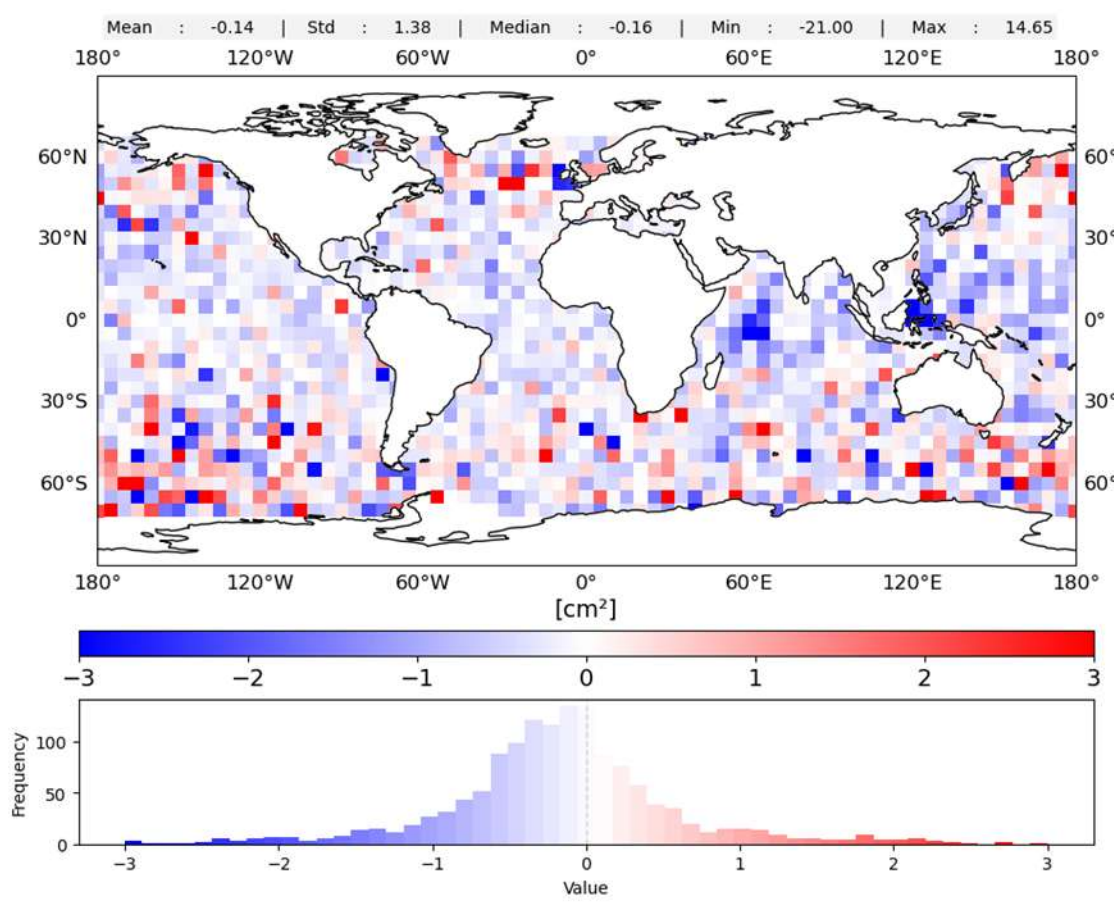

Figure B9. KaRIn/S3A coherence gain from the U-Net v2.0 denoising filter.